\documentclass[twocolumn,traditabstract,longauth]{aa}
\usepackage{amssymb}
\usepackage{mathptmx}
\usepackage{natbib}
\bibpunct{(}{)}{;}{a}{}{,}
\usepackage[]{graphicx}
\usepackage{lscape}
\usepackage{txfonts}
\usepackage{verbatim}
\usepackage{url}
\defcitealias{nesvadba18}{N18}
\defcitealias{canameras15}{C15}
\defcitealias{canameras17}{C17}
\begin{document}

\title{\textit{Planck}'s dusty GEMS. VI. Multi-\textit{J} CO excitation and interstellar medium conditions in dusty starburst galaxies at 
\textit{z}~=~2--4 \thanks{Based on IRAM data obtained with programs 082-12, D09-12, 065-13, 094-13, 223-13, 108-14, and 217-14.}}

\author{R.~Ca\~nameras\inst{1}, C.~Yang\inst{2}, N.~P.~H.~Nesvadba\inst{3}, A.~Beelen\inst{3}, R.~Kneissl\inst{2,4}, 
S.~Koenig\inst{5}, E.~Le~Floc'h\inst{6}, M.~Limousin\inst{7}, S.~Malhotra\inst{8}, A.~Omont\inst{9}, D.~Scott\inst{10}}

\institute{
Dark Cosmology Centre, Niels Bohr Institute, University of Copenhagen, Juliane Maries Vej 30, DK-2100 Copenhagen, Denmark \\
{\tt e-mail: canameras@dark-cosmology.dk}
\and
European Southern Observatory, ESO Vitacura, Alonso de Cordova 3107, Vitacura, Casilla 19001 Santiago, Chile
\and
Institut d'Astrophysique Spatiale, CNRS, univ. Paris-Sud, Universit\'e Paris-Saclay, B\^at. 121, 91405 Orsay France
\and
Atacama Large Millimeter/submillimeter Array, ALMA Santiago Central Offices, Alonso de Cordova 3107, Vitacura, Casilla 
763-0355, Santiago, Chile
\and
Chalmers University of Technology, Onsala Space Observatory, Onsala, Sweden
\and
Laboratoire AIM, CEA/DSM/IRFU, CNRS, Universit\'e Paris-Diderot, B\^at. 709, 91191 Gif-sur-Yvette, France
\and
Aix-Marseille Universit\'e, CNRS, CNES, LAM, Marseille, France
\and
School of Earth and Space Exploration, Arizona State University, Tempe, AZ 85287, USA
\and
UPMC Universit\'e Paris 06, UMR 7095, Institut d'Astrophysique de Paris, 75014 Paris, France
\and
Department of Physics and Astronomy, University of British Columbia, 6224 Agricultural Road, Vancouver, 6658 British 
Columbia, Canada}

\titlerunning{\textit{Planck}'s dusty GEMS. VI. Molecular gas properties}

\authorrunning{R. Ca\~nameras et al.} \date{Received / Accepted}

\abstract{We present an extensive CO emission-line survey of the {\it Planck}'s dusty Gravitationally Enhanced subMillimetre 
Sources, a small set of 11 strongly lensed dusty star-forming galaxies at $z = 2$--4 discovered with {\it Planck} and {\it Herschel} 
satellites, using EMIR on the IRAM 30-m telescope. We detected a total of 45 CO rotational lines from $J_{\rm up}=3$ to $J_{\rm up}=11$, 
and up to eight transitions per source, allowing a detailed analysis of the gas excitation and interstellar medium conditions within these 
extremely bright ($\mu L_{\rm FIR} = 0.5$--$3.0 \times 10^{14}$~L$_{\odot}$), vigorous starbursts. The peak of the CO 
spectral-line energy distributions (SLEDs) fall between $J_{\rm up}=4$ and $J_{\rm up}=7$ for nine out of 11 sources, in the same range as
other lensed and unlensed submillimeter galaxies (SMGs) and the inner regions of local starbursts. We applied radiative transfer models 
using the large velocity gradient approach to infer the spatially-averaged molecular gas densities, 
$n_{\rm H_2} \simeq 10^{2.6}$--$10^{4.1}$~cm$^{-3}$, and kinetic temperatures, $T_{\rm k} \simeq 30$--1000~K. In five sources, we find evidence 
of two distinct gas phases with different properties and model their CO SLED with two excitation components. The warm (70--320~K) and 
dense gas reservoirs in these galaxies are highly excited, while the cooler (15--60~K) and more extended low-excitation components 
cover a range of gas densities. In two sources, the latter is associated with diffuse Milky Way-like gas phases of density
$n_{\rm H_2} \simeq 10^{2.4}$--$10^{2.8}$~cm$^{-3}$, which provides evidence that a significant fraction of the total gas masses of dusty 
starburst galaxies can be embedded in cool, low-density reservoirs. The delensed masses of the warm star-forming molecular gas 
range from 0.6 to $12 \times 10^{10}$~M$_{\odot}$. Finally, we show that the CO line luminosity ratios are consistent with those predicted 
by models of photon-dominated regions (PDRs) and disfavor scenarios of gas clouds irradiated by intense X-ray fields from active galactic 
nuclei. By combining CO, [\ion{C}{i}] and [\ion{C}{ii}] line diagnostics, we obtain average PDR gas densities significantly higher 
than in normal star-forming galaxies at low-redshift, as well as far-ultraviolet radiation fields 10$^2$ to 10$^4$ times more intense 
than in the Milky Way. These spatially-averaged conditions are consistent with those in high-redshift SMGs and in a range of low-redshift 
environments, from the central regions of ultra-luminous infrared galaxies and bluer starbursts to Galactic giant molecular clouds.}


\maketitle

\section{Introduction}
\label{sec:intro}

Massive, dusty, star-forming galaxies at high-redshift exhibit enhanced star-formation rates \citep[between a few hundred and 
1000~M$_{\odot}$~yr$^{-1}$, for example,][]{casey14} compared to those measured in low-redshift nuclear ultra-luminous infrared 
galaxies (ULIRGs), and account for significant fractions of the cosmic energy budget from star formation at $z \sim 2$--4 
\citep[e.g.,][]{hauser01,magnelli13,dunlop17}. Their intense star formation is driven by deep gravitational potentials and
higher gas masses \citep[e.g.,][]{tacconi08,ivison11}, gas fractions \citep[e.g.,][]{daddi10,tacconi10}, and gas mass surface 
densities \citep[e.g.,][]{riechers13,canameras17b} than in the local Universe. The bulk of their molecular gas reservoirs 
directly fueling star formation are dense and generally highly turbulent \citep{lehnert09,solomon05,carilli13}. Determining 
the physical conditions of the molecular gas and overall interstellar medium (ISM) within these galaxies is crucial to 
understand the timescales over which their extreme star-formation activity can be sustained, and to characterize the feedback 
processes underlying the most rapid phase of stellar mass assembly.

The gas mass census of high-redshift star-forming galaxies has been mostly obtained from observations of carbon monoxide 
(CO) line emission, since CO is the most abundant molecule in the interstellar medium after H$_{\rm 2}$ and emits rotational 
transitions observable with space- and ground-based (sub)millimeter receivers. CO molecules form within photon-dominated 
regions (PDRs), in the outer layers of star-forming clouds between \ion{H}{ii} regions and the inner prestellar cores, 
where the incident far-ultraviolet (FUV) radiation fields from young stellar populations are sufficiently attenuated, and 
hydrogen is mainly in molecular form \citep{hollenbach97}. Combining CO line diagnostics and PDR models therefore enables us
to probe the density and radiation fields in these environments \citep{kaufman99,lepetit06,rawle14}. Moreover, observing the 
CO spectral line energy distributions (SLEDs) allows us to characterize the population of multiple rotational levels and to 
infer the underlying gas density and kinetic temperature using radiative transfer models \citep[e.g.,][]{weiss05,papadopoulos10}.

Deriving the total molecular gas masses from CO relies on measurements of the ground-state rotational transition, together 
with the choice of a CO-to-H$_{\rm 2}$ conversion factor, $\alpha_{\rm CO}$ \citep[see][for a review]{bolatto13}. At high redshift,
detecting CO(1--0) is challenging due to its low fluxes and redshifted frequencies and the measured line fluxes can be 
significantly underestimated due to the cosmic microwave background radiation \citep{dacunha13,zhang16}, so the overall gas 
properties are often deduced from higher-$J$ CO transitions. However, the conversion factors between mid-$J$ and CO(1--0) are 
highly uncertain, due to the wide range of gas excitations over the SMG population \citep{carilli13,sharon16}. Early studies 
assumed local thermodynamic equilibrium, but subthermally excited gas reservoirs were subsequently observed in several SMGs 
\citep[e.g.,][]{hainline06}. More recently, \citet{narayanan14} have suggested to use the star-formation surface density as a 
proxy of the overall excitation, but also show that this quantity poorly describes the gas conditions in the most intense 
starbursts. Given these conversion uncertainties, modeling individual CO SLEDs becomes mandatory to obtain robust mass estimates 
for the different gas phases in dusty star-forming galaxies, and to better constrain the high-redshift mode of star formation.

In the local Universe, similar studies have highlighted the presence of multiple gas components with a range of densities 
and temperatures in normal star-forming galaxies \citep[e.g.,][]{liu15}, LIRGs \citep[e.g.,][]{lu17}, and ULIRGs 
\citep[e.g.,][]{pearson16}. These gas phases have been detected in the most massive galaxies at $z \sim 1$--3 
\citep{ivison10,harris10}, with also hints of diffuse reservoirs that could be distributed over large scales and not directly 
related to the on-going star formation \citep{dannerbauer09,danielson11}. The presence of such low density ISM components 
could have important implications for our understanding of the stellar mass build-up of massive galaxies, but evidence is 
still scarce because even small mass fractions of warm, excited gas are sufficient to dominate the spatially-integrated 
luminosities \citep[see, e.g.,][]{rangwala11}.

Probing the ISM conditions within high-redshift dusty star-forming galaxies using detailed analyses of the CO gas excitation 
is greatly facilitated through observing the most strongly lensed systems \citep[e.g.,][]{weiss07,cox11}, for which the magnified 
emission from intrinsically faint high-$J$ CO lines beyond the turnover of the SLED exceeds the sensitivity limit of single-dish 
telescopes and interferometers \citep[e.g.,][]{yang17}. The flux boost due to gravitational lensing is crucial for obtaining line 
detections with sufficient signal-to-noise ratios, for a wide range of gas excitations, and for intrinsically fainter SMGs than 
those from unlensed samples. This allows us, for example, to investigate the fundamental differences between galaxies in the 
starburst and steady disk modes of star formation.

\begin{table*}
\centering
\begin{tabular}{lcccccccc}
\hline
\hline
Source & RA & Dec & $z_{\rm spec}$ & $\mu_{\rm dust}$ & $\mu_{\rm gas}$ & $\Delta \mu / \mu$ & $L_{\rm FIR}$ & $r_{\rm 1/2}$ \\
 & (J2000) & (J2000) & & & & & [10$^{12}$ L$_{\odot}$] & [kpc] \\
\hline
PLCK\_G045.1+61.1 & 15:02:36.04 & +29:20:51 & 3.4286 $\pm$ 0.0003 & 19.1 $\pm$ 1.2 $^a$ & 15.5 $\pm$ 0.7 & $<10$\% & 4.4 $\pm$ 0.3 & ... \\
PLCK\_G080.2+49.8 & 15:44:33.25 & +50:23:45 & 2.5984 $\pm$ 0.0001 & 14.7 $\pm$ 0.8 & 15.9 $\pm$ 1.5 & $<30$\% & 3.1 $\pm$ 0.2 & ... \\
PLCK\_G092.5+42.9 & 16:09:17.76 & +60:45:21 & 3.2557 $\pm$ 0.0002 & 15.4 $\pm$ 1.0 & 12.0 $\pm$ 0.6 & $<10$\% & 16.1 $\pm$ 1.2 & ... \\
PLCK\_G102.1+53.6 & 14:29:17.98 & +59:21:09 & 2.9170 $\pm$ 0.0003 & 7.1 $\pm$ 0.2 & 6.9 $\pm$ 0.3 & $<5$\% & 5.0 $\pm$ 0.3 & ... \\
PLCK\_G113.7+61.0 & 13:23:02.88 & +55:36:01 & 2.4166 $\pm$ 0.0002 & 11.2 $\pm$ 0.7 & 9.7 $\pm$ 0.5 & $<5$\% & 8.8 $\pm$ 0.7 & ... \\
PLCK\_G138.6+62.0 & 12:02:07.68 & +53:34:40 & 2.4420 $\pm$ 0.0002 & $\sim$20 $^b$ & $\sim$20 $^b$ & ... & 4.5 $\pm$ 2.3 & ... \\
PLCK\_G145.2+50.9 & 10:53:22.56 & +60:51:49 & 3.5487 $\pm$ 0.0003 & 7.6 $\pm$ 0.5 & 8.9 $\pm$ 0.5 & $<10$\% & 28.7 $\pm$ 2.2 & ... \\
PLCK\_G165.7+67.0 & 11:27:14.60 & +42:28:25 & 2.2362 $\pm$ 0.0003 & 29.4 $\pm$ 5.9 & 24.1 $\pm$ 4.8 & $<30$\% & 3.5 $\pm$ 0.7 & 1.4 \\
PLCK\_G200.6+46.1 & 09:32:23.67 & +27:25:00 & 2.9726 $\pm$ 0.0004 & $\sim$15 $^b$ & $\sim$15 $^b$ & ... & 3.8 $\pm$ 1.3 & ... \\
PLCK\_G231.3+72.2 & 11:39:21.60 & +20:24:53 & 2.8589 $\pm$ 0.0003 & 7.9 $\pm$ 0.3 & 6.0 $\pm$ 0.5 & $<10$\% & 9.5 $\pm$ 0.5 & ... \\
PLCK\_G244.8+54.9 & 10:53:53.04 & +05:56:21 & 3.0054 $\pm$ 0.0001 & 21.8 $\pm$ 0.6 $^c$ & 22.0 $\pm$ 1.3 $^c$ & $<5$\% & 12.2 $\pm$ 0.4 & 0.6 \\
\hline
\end{tabular}
\caption{Properties of the {\it Planck}'s dusty GEMS that were part of our CO emission-line survey with EMIR. The source 
positions are those used to point the IRAM 30-m telescope for the single-dish spectroscopic observations. Here $z_{\rm spec}$ is 
the best spectroscopic redshift inferred from single Gaussian fits to the lowest-$J$ CO emission lines detected with EMIR,
between $J_{\rm up}=3$ and $J_{\rm up}=5$. The quantity $\mu_{\rm dust}$ is the luminosity-weighted magnification factor of the
880-$\mu$m dust continuum, mostly obtained from the highest resolution SMA maps, either in the EXT or VEXT configuration. 
$\mu_{\rm gas}$ is the luminosity-weighted magnification factor of the gas component in the GEMS derived from our CO $J_{\rm up}=4$--6 
interferometry using IRAM/PdBI. Both factors are inferred from detailed strong lensing models that will be presented in a 
forthcoming paper (Ca\~nameras et al. 2018c, in prep.), with errors showing their $\pm 1\sigma$ statistical uncertainties. 
Expected systematics on the magnification factors are listed in the column headed $\Delta \mu / \mu$ (see further details in the 
text). The quantity $L_{\rm FIR}$ is the delensed value of the total FIR luminosity (8--1000~$\mu$m) presented in C15, including 
statistical uncertainties on $\mu$. Finally, $r_{\rm 1/2}$ is the intrinsic half-light radius of the dust continuum from the source.
{\bf Notes.} \tablefoottext{a}{Computed from the ALMA 0.7-mm continuum map at 0.42\arcsec~$\times$~0.28\arcsec\ beam size 
presented in \citet{nesvadba16}.}
\tablefoottext{b}{Rough estimate of $\mu$ from the scaling relation between $L_{\rm FIR}$ and $T_{\rm dust}$ in unlensed SMGs 
(see C15 for details), for the sources without detailed lensing models.}
\tablefoottext{c}{The magnification factors of the Ruby are inferred from the ALMA very extended baseline observations of 
the 3-mm dust continuum and CO(4--3) line emission presented in (Ca\~nameras et al., 2017a).}}
\label{tab:prop}
\end{table*}

Here we present IRAM 30-m telescope observations of several CO rotational lines in the {\it Planck}'s dusty Gravitationally 
Enhanced subMillimetre Sources \citep[GEMS,][hereafter C15]{canameras15}, a small set of extremely bright strongly lensed dusty 
star-forming galaxies identified with {\it Planck}. This sample was obtained from the {\it Herschel}/SPIRE 250-, 350- and 
500-$\mu$m follow-up photometry of high-redshift source candidates identified with {\it Planck}/HFI over the 50\% of the sky 
with the lowest contamination from Galactic cirrus \citep[see further details in][]{planck15}. The {\it Planck}'s dusty GEMS 
were then selected as the brightest isolated point sources in SPIRE maps, with spectral energy distribution (SED) peaking 
either at 350 or 500~$\mu$m and with 350-$\mu$m flux density above 400~mJy, following the predicted submm number counts of 
\citet{negrello07,negrello10}. Given the unprecedented sky coverage of the original sample identified with {\it Planck}, 
the resulting 11 sources are amongst the brightest high-redshift dusty star-forming galaxies on the sky outside of the Galactic 
plane. Their 350-$\mu$m flux densities lie between 294~mJy and 1054~mJy, thereby extending other samples of strongly lensed 
SMGs from {\it Herschel} and the South-Pole Telescope toward higher fluxes \citep[e.g.,][]{vieira13,wardlow13,bussmann13}. 
The overall dust heating in these sources is dominated by intense star formation, with minor contributions from central active 
galactic nuclei (AGN) to the overall FIR luminosities \citepalias{canameras15}.

All {\it Planck}'s dusty GEMS are directly observable from the northern hemisphere, which allowed us to measure robust 
spectroscopic redshifts between $z=2.236$ and $z=3.549$ with the EMIR broadband receiver on the IRAM 30-m telescope 
\citepalias{canameras15}. Dust continuum interferometry at 880~$\mu$m with the SMA previously revealed that the GEMS are either 
single or multiple compact sources or elongated arcs at subarcsec resolution, and that they are systematically aligned with 
foreground mass concentrations detected with optical and near-infrared imaging \citep[\citetalias{canameras15};][]{frye18}. We 
published detailed lensing models of individual GEMS as part of our overall follow-up program. Firstly, in \citet{canameras17b}, 
the analysis of ALMA 0.1\arcsec$-$resolution observations of PLCK\_G244.8+54.9 (hereafter the ``Ruby'') showed that the submm 
emission forms a nearly complete Einstein ring of 1.4\arcsec\ diameter around a red massive foreground galaxy detected with 
{\it HST}/WFC3. VLT/X-Shooter spectroscopy revealed that the main deflector is one of the most distant strong lensing galaxies 
known to date, at $z=1.525$. Secondly, in Ca\~nameras et al. (2018a, A\&A accepted), we have extensively studied the 
foreground mass distribution of PLCK\_G165.7+67.0 (the ``Emerald'') which forms a giant submm arc at $z=2.236$, behind two 
small groups of galaxies embedded within a massive cluster at $z=0.351$. Similarly, we derived lensing models for seven of the 
nine remaining sources in the sample, based on our SMA dust continuum and PdBI subarcsec line interferometry of the GEMS, and 
optical and near-infrared imaging and spectroscopy of the foreground deflectors (see Table~\ref{tab:prop}). In addition to 
providing delensed source properties, these models are crucial for constraining the impact of possible differential lensing 
effects on the shape of the CO ladder and resulting gas properties. They will be further discussed in a forthcoming paper 
(Ca\~nameras et al. 2018c, in prep.).

After detecting two CO or [\ion{C}{i}] transitions per source as part of our redshift search in the 3-mm  band, we pursued 
the millimeter emission-line survey with EMIR in order to obtain further information on the gas conditions and gas heating 
mechanisms in these strongly gravitationally lensed high-redshift galaxies. A detailed analysis of the atomic gas and ISM 
properties deduced from the [\ion{C}{i}] ${\rm ^3P_1}$--${\rm ^3P_0}$ and ${\rm ^3P_2}$--${\rm ^3P_1}$ fine-structure lines 
are discussed in a companion paper (Nesvadba et al. 2018, A\&A submitted, hereafter \citetalias{nesvadba18}). Here we focus 
on the characterization of spatially-integrated molecular gas reservoirs over the entire sample, using our CO line survey 
and modeling of the CO ladder from $J_{\rm up}=3$ to $J_{\rm up}=11$.

Our paper is organized in the following way: We start with a description of the single-dish observations and data reduction 
in Sect.~\ref{sec:obs}, and characterize the overall properties of CO line emission in the sample in Sect.~\ref{sec:coprop}.
In Sect.~\ref{sec:prop} we describe the magnification factor estimates and discuss the possible differential lensing effects
and our choice of the CO-to-H$_{\rm 2}$ conversion factor. We present our analysis of the CO excitation and resulting ISM conditions 
in Sect.~\ref{ssec:lvg}, and characterize the number of gas components and their total masses in Sect.~\ref{ssec:mgas}. 
In Sect.~\ref{ssec:pdr}, we combine line diagnostics from different species to derive PDR models. We then discuss the ISM 
properties obtained with the two approaches, our constraints on the sizes of the gas reservoirs and other heating mechanisms 
in Sect.~\ref{sec:discu} and, finally, we conclude with a summary of our analysis in Sect.~\ref{sec:summary}.

Throughout the paper we adopt a flat $\Lambda$CDM cosmology with $H_0 = 67.8 \pm 0.9$~km~s$^{-1}$~Mpc$^{-1}$, 
$\Omega_{\rm M}=0.308\pm0.012$, and $\Omega_{\Lambda}=1-\Omega_{\rm M}$ \citep{planck16}. All observed transitions are rotational 
transitions of the $^{12}$CO isotope, which we refer to as ``CO''.

\section{Observations and data reduction}
\label{sec:obs}

%\subsection{IRAM 30-m EMIR spectroscopy}
%\label{ssec:emir}

High-resolution spectra of the CO lines in the GEMS were obtained with the IRAM 30~meter telescope as part of an 
extended survey of the brightest atomic and molecular emission lines conducted in the 3-mm, 2-mm, 1.3-mm and 0.8-mm 
bands. Observations were carried out with the eight mixer receiver broadband instrument \citep[EMIR,][]{carter12}, during 
several programs between September 2012 and June 2014 (PI: N. Nesvadba, see Table~\ref{tab:obslog}). The blind redshift 
search in the 2-mm and 3-mm bands described in C15 was performed during programs 082-12, D09-12 and 094-13, while program 
223-13 was intended to extend our characterization of the CO ladder to higher-$J$. EMIR offers a total bandwidth of up to 
16-GHz separated into 4-GHz sidebands, in each of the two orthogonal polarizations. We used both the FTS or WILMA backends, 
covering 16 and 8~GHz of total bandwidth, with a spectral resolution of 195~kHz and 2~MHz, respectively. During the 
integration, the sky emission was subtracted using a wobbler switching of 60\arcsec, larger than the telescope primary 
beam (from 7.5\arcsec\ at 340~GHz to 29\arcsec\ at 86~GHz) and larger than the angular size of the sources. We 
calibrated the pointing every two hours using radio-loud quasars at small angular separation from the GEMS, to obtain an
rms pointing accuracy below 3\arcsec\ \citep{greve96}. We focused the telescope after significant variations in the 
atmospheric temperature including sunrise and sunset, and every 4~hours in stable conditions. The observations were 
carried out under good to average weather conditions using a standard calibration mode, and we flagged the scans taken 
at high precipitable water vapor (PWV~$>$~8--10~mm at 3~mm).

The scans were reduced using the {\tt CLASS} software package from the {\tt GILDAS}\footnote{\url{http://www.iram.fr/IRAMFR/GILDAS}} 
distribution delivered by IRAM. We inspected all individual 30-s scans in each polarization, and rejected those with 
spikes falling at the frequency of the line and those with poor baselines (most of them are also flagged as high PWV
scans). After subtracting the baselines in individual scans using first-order polynomials, we coadded and smoothed the 
spectra. The rms values of the resulting spectra were measured with {\tt CLASS} on baseline channels. The tuning 
frequencies, integration times and rms noise levels for each spectrum are summarized in Table~\ref{tab:obslog}. 
We used the telescope efficiencies measured during the relevant calibration campaign to convert the peak antenna 
temperatures, $T^*_{\rm a}$, to flux density units.

\begin{figure*}
\centering
\includegraphics[height=0.20\textwidth]{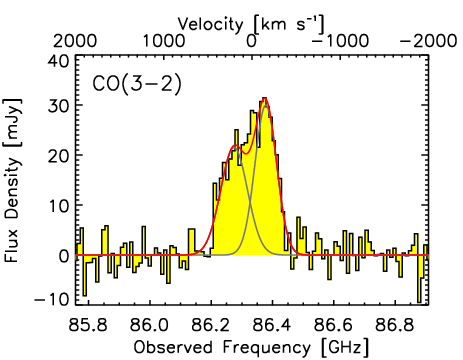}
\includegraphics[height=0.20\textwidth]{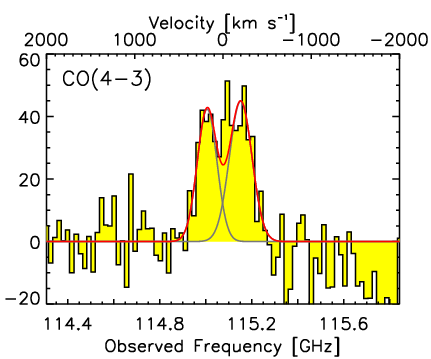}
\includegraphics[height=0.20\textwidth]{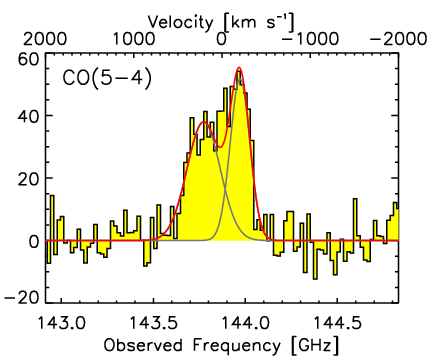} 
\includegraphics[height=0.20\textwidth]{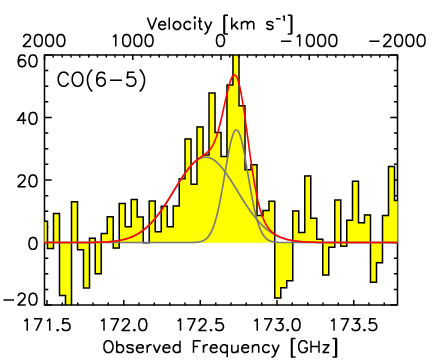}

\includegraphics[height=0.20\textwidth]{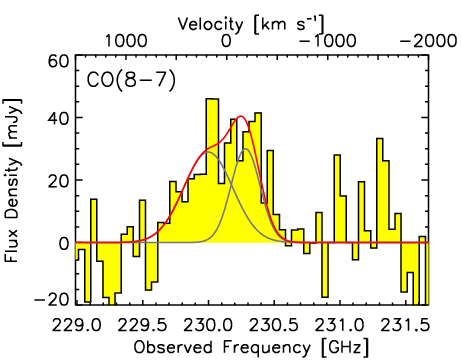}
\includegraphics[height=0.20\textwidth]{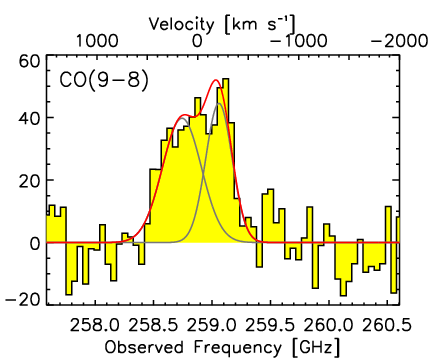}
\includegraphics[height=0.20\textwidth]{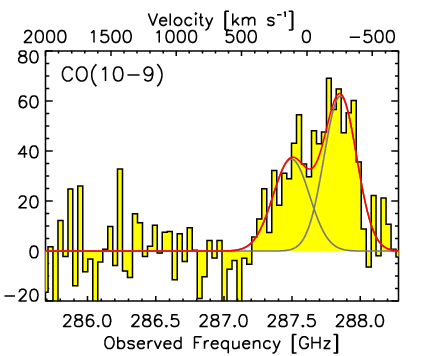}
\includegraphics[height=0.20\textwidth]{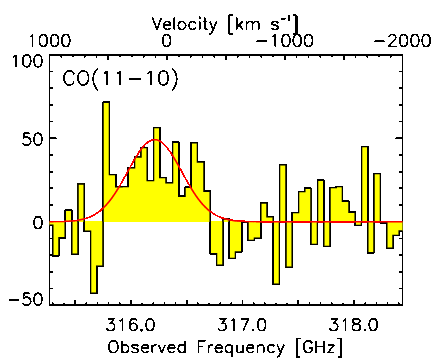}
\caption{Single-dish spectra of the CO rotational emission lines in PLCK\_G244.8+54.9 observed with EMIR. 
The continuum-subtracted and binned spectra were fitted with two Gaussian components using the {\tt CLASS} package 
of {\tt GILDAS}, by fixing the peak velocities to those measured on the CO(3--2) spectrum. The best-fit line profiles
are plotted as red curves, with the individual spectral components overlaid in gray. Velocity offsets are defined relative 
to the best spectroscopic redshift $z=3.0054 \pm 0.0001$ presented in Table~\ref{tab:prop}. The resulting line properties 
are listed in Table~\ref{tab:linefit} and EMIR spectra of other {\it Planck}'s dusty GEMS are shown in Appendix~A.}
\label{fig:lines1}
\end{figure*}

We used scans from either the FTS or WILMA backend, depending on the number of bad scans and line S/N. Typically, 
WILMA provided a higher baseline stability and was favored for low S/N spectra. Profiles and non-detections of all 
CO emission lines observed with EMIR are shown in Fig.~\ref{fig:lines1} and Appendix~A.

Furthermore, the CO(1--0) transition in five of the GEMS was observed with the Green Bank telescope (GBT) by 
\citet[][]{harrington18}. We will discuss the impact of including the measured line fluxes in our analysis of the CO gas 
excitation and the implications for the number of molecular gas phases in these intense starbursts, under the assumption 
that relative flux calibrations between IRAM and GBT spectra are reliable and that the differences in beam sizes play a 
minor role.

\section{Characterizing the CO line emission}
\label{sec:coprop}

We now present our results on the properties of the star-forming gas reservoirs in the GEMS, through searching 
for the peak of the CO ladder and the characterization of the spatially-integrated CO line ratios.

\subsection{CO line profiles of individual sources}
\label{ssec:coprof}

By combining all observing runs, we detected 45 CO lines with $J_{\rm up} \geq 3$ out of the 48 observed lines, and up to high 
rotational levels (maximum of $J_{\rm up}=11$) that trace the warm and dense regime of the bulk of the molecular gas. Our data 
set includes between two and eight CO lines per source. We fitted each baseline-subtracted and coadded spectrum using a Gaussian 
function and measured the central frequencies of the lines, as well as the redshifts, full-widths-at-half-maximum (FWHM), and 
velocity-integrated fluxes. Results are reported in Table~\ref{tab:linefit}, together with the uncertainties provided by the 
{\tt CLASS} fitting routine. Line fluxes are uncorrected for lensing magnification. Seven out of the 11 GEMS exhibit single 
Gaussian profiles and we only use a double Gaussian for PLCK\_G045.1+61.1, PLCK\_G092.5+42.9, PLCK\_G145.2+50.9, and 
PLCK\_G244.8+54.9, where this additional spectral component is robustly detected (see Figures~\ref{fig:lines1}, 
\ref{fig:lines2}, \ref{fig:lines3}, and \ref{fig:lines4}). 

We computed flux upper limits for the lines without 3$\sigma$ detections following the approach adopted in \citet{rowlands15}. 
We obtained a broad range of line FWHM values, from 200 to 750~km~s$^{-1}$, similar to lensed and unlensed SMGs in the literature 
\citepalias[see also][]{canameras15}, and very high fluxes up to 37~Jy~km~s$^{-1}$, uncorrected for gravitational magnification.
The best spectroscopic redshifts of each source listed in Table~\ref{tab:prop} were measured from single Gaussian fits to the 
two lowest-$J$ CO lines detected with EMIR, typically either $J_{\rm up}=3$ and 4 or $J_{\rm up}=4$ and 5.

\subsection{CO spectral line energy distributions}
\label{ssec:cosled}

\begin{figure*}
\centering
\includegraphics[width=0.9\textwidth]{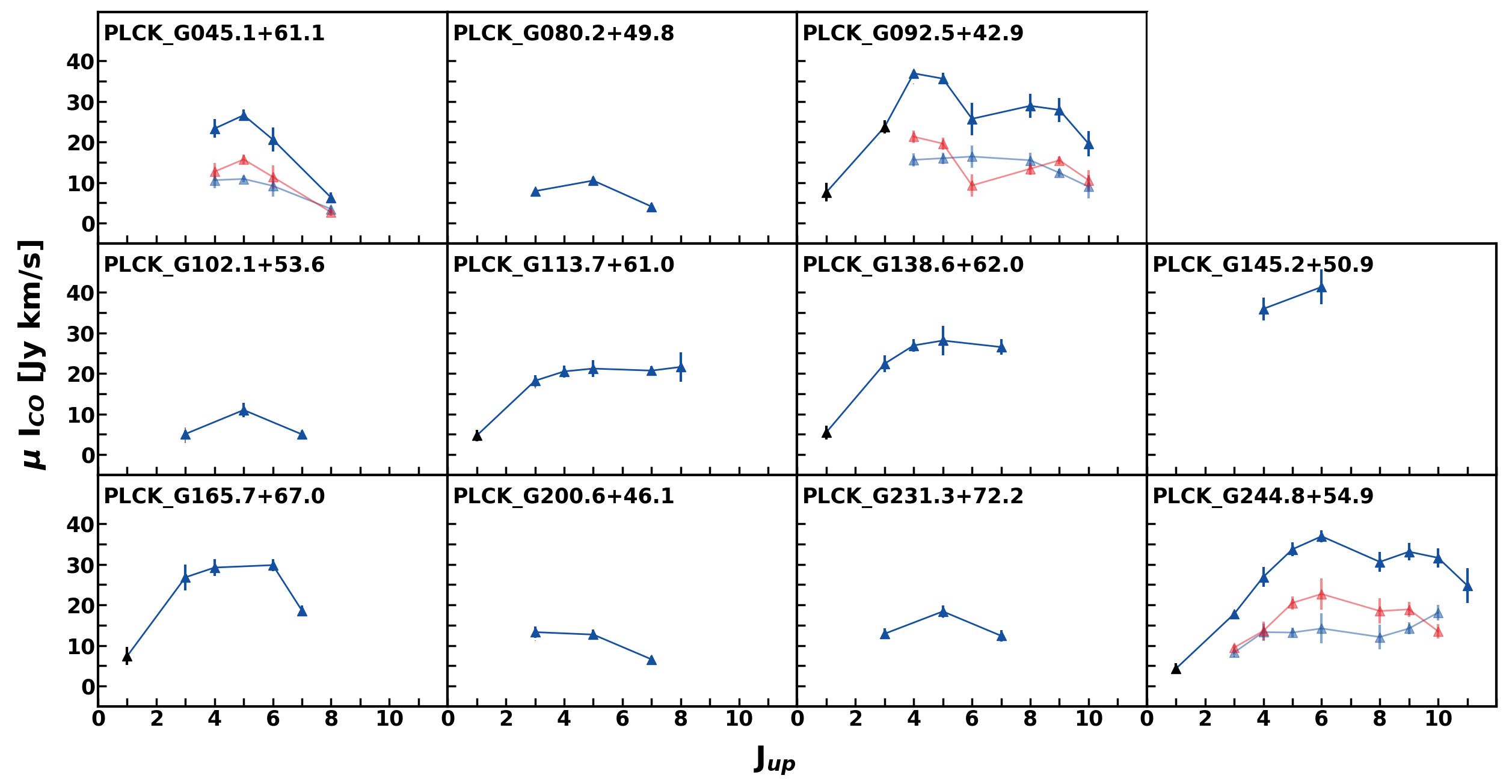}
\caption{CO spectral-line energy distributions of the {\it Planck}'s dusty GEMS. Blue triangles show the velocity-integrated 
line fluxes uncorrected for gravitational magnification, as measured in this work with EMIR on the IRAM 30-m telescope. All 
CO(1--0) fluxes are taken from \citet[][black triangles]{harrington18} and CO(3--2) for PLCK\_G092.5+42.9 is from
\citet{harrington16}. For PLCK\_G045.1+61.1, PLCK\_G092.5+42.9, and PLCK\_G244.8+54.9, light blue and red triangles show 
the SLEDs obtained for the blue and red kinematic components, respectively, for the transitions with S/N values
sufficiently high to obtain a robust separation of the two components.}
\label{fig:cogrid}
\end{figure*}

Figure~\ref{fig:cogrid} shows the individual CO SLEDs of the 11 GEMS, in units of velocity-integrated flux densities. 
This plot demonstrates that our survey covers the peak of the SLED for nine out of 11 sources, and that this peak
systematically falls between $J_{\rm up}=4$ and $J_{\rm up}=7$. This is crucial to shed light on the gas conditions in the
GEMS, since the location of the maximum on the CO excitation ladder directly depends on the gas kinetic temperature and 
density. For this reason, the only two CO transitions detected in PLCK\_G145.2+50.9 and the flat ladder of 
PLCK\_G113.7+61.0 in the range $J_{\rm up}=4$--8 will, to a certain extent, prevent us from deriving detailed gas properties 
in these sources. The SLEDs of PLCK\_G092.5+42.9 and PLCK\_G244.8+54.9, the two brightest GEMS with well-sampled SLEDs 
up to $J_{\rm up}=10$--11 exhibit double peaks that strongly suggest the presence of two distinct gas excitation components. 
We will provide further evidence of these multiple gas phases using detailed excitation models in Sect.~\ref{sec:coexc}.

We also determined the shape of the SLED for individual ``blue'' and ``red'' components (named according to their position
relative to the systemic redshifts of Table~\ref{tab:prop}) in PLCK\_G045.1+61.1, PLCK\_G092.5+42.9 and PLCK\_G244.8+54.9,
the three GEMS where two spectral components are detected over several CO lines. To obtain a robust separation, we fixed 
the central velocities of the individual components to those measured on the lowest-$J$ transition with the highest S/N 
value, typically either CO(3--2) or CO(4--3) (see Table~\ref{tab:linefit}). The two Gaussians are centered at about --420 
and 30~km~s$^{-1}$, --120 and 170~km~s$^{-1}$, --210 and 150~km~s$^{-1}$ (with respect to the spectroscopic redshifts of 
Table~\ref{tab:prop}) for PLCK\_G045.1+61.1, PLCK\_G092.5+42.9 and PLCK\_G244.8+54.9, respectively. Varying only the 
FWHM and peak flux of each kinematic component enables us to properly fit the profiles of all mid- and high-$J$ lines, 
apart from the CO(11--10) transition in PLCK\_G244.8+54.9, which has a lower S/N value. The spectral components in
PLCK\_G244.8+54.9 illustrated in Fig.~\ref{fig:lines1} have central velocities consistent with those of the two regions 
identified in \citet{canameras17b} from ALMA 0.1\arcsec$-$resolution imaging of the CO(4--3) line emission. The resulting CO 
SLEDs for the blue and red kinematic components are shown in Fig.~\ref{fig:cogrid}. For PLCK\_G045.1+61.1, both SLEDs are 
very similar to the one derived from integrated CO fluxes, and peak at $J_{\rm up}=5$. In PLCK\_G092.5+42.9, the measured 
fluxes of the blue component are nearly constant between $J_{\rm up}=4$ and $J_{\rm up}=8$, implying that the variations of 
the global SLED are mainly driven by the red component. The relative gas excitation in PLCK\_G244.8+54.9 is less obvious, 
with the red component having a larger contribution to the main peak. We will further investigate their individual 
properties in Sect.~\ref{ssec:lvg}.

The relative variations between sources are shown in Fig.~\ref{fig:normsled} by plotting the CO SLEDs normalized by the
CO(3--2) velocity-integrated fluxes. We also show how the GEMS compare with other well-studied starburst galaxies in the 
literature, such as the local starburst Arp~220 \citep{wiedner02,rangwala11}, the central region of M82 \citep{weiss05},
and the Cosmic Eyelash \citep{danielson11}. Overall, the variety of CO ladders is consistent with the wide range of CO 
excitations found over the SMG population \citep{casey14}. The CO ladder of the most excited GEMS, the Ruby, peaks at 
$J_{\rm up}=6$, similarly to the local starburst M82 (Fig.~\ref{fig:normsled}). This is in line with the detailed analysis 
of \citet{canameras17b}, which showed that star-forming regions within this maximal starburst could resemble the densest 
inner parts of molecular regions within low-redshift galaxies. 

The CO(1--0) fluxes are comparatively higher than those from local starbursts, assuming that the relative flux calibrations 
between our CO line survey with EMIR and GBT observations from \citet{harrington18} have uncertainties of up to $\sim$30\%. 
This suggests that the GEMS host a low-excitation gas component distinct from the excited gas phases, as further discussed in 
Sect.~\ref{ssec:mgas}. Moreover, the fact that the SLED of the most excited sources in the sample peak at $J_{\rm up} \leq 6$, and 
compare well with the gas excitation in Arp 220, provides further evidence that the GEMS are intense starbursts that do not host 
a powerful AGN \citep[see, in contrast, the turnover at $J_{\rm up}=10$ for the $z=3.9$ QSO APM 08279,][]{weiss07,bradford11}. We 
will constrain the maximal contribution from an X-ray dominated region to the overall gas heating in Sect.~\ref{ssec:agn}. 

Sources in local thermal equilibrium (LTE) with thermalized CO rotational levels exhibit SLEDs rising as the square of 
$J_{\rm up}$. Figure~\ref{fig:normsled} shows that only PLCK\_G092.5+42.9, PLCK\_G102.1+53.6, and PLCK\_G244.8+54.9 have 
CO transitions nearly in thermal equilibrium up to $J_{\rm up} \sim 4$, consistent with the presence of subthermally excited
reservoirs, as also found, for example, in \citet{harris10}.

\subsection{CO line luminosities}
\label{ssec:lumcorr}

To quantify the total energy radiated by each CO transition we converted the fluxes to line luminosities, in L$_{\odot}$, 
using the relation from \citet{solomon92}:
\begin{equation}
L_{\rm line} = 1.04 \times 10^{-3} I_{\rm line} \nu_{\rm obs} D_{\rm L}^2
\end{equation}
where $I_{\rm line}$ is the velocity-integrated flux in Jy~km~s$^{-1}$, $\nu_{\rm obs}$ the observed frequency of the line in 
GHz and $D_{\rm L}$ the luminosity distance in Mpc. We also computed line luminosities in K~km~s$^{-1}$~pc$^2$ using:

\begin{equation}
L'_{\rm line} = 3.25 \times 10^{7} I_{\rm line} \nu_{\rm obs}^{-2} D_{\rm L}^2 (1+z)^{-3}
\end{equation}
All observed line luminosities (uncorrected for the gravitational magnification) are reported in Table~\ref{tab:linefit}. 

\begin{figure*}
\centering
\includegraphics[width=0.7\textwidth]{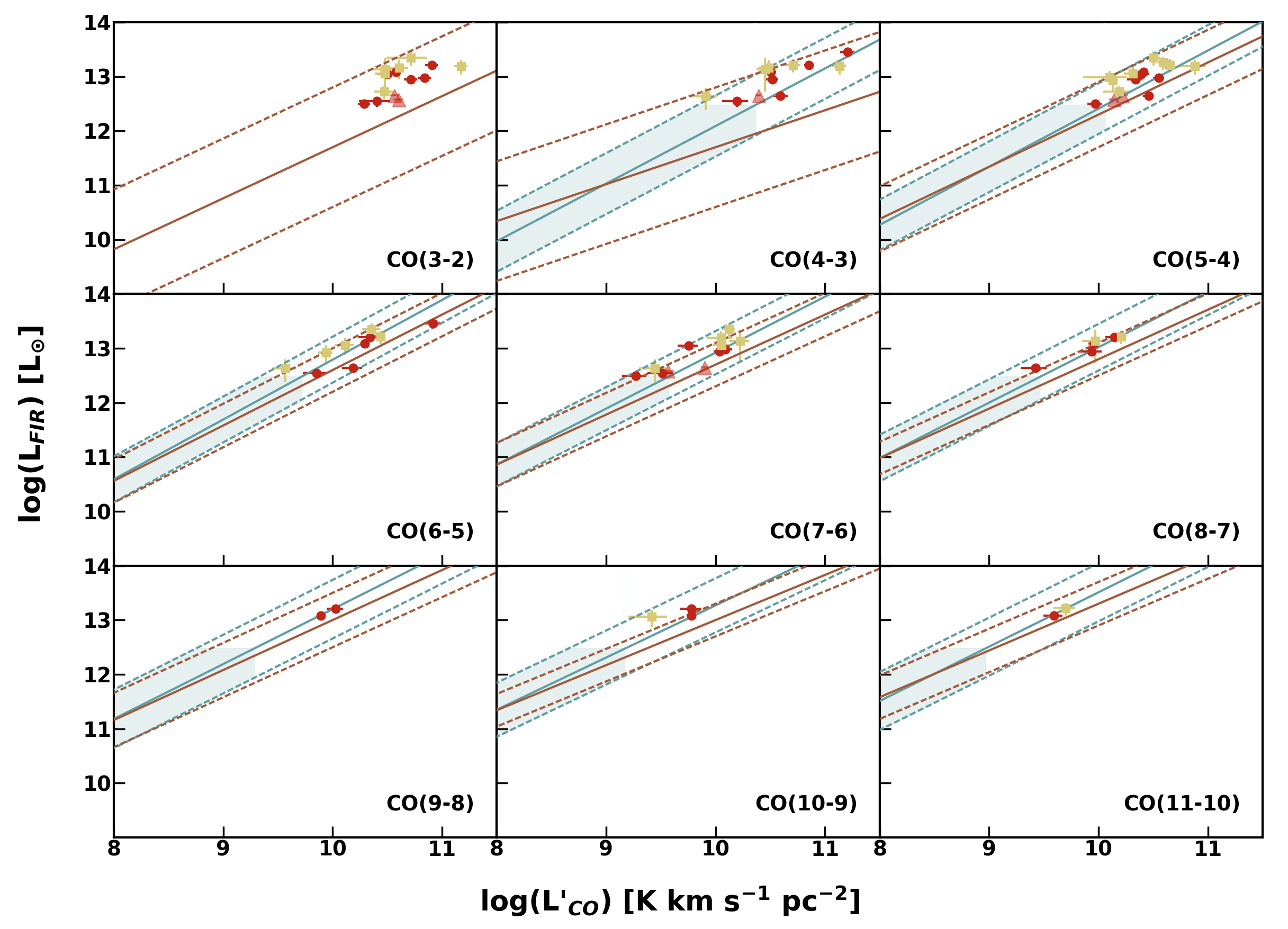}
\caption{Relations between $L_{\rm FIR}$ and $L'_{\rm CO}$ for the nine CO transitions detected at least in one of the {\it Planck}'s 
dusty GEMS. Red points show results from this work, corrected for the gravitational magnification and with error bars including
the uncertainties on $\mu$ from Table~\ref{tab:prop}. Red triangles mark the position of PLCK\_G138.6+62.0 and PLCK\_G200.6+46.1 
which have more uncertain magnification factors. Yellow squares are the high-redshift strongly lensed SMGs from the H-ATLAS survey, 
also corrected for strong lensing magnification \citep[][]{yang17}. Solid blue lines indicate the best linear fits obtained by
\citet{liu15} for a diverse sample of local galaxies and individual star-forming regions with luminosities in the range 
$10^8 \leq L_{\rm FIR} \leq 10^{12}$~L$_{\odot}$, as shown by the blue shaded regions. Dashed blue lines show their 
$\pm 2 \sigma$ dispersions. Brown lines represent similar relations for low-redshift ULIRGs 
with $10^{11} \leq L_{\rm FIR} \leq 3 \times 10^{12}$~L$_{\odot}$ presented in \citet{kamenetzky16}. The properties of high-redshift 
SMGs are consistent with local relations and extend their validity over five orders of magnitude in FIR luminosity.}
\label{fig:lumcorr}
\end{figure*}

In Figure~\ref{fig:lumcorr}, we plot the relations between $L_{\rm FIR}$ and $L'_{\rm CO}$ for all CO transitions detected 
in at least one GEMS, and the best-fitting linear relations obtained for local galaxies and individual star-forming regions 
\citep{liu15} as well as for low-redshift ULIRGs only \citep{kamenetzky16}. The two low-redshift comparison samples cover 
FIR luminosities up to about $10^{12}$~L$_{\odot}$. Moreover, ULIRGs span a particularly small dynamical range of less than two
orders of magnitude in luminosity, which leads to differences in the high-$J$ $L_{\rm FIR}$--$L'_{\rm CO}$ correlations between
both samples, although the measured slopes appear to be consistently sublinear \citep[e.g.,][]{greve14,kamenetzky16}.
The GEMS closely follow the low-redshift correlations from \citet{liu15} and cover the same regime as other samples of 
high-redshift dust-obscured star-forming galaxies \citep{yang17}. When comparing to local ULIRGs, the CO transitions with 
$J_{\rm up} \geq 5$ that are thought to arise from the denser and warmer gas components directly related to the on-going 
star-formation activity appear to best follow the local trends. For $J_{\rm up} \leq 4$, the lower number of sources 
available in the catalog of \citet[][]{kamenetzky16} results in more scattered relations, which complicates the 
interpretation.

We derived the spatially-integrated CO line luminosity ratios for the five GEMS with both CO(1--0) 
\citep[from][]{harrington18}, and mid-$J$ flux measurements, finding $r_{\rm 32/10} = 0.42 \pm 0.04$, 
$r_{\rm 43/10} = 0.30 \pm 0.05$ and $r_{\rm 54/10} = 0.22 \pm 0.05$. These values are 
20--30\% lower than ratios obtained in \citet{bothwell13} for unlensed SMGs. The $r_{\rm 32/10}$ ratio has been particularly 
well constrained for high-redshift dusty star-forming galaxies, with measurements in the range $r_{\rm 32/10} = 0.40$--0.65 
\citep{harris10,tacconi10,carilli13,greve14,genzel15,daddi15,sharon16,yang17}, and our estimate therefore remains consistent 
with the typical values found in the literature. We further interprete these line ratios in terms of gas phases following our 
gas excitation analysis in Sect.~\ref{ssec:mgas}.

\section{Global gas properties of the GEMS}
\label{sec:prop}

\subsection{Correcting for the lensing magnification}
\label{ssec:lens}

\subsubsection{Magnification factor estimates}
\label{sssec:mu}

For each source, we used our best-fitting lens model with {\tt LENSTOOL} \citep{jullo07} to derive amplification maps with the 
same sampling as the subarcsec resolution CO line and  dust continuum maps from our SMA, PdBI, and ALMA follow-up interferometry. 
The details of our lens modeling approach are described, for example, in \citet[]{canameras17a} and Ca\~nameras et al. 
(2018a, A\&A accepted). We then used these maps to compute the luminosity-weighted gravitational magnification 
factors of the gas and dust components, in order to correct the observed velocity-integrated CO fluxes and line luminosities. 
We accounted for the frequency variations of the beam size by using the following parametrization of the observed half-power 
beam width\footnote{\url{http://www.iram.es/IRAMES/mainWiki/Iram30mEfficiencies}} (HPBW): 
${\rm HPBW/[arcsec]=2460 \times (\nu/[GHz])^{-1}}$. We only considered pixels where the CO line or dust continuum emission is 
detected above $4 \sigma$, and we rejected a small fraction of 5--10 pixels that lie on top of the critical lines and have 
artificially high magnification factors above 200.

\begin{figure}
\centering
\includegraphics[width=0.5\textwidth]{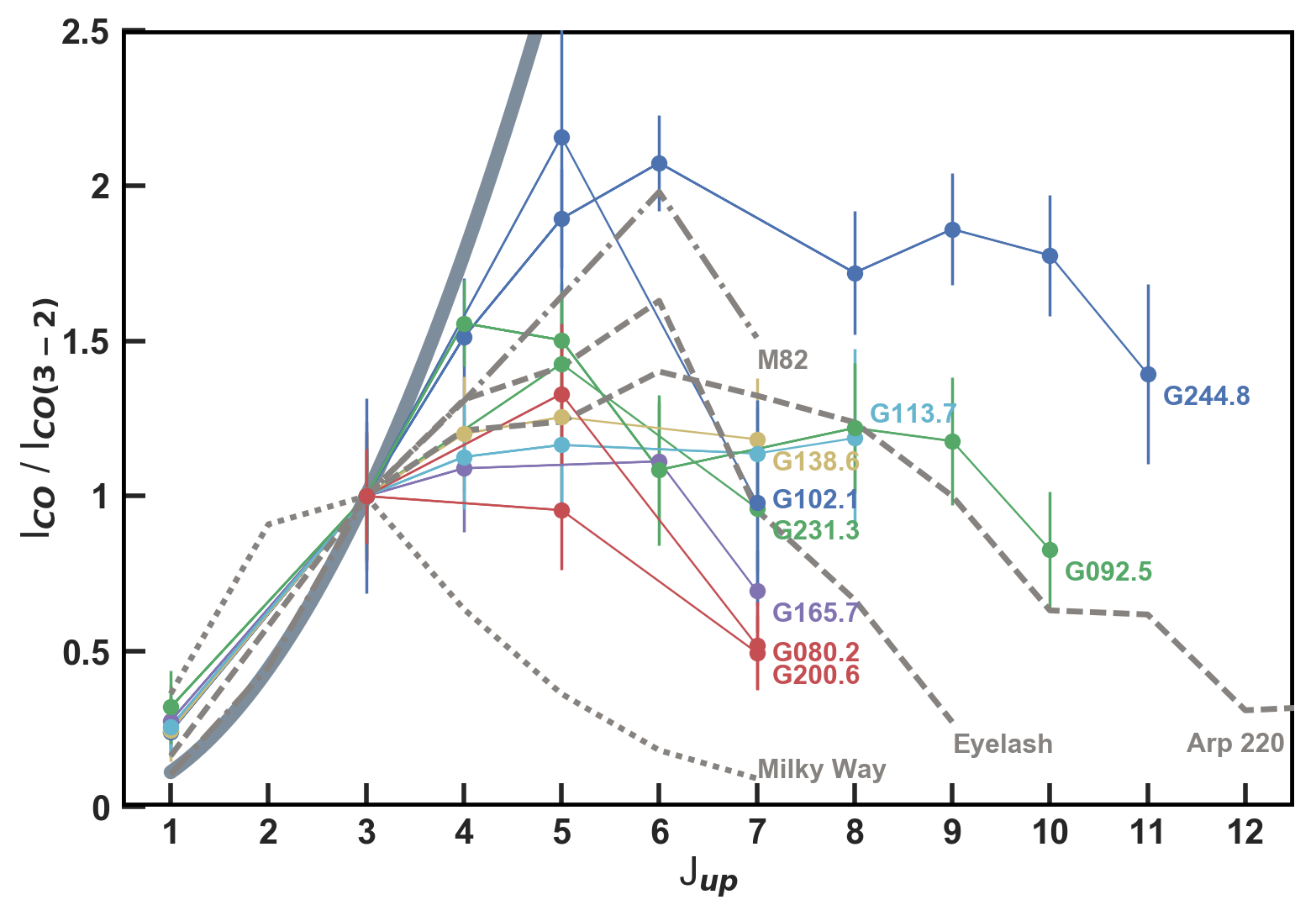}
\caption{Observed CO SLEDs normalized by the CO(3--2) line fluxes for the GEMS with CO(3--2) detections \citep[the CO(1--0) 
fluxes are from][]{harrington18}, the local starburst Arp~220 \citep{wiedner02,rangwala11}, the central region of M82 
\citep{weiss05} and the Cosmic Eyelash \citep{danielson11}. For comparison, we also show the CO SLED spatially-integrated 
over the inner disk of the Milky Way \citep[dotted line,][]{fixsen99}, and the expected trend for optically thick gas in local 
thermodynamic equilibrium (thick gray line).}
\label{fig:normsled}
\end{figure}

Table~\ref{tab:prop} summarizes the resulting magnification factors for each source and their 1$\sigma$ statistical 
uncertainties. The luminosity-weighted magnification factors of the gas component were inferred from our high-resolution 
line imaging with the PdBI, for mid-$J$ CO transitions between $J_{\rm up}=4$ and $J_{\rm up}=6$. Unless otherwise stated, 
for the cold dust component we used SMA continuum maps at 880-$\mu$m either in the EXT or VEXT configuration, at 
similar beam size as our PdBI observations. The CO maps of PLCK\_G092.5+42.9, PLCK\_G102.1+53.6, and PLCK\_G231.3+72.2 
have significantly lower resolutions than those from the SMA, and we computed $\mu_{\rm dust}$ from the 2-mm dust continuum 
extracted in line-free baseline channels from the PdBI data cubes to avoid beam effects on the magnification factor 
estimates. We also quote the systematic errors on $\mu$ induced by the choice of the mass profile of the main deflector 
or the inclusion of photometrically-selected multiple image systems, as inferred from the alternative models 
that will be further discussed in a forthcoming paper (Ca\~nameras et al. 2018c, in prep.). The luminosity-weighted 
magnification factors of the Ruby were computed within the beam of the 30-m telescope in the same way as for the other 
GEMS, based on the detailed lensing analysis of \citet{canameras17a} that shows local variations of $\mu$ by up to 
a factor of 2 for individual clumps in this source.

PLCK\_G200.6+46.1 and PLCK\_G138.6+62.0 are the only unresolved sources in our follow-up submm interferometry of the 
whole sample. The main lens galaxies in these systems also lack spectroscopic redshift measurements, which precludes the 
calculation of detailed lensing models. Since most results rely on line ratios, we nonetheless included these GEMS in 
the overall analysis and used their position in the $L_{\rm FIR}$--$T_{\rm dust}$ parameter space relative to the sequence 
of unlensed SMGs to derive their gravitational magnification \citepalias[see figure~5 and justification in][]{canameras15}. 
This method, although crude, provides estimates of $\mu$ which are within 25\% of the magnification factors obtained from 
detailed modeling for six sources (Table~\ref{tab:prop}), and within a factor 2 for the remaining three.

\subsubsection{Constraints on the differential magnification}
\label{sssec:diff}

Previous studies have cautioned that differential magnification effects can become an important source of systematic 
uncertainties when characterizing the molecular gas properties of strongly lensed dusty star-forming galaxies 
\citep[e.g.,][]{blain99b}. This effect can play an important role when comparing fluxes from mid-$J$ CO lines with 
CO(1--0) because these transitions have critical densities varying by 1--3 orders of magnitude and therefore trace
different gas components in the sources. $J_{\rm up}>2$ CO lines trace warm molecular regions which are expected to be compact 
in high-redshift starbursts, under the form of late-stage mergers akin to the nucleus of Arp~220 \citep{scoville17} or 
massive star-forming clumps distributed over gas-rich disks \citep[e.g.,][]{swinbank10,thomson15}, while CO(1--0) is 
thought to arise from the cool and more diffuse ISM \citep[e.g.,][]{ivison11}. When observing strongly lensed galaxies, 
one of these components can benefit from a higher gravitational magnification, depending on the position of the source 
plane caustic lines. Simulations showed that uncertainties on the CO(6--5)/CO(1--0) line flux ratio due to differential 
magnification can reach up to 30\% \citep{serjeant12} and therefore distort the observed CO SLEDs, but the amplitude of
this effect essentially depends on the actual distribution of the low and high-excitation gas phases in these systems. 
We quantified differential magnification over the sample using two different approaches.

Firstly, we used our detailed lensing models to quantify the level of differential lensing between mid-$J$ CO lines and 
dust continuum of individual {\it Planck}'s dusty GEMS. IRAM/PdBI and ALMA subarcsec resolution line-imaging constrain 
the morphology of the CO line emission in all sources and allow us to compare with the resolved dust continuum maps to 
search for variations of the luminosity-weighted magnification factors between the two components. The results of 
Table~\ref{tab:prop} demonstrate that values of $\mu_{\rm gas}$ and $\mu_{\rm dust}$ are within 1$\sigma$ for 
PLCK\_G080.2+49.8, PLCK\_G102.1+53.6, PLCK\_G165.7+67.0 and PLCK\_G244.8+54.9, and differ by at most 15--30\% for other 
sources. These variations are comparable with other measurement uncertainties, including IRAM flux calibration 
uncertainties, and might be partly due to residual differences in the beam dimensions of about 20\% (e.g. for 
PLCK\_G045.1+61.1 and PLCK\_G092.5+42.9), despite our efforts in computing the luminosity-weighted magnification factors 
at similar angular resolutions.

The simulation study presented in \citet{serjeant12} suggests that differential magnification effects between [\ion{C}{ii}] 
and FIR bolometric emissions are minor, for ISM configurations resembling that of the Cosmic Eyelash. [\ion{C}{ii}] and 
CO(1--0) line emissions also present similar magnifications. This suggests that if the sources have configurations similar 
to the Cosmic Eyelash, which is broadly supported by our multiwavelength analysis of PLCK\_G165.7+67.0 (Ca\~nameras et al. 
2018a, A\&A accepted), the $\mu_{\rm dust}$ value can serve as a proxy for the magnification of the low-density CO(1--0) 
gas reservoirs. Under these assumptions, our high-resolution interferometry therefore suggests that differential lensing 
effects between CO(1--0) and mid-$J$ CO lines are minor in this sample. The effect is most likely negligible between mid-
and high-$J$ CO transitions that both trace the compact sites of star formation, as demonstrated by \citet{rybak15} for 
the well-studied dusty starburst SDP~81. 

Secondly, we compared the CO line profiles from our IRAM survey to show that differential lensing is not producing a 
significant bias between different rotational levels. Values given in Table~\ref{tab:linefit} show that for each source, 
most line FWHMs measured with single component Gaussian fits are within the 1$\sigma$ uncertainties, suggesting that 
$J_{\rm up}>3$ transitions consistently trace the same intrinsic gas kinematics. For PLCK\_G045.1+61.1, PLCK\_G092.5+42.9, 
PLCK\_G145.2+50.9, and PLCK\_G244.8+54.9, we also detect the same number of spectral components from $J_{\rm up}=3$ to 
$J_{\rm up}=10$, on the transitions with sufficient signal-to-noise ratios (Fig.~\ref{fig:lines1}, \ref{fig:lines2}, 
\ref{fig:lines3} and \ref{fig:lines4}). More puzzling is the line profile of CO(6--5) in PLCK\_G092.5+42.9, where the two spectral 
components are robustly detected but with very different flux ratios compared to other transitions. However, since the 
profiles of CO(8--7) and CO(9--8) lines in this source are consistent with those of CO(4--3) and CO(5--4), we conclude 
that different gas excitation properties within the blue and red components are more likely to produce different flux 
ratios on a single transition than differential lensing effects. For these reasons, we assumed that differential lensing 
is not likely to induce major distortions of the CO SLEDs and hence we ignored this effect for the gas excitation 
analysis.

\subsection{The $\alpha_{\rm CO}$ conversion factor}
\label{ssec:aCO}

\begin{figure}
\centering
\includegraphics[width=0.48\textwidth]{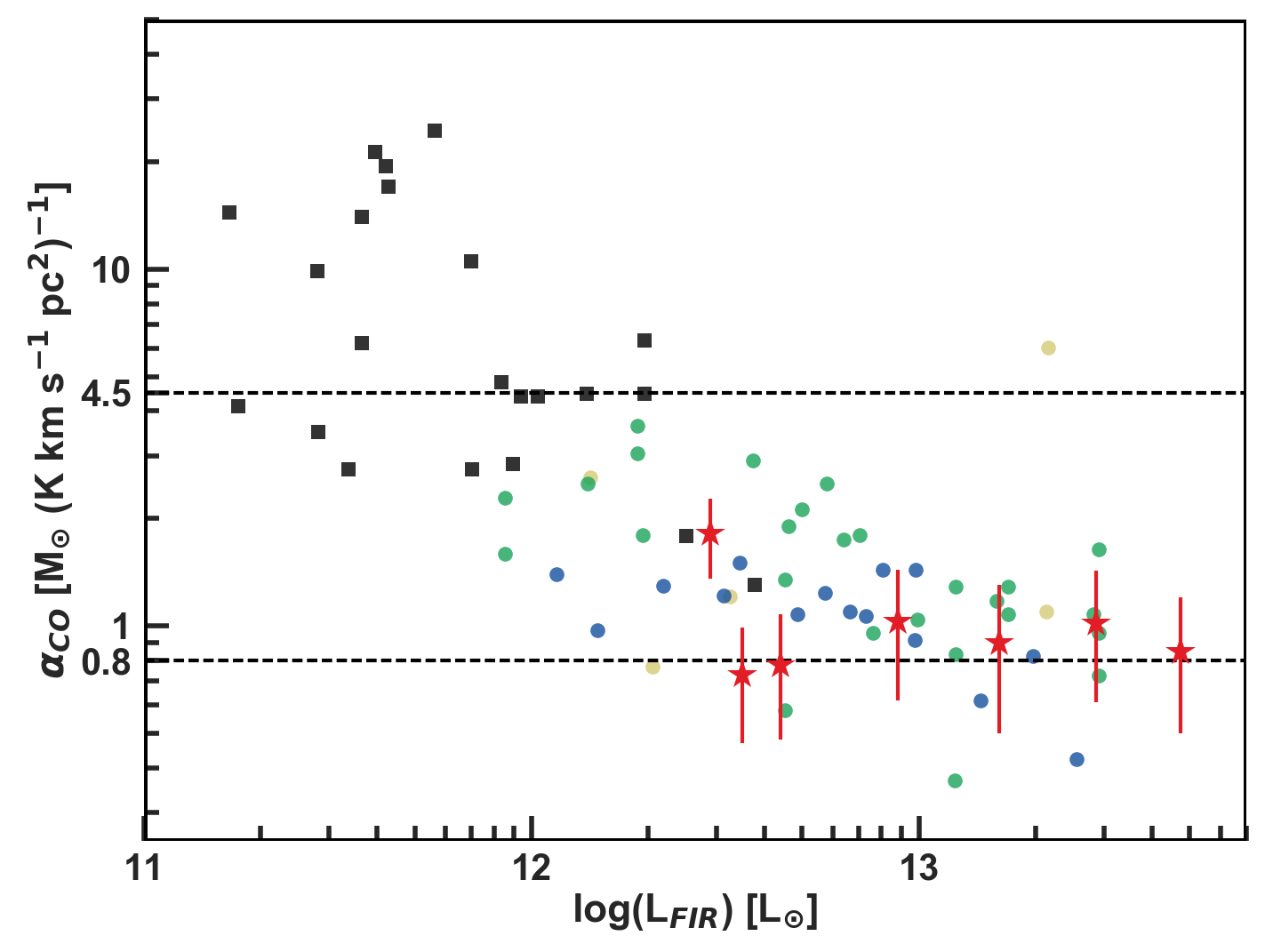}
\caption{CO luminosity to gas mass conversion factor, $\alpha_{\rm CO}$, versus FIR luminosity, for the seven 
{\it Planck}'s dusty GEMS with direct $\alpha_{\rm CO}$ measurements using the molecular gas masses inferred from the 
[\ion{C}{i}](1-0) line fluxes of Nesvadba et al. (2018, A\&A submitted, red stars). All but one GEMS are consistent 
with the usual ULIRG value $\alpha_{\rm CO} = 0.8$~M$_{\odot}$~(K~km~s$^{-1}$~pc$^2$)$^{-1}$ (lower dashed line). Other points 
show $\alpha_{\rm CO}$ estimates in the literature, for the strongly lensed SMGs from the SPT survey (blue points), 
high-redshift unlensed dust-obscured starbursts (green points), QSOs (yellow points) and main sequence galaxies 
\citep[black squares,][and references therein]{aravena16}. The relative uncertainties on $\alpha_{\rm CO}$ obtained 
for these high-redshift samples are comparable to those of the GEMS.}
\label{fig:aco}
\end{figure}

Deriving the molecular hydrogen masses of the GEMS from our analysis of the CO gas excitation relies on our assumption
about the $\alpha_{\rm CO}$ conversion factor, which can induce major uncertainties in high-redshift studies despite the 
overall consensus reached for local galaxy populations \citep[see][]{bolatto13}. However, the $^3$P$_1$--$^3$P$_0$ 
fine-structure line of atomic carbon has also proven to be a reliable proxy of the total molecular gas content in 
galaxies. Due to its low critical density \citep[$n_{\rm crit} \simeq 500$~cm$^{-3}$,][]{carilli13} it is easily 
thermalized in molecular clouds and originates from the extended and low-density gas reservoirs that contain the bulk 
of the molecular gas mass, similarly to CO(1--0), both in nearby \citep[e.g.,][]{papadopoulos04,jiao17} and high-redshift 
ULIRGs \citep[e.g.,][]{alaghband13}. We can therefore invert the problem by using the detailed analysis of atomic carbon
in the GEMS from \citetalias{nesvadba18} to derive $\alpha_{\rm CO}$ for the seven sources with [\ion{C}{i}](1--0) line 
detections with EMIR, assuming that it traces the same gas components as CO(1--0).

For these seven GEMS, we used the [\ion{C}{i}]-inferred H$_{\rm 2}$ masses from \citetalias{nesvadba18}, obtained with the 
relation from \citet{papadopoulos04} \citep[see also equation~2 in][]{wagg06} and for a common value of the carbon abundance,
$X_{\rm [\ion{C}{i}]}=3 \times 10^{-5}$ \citep{weiss05}, an excitation factor, $Q_{\rm 10} \simeq 0.5$, and the Einstein coefficient,
$A_{\rm 10}=7.93 \times 10^{-8}$~s$^{-1}$. We then used the CO(1--0) line luminosities, $L'_{\rm CO(1-0)}$, either measured 
in \citet{harrington18} or converted from the mid-$J$ transitions detected with EMIR using the average $r_{\rm 32,10}$ and 
$r_{\rm 43,10}$ conversion factors measured over the sample in Sect.~\ref{ssec:lumcorr}. Comparing $M^{\rm [\ion{C}{i}]}({\rm H_2})$ 
and $L'_{\rm CO(1-0)}$ results in the values of $\alpha_{\rm CO}$ shown in Table~\ref{tab:mass} and Figure~\ref{fig:aco}.
For all but one source, we obtain low conversion factors consistent within 1$\sigma$ with 
$\alpha_{\rm CO} \sim 0.8$~M$_{\odot}$~(K~km~s$^{-1}$~pc$^2$)$^{-1}$, the value measured for local ULIRGs and widely used 
for luminous dusty starbursts at high redshift \citep[see also][]{bolatto13}. PLCK\_G080.2+49.8 is the only GEMS
for which we measure a higher conversion factor inconsistent with the ULIRG value, which is perhaps not surprising 
since the local star-formation properties within this source are more akin to high-redshift main-sequence galaxies 
than other extreme starbursts in our sample \citepalias{nesvadba18}.

These results are nevertheless strongly dependent on our choice of the carbon abundance. Enhanced abundances with respect
to those over the Galactic plane have been measured in high-redshift starbursts 
\citep[$X_{\rm [\ion{C}{i}]} \simeq 4$--$5 \times 10^{-5}$,][]{weiss05,danielson11,alaghband13}. We refer to \citetalias{nesvadba18} 
for a discussion of the physical origin and implications of a possible enhanced carbon abundance in the GEMS, and emphasize 
that this would lower our estimates of $\alpha_{\rm CO}$ and favor the use of a low ULIRG-like factor. Given these 
considerations, in Sect.~\ref{ssec:mgas} we will therefore take a common factor of 0.8~M$_{\odot}$/(K~km~s$^{-1}$~pc$^2$)
to convert the CO line luminosities of the 11 sources to gas masses. This is in line with results obtained in C15 from 
dust mass measurements and assuming solar-like metallicities. Moreover, Figure~\ref{fig:aco} illustrates that this choice
is also consistent with independent estimates for other strongly lensed SMGs in the literature \citep[e.g.,][]{aravena16}.

\begin{table*}
\centering 
\begin{tabular}{lcccccc}
\hline
\hline
\rule{0pt}{3ex} Source & $\alpha_{\rm CO}$ & $\mu M^{\rm CO}$(H$_{\rm 2}$) & $M^{\rm CO}$(H$_{\rm 2}$) & $\mu M_{\rm blue}^{\rm CO}$(H$_{\rm 2}$) & $\mu M_{\rm red}^{\rm CO}$(H$_{\rm 2}$) & $\delta_{\rm GDR}$  \\
\rule{0pt}{3ex} & [M$_{\odot}$ (K~km~s$^{-1}$~pc$^2$)$^{-1}$] & [10$^{11}$ M$_{\odot}$] & [10$^{10}$ M$_{\odot}$] & [10$^{11}$ M$_{\odot}$] & [10$^{11}$ M$_{\odot}$] &  \\
\hline
PLCK\_G045.1+61.1 & 0.78 $\pm$ 0.30 & 8.6 $\pm$ 0.4 & 5.5 $\pm$ 0.5 & 4.3 $\pm$ 1.2 & 3.9 $\pm$ 1.2 & 220 $\pm$ 18 \\
PLCK\_G080.2+49.8 & 1.82 $\pm$ 0.46 & 1.0 $\pm$ 0.1 & 0.6 $\pm$ 0.1 & ... & ... & 22 $\pm$ 5 \\
PLCK\_G092.5+42.9 & 0.90 $\pm$ 0.40 & 8.3 $\pm$ 0.4 & 6.9 $\pm$ 0.7 & 7.5 $\pm$ 1.4 & 5.7 $\pm$ 1.5 & 273 $\pm$ 34 \\
PLCK\_G102.1+53.6 & ... & 1.6 $\pm$ 0.2 & 2.3 $\pm$ 0.4 & ... & ... & 84 $\pm$ 13 \\
PLCK\_G113.7+61.0 & 1.03 $\pm$ 0.41 & 5.8 $\pm$ 0.2 & 6.0 $\pm$ 0.5 & ... & ... & 262 $\pm$ 30 \\
PLCK\_G138.6+62.0 & 0.85 $\pm$ 0.35 & 4.5 $\pm$ 0.2 & 2.2 $\pm$ 0.7 & ... & ... & 15 $\pm$ 8 \\
PLCK\_G145.2+50.9 & 1.02 $\pm$ 0.41 & 10.5 $\pm$ 0.4 & 11.8 $\pm$ 1.1 & ... & ... & 111 $\pm$ 9 \\
PLCK\_G165.7+67.0 & 0.73 $\pm$ 0.26 & 8.1 $\pm$ 0.2 & 3.4 $\pm$ 0.8 & ... & ... & 265 $\pm$ 68 \\
PLCK\_G200.6+46.1 & ... & 4.2 $\pm$ 0.3 & 2.8 $\pm$ 1.0 & ... & ... & 133 $\pm$ 74 \\
PLCK\_G231.3+72.2 & ... & 2.5 $\pm$ 0.3 & 4.2 $\pm$ 0.9 & ... & ... & 96 $\pm$ 16 \\
PLCK\_G244.8+54.9 & ... & 5.7 $\pm$ 0.3 & 2.6 $\pm$ 0.3 & 3.4 $\pm$ 1.1 & 3.0 $\pm$ 1.0 & 203 $\pm$ 19 \\
\hline
\end{tabular}
\caption{Estimates of the CO-to-H$_{\rm 2}$ conversion factors, gas masses and gas-to-dust ratios. The $\alpha_{\rm CO}$ column 
lists the CO luminosity to gas mass conversion factors, expressed in units of M$_{\odot}$/(K~km~s$^{-1}$~pc$^2$) and deduced from 
the [\ion{C}{i}]-inferred molecular gas masses for the sources with [\ion{C}{i}](1--0) detections (Nesvadba et al. 2018, A\&A 
submitted, see further details in the text). The molecular hydrogen masses, $\mu\ M^{\rm CO}$(H$_{\rm 2}$), are derived from the 
modeled CO(1--0) luminosities assuming $\alpha_{\rm CO} = 0.8$~M$_{\odot}$/(K~km~s$^{-1}$~pc$^2$). Intrinsic values are corrected for 
the magnification factors $\mu_{\rm gas}$ of Table~\ref{tab:prop}. We also list the gas masses inferred for the individual kinematic 
components in three {\it Planck}'s dusty GEMS. The quantities $\delta_{\rm GDR}$ are the gas-to-dust mass ratios deduced from the 
dust masses presented in C15, corrected for $\mu_{\rm dust}$, and the total molecular gas masses, $M^{\rm CO}$(H$_{\rm 2}$). Errors 
include uncertainties on the line fluxes and magnification factors.}
\label{tab:mass}
\end{table*}

\section{Properties of the CO gas excitation}
\label{sec:coexc}

We now further investigate the CO excitation within each of the GEMS and deduce the physical gas properties using two 
independent radiative transfer analyses that rely either on the entire SLEDs or a range of measured CO line ratios.

\subsection{Large velocity gradient models}
\label{ssec:lvg}

\begin{figure*}
\includegraphics[width=.5\textwidth]{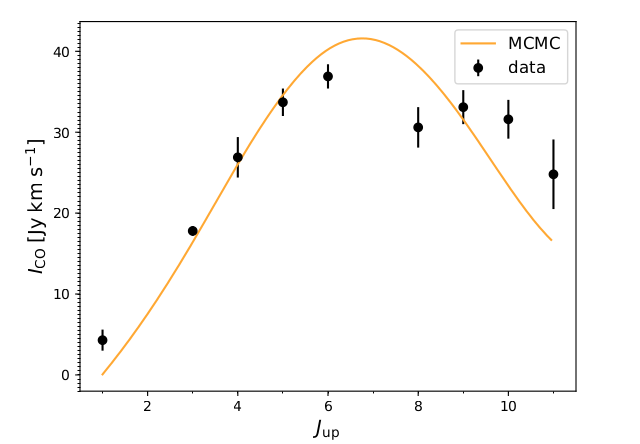}
\includegraphics[width=.5\textwidth]{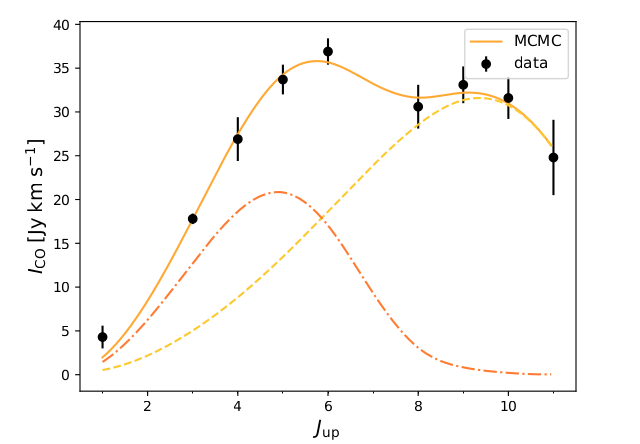}

\includegraphics[width=.5\textwidth]{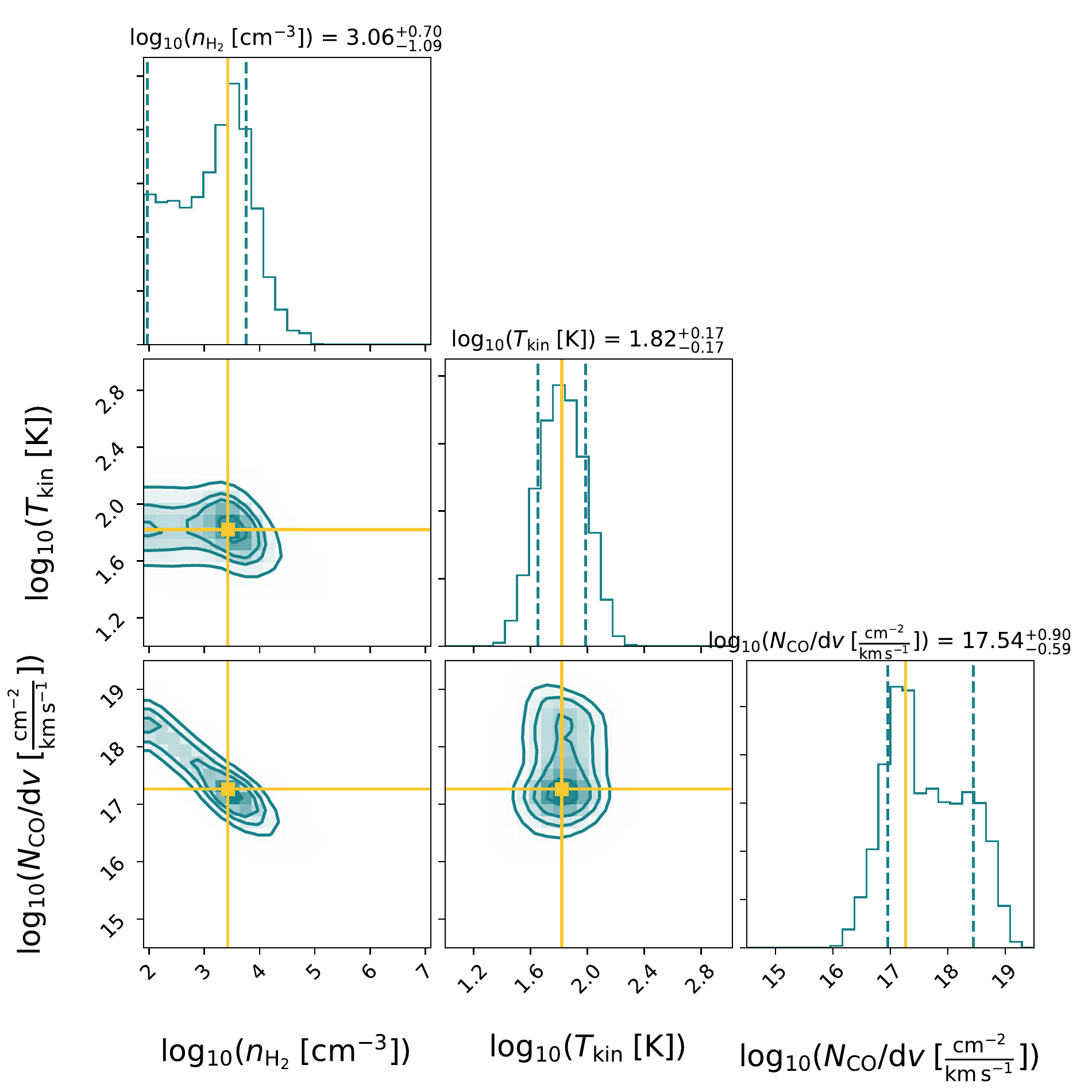}
\includegraphics[width=.5\textwidth]{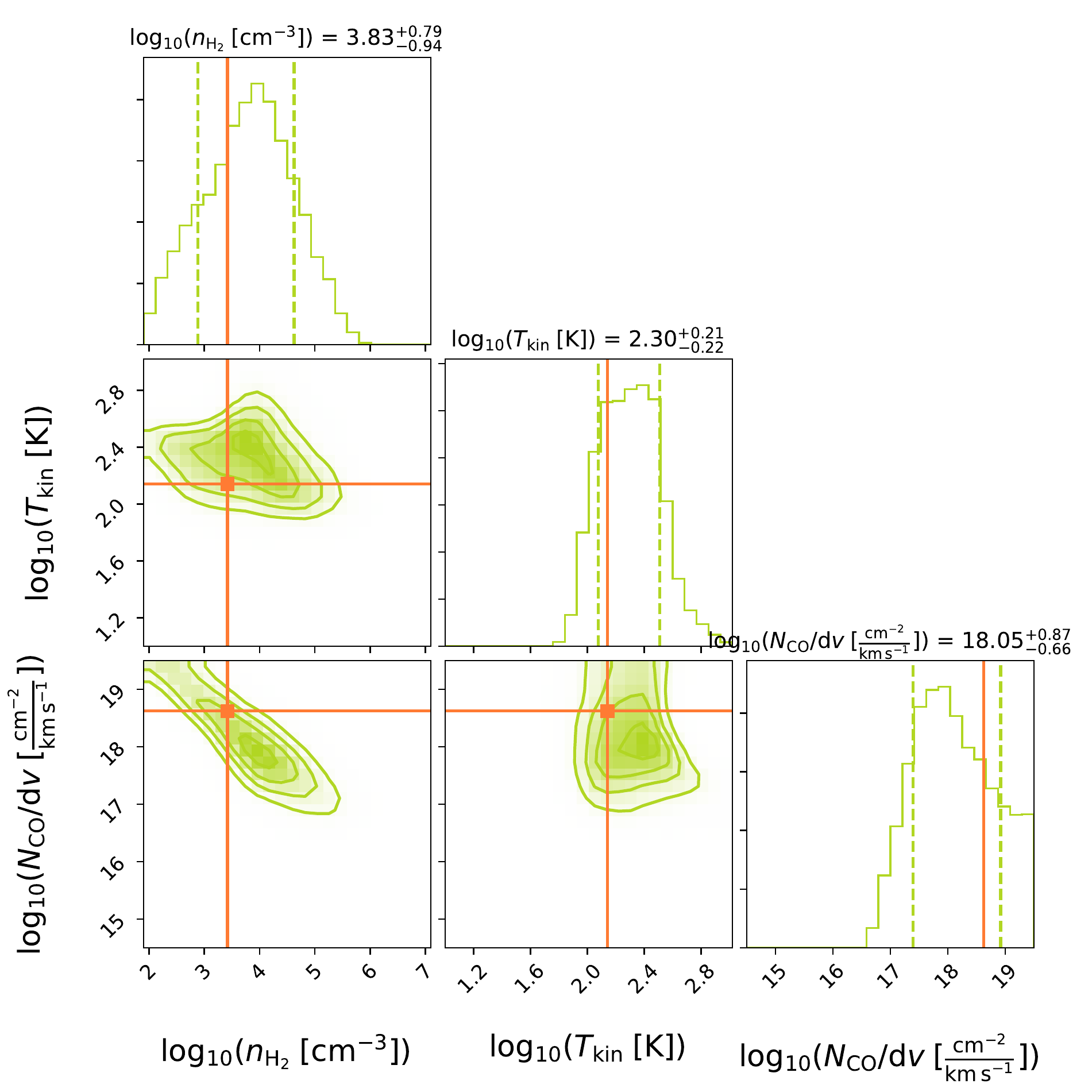}
\caption{{\it Top left}: Observed CO SLED of PLCK\_G244.8+54.9, the Ruby (black points), without correcting 
the velocity-integrated fluxes for gravitational magnification. The orange line shows the best-fit model with {\tt RADEX} 
using a single gas excitation component, and illustrates a case where this simple model poorly reproduces the CO ladder for
$J_{\rm up} \gtrsim 6$. {\it Top right}: Best-fit model from {\tt RADEX} using two gas excitation components (solid orange
line). The dot-dashed orange line shows the low excitation component, which is assumed to be cooler and more extended 
than the high excitation component (dashed yellow line) in the analysis. {\it Bottom}: one-dimensional and joint two-dimensional 
posterior probability distributions of $n_{\rm H_2}$, $T_{\rm k}$ and $N_{\rm CO}/dv$, obtained from our MCMC sampling of the 
{\tt RADEX} parameter space in the two-component model, for the low (left) and high (right) excitation components. Contours 
increase in steps of 0.5$\sigma$. Yellow and orange solid lines show the maximum posterior probability of each parameter, 
while dotted lines mark the $\pm 1 \sigma$ range in the distributions. The resulting parameter values are listed above the 
corresponding histograms.}
\label{fig:g244}
\end{figure*}

\subsubsection{Modeling approach}

The large velocity gradient (LVG) approach \citep[e.g.,][]{young91} is commonly used to model the molecular gas excitation 
in galaxies having optically thick gas reservoirs that are not thermalized, as is the case for the GEMS (see 
Fig.~\ref{fig:normsled}). LVG models can predict the shape of the CO ladder by computing the collisional excitation of CO 
molecules for a range of gas physical conditions, assuming that turbulent motions within star-forming clouds result in 
velocity gradients in order to compute the escape probability of optically thick CO emission. We derived radiative transfer LVG 
models for the GEMS using the Markov chain Monte Carlo (MCMC) implementation of {\tt RADEX} \citep{vandertak07} presented in
\citet{yang17}\footnote{\url{https://github.com/yangcht/radex_emcee}}. This one-dimensional code assumes spherical symmetry 
and describes the velocity gradients, $dv/dr$, in the expanding sphere approximation. It performs an MCMC sampling of the 
parameter space that includes the molecular hydrogen density, $n_{\rm H_2}$, the gas kinetic temperature, $T_{\rm k}$, the CO 
column density per unit velocity gradient, $N_{\rm CO}/dv$, and the size of the emitting region, and samples the posterior 
probability distributions functions from the CO line fluxes modeled by {\tt RADEX}.

Following the CO excitation analysis of strongly-lensed SMGs from the H-ATLAS survey presented in \citet{yang17}, we assumed
flat prior probabilities for the input parameters within the following ranges: $n_{\rm H_2}=10^{1.5}$--$10^{7.0}$~cm$^{-3}$, 
$T_{\rm k}=T_{\rm CMB}-10^3$~K and $N_{\rm CO}/dv=10^{15.5}$--$10^{19.5}$~cm$^{-2}$~km$^{-1}$~s. Here $T_{\rm CMB}$ is the CMB temperature 
at the redshift of the source \citep[see further justifications and references in][]{yang17}. The size of the CO-emitting 
region is also a free parameter in {\tt RADEX} that acts as a normalization factor on the overall SLEDs, but the size 
estimates from the LVG models are quite uncertain given the dependency on both the magnification factor and the beam filling 
factor. We therefore refer to Sect.~\ref{sec:discu} for a dicussion of the physical extent of the molecular gas reservoirs 
and mainly discuss the best-fitting values of $n_{\rm H_2}$, $T_{\rm k}$ and $N_{\rm CO}/dv$ that fully determine the shape of 
the SLEDs.
 
For each source, we sampled the posterior probability distribution functions using 1000 MCMC iterations. As an example, we 
show the one-dimensional and joint two-dimensional posteriors of each parameter for PLCK\_G244.8+54.9 in Figure~\ref{fig:g244}. 
The best-fitting values of $n_{\rm H_2}$, $T_{\rm k}$ and $N_{\rm CO}/dv$ are taken from the maximum of the joint probability 
distribution and are listed in Table~\ref{tab:radout} together with 1$\sigma$ uncertainties. We also plot the modeled CO 
SLEDs from the best-fitting estimates of the gas densities and kinetic temperatures together with our CO flux measurements 
with EMIR in Figs.~\ref{fig:g244}, \ref{fig:sled2}, and \ref{fig:sled3}.

\subsubsection{Integrated physical properties}

The well-sampled mid-$J$ regime of the CO SLEDs provides robust constraints on the gas density and temperature for all GEMS.
For a single gas excitation component, we obtain molecular hydrogen densities ranging between 10$^{2.6}$ and 10$^{4.1}$~cm$^{-3}$,
gas temperatures log$(T_{\rm k})=1.5$--3.0~K and $N_{\rm CO}/dv = 10^{16-17.5}$~cm$^{-2}$~km$^{-1}$~s. These conditions are very
similar to those within high-redshift SMGs from the H-ATLAS sample \citep{yang17} and cover the same density and temperature
regimes as local ULIRGs, whose CO ladders also peak between $J_{\rm up}=4$ and $J_{\rm up}=7$ \citep[e.g.,][]{weiss05,rosenberg15}.

\begin{figure*}
\centering
\includegraphics[width=0.3\textwidth]{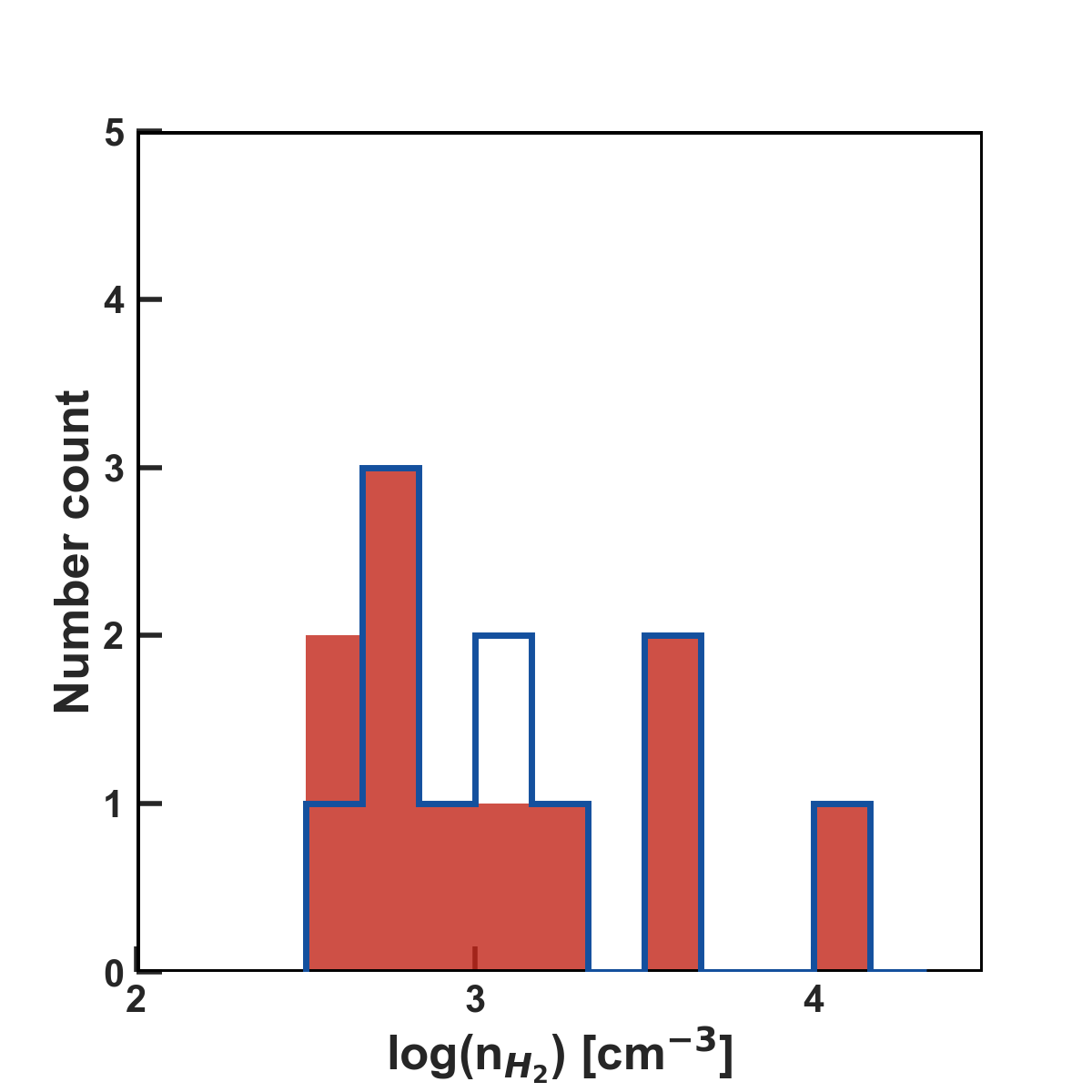}
\includegraphics[width=0.3\textwidth]{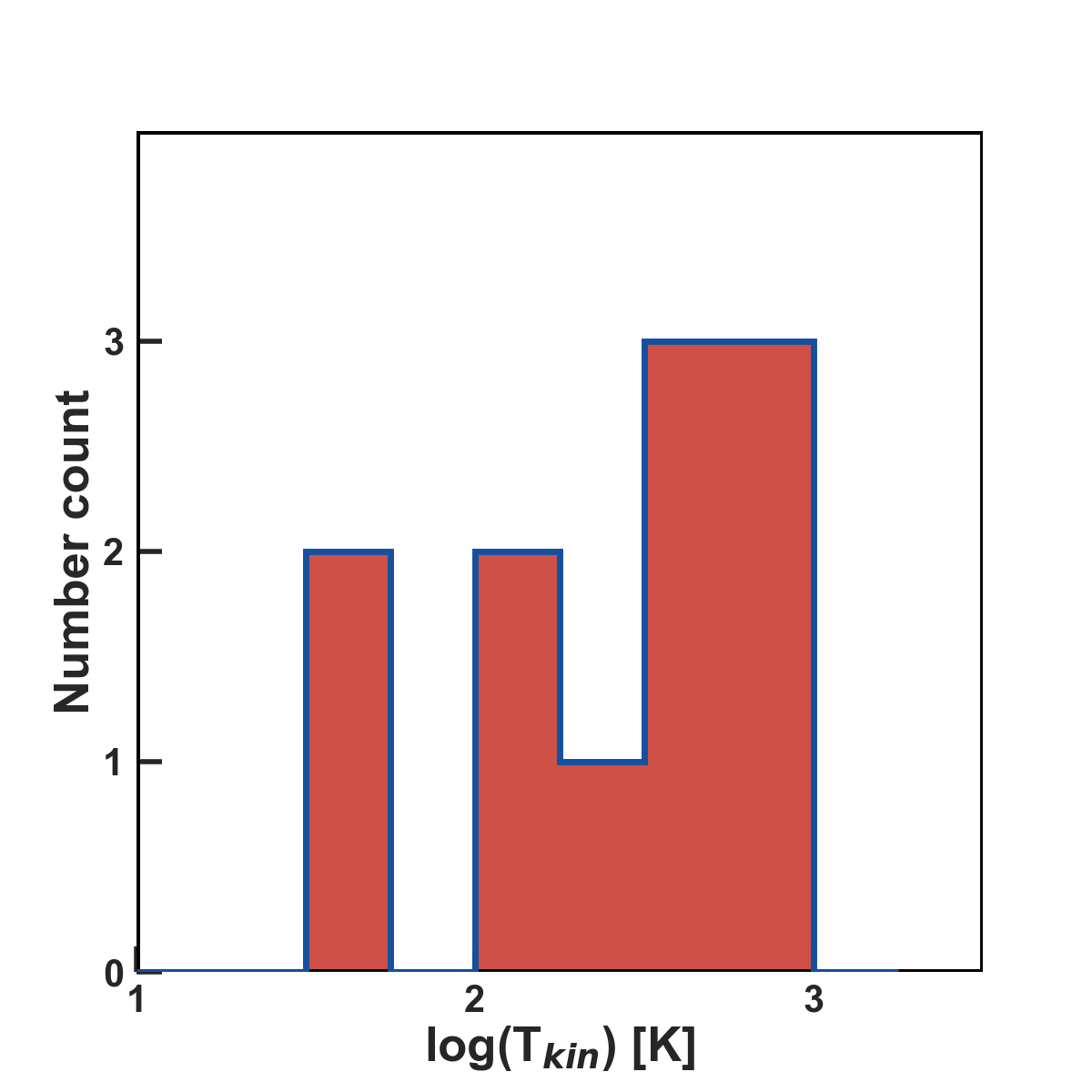}
\includegraphics[width=0.3\textwidth]{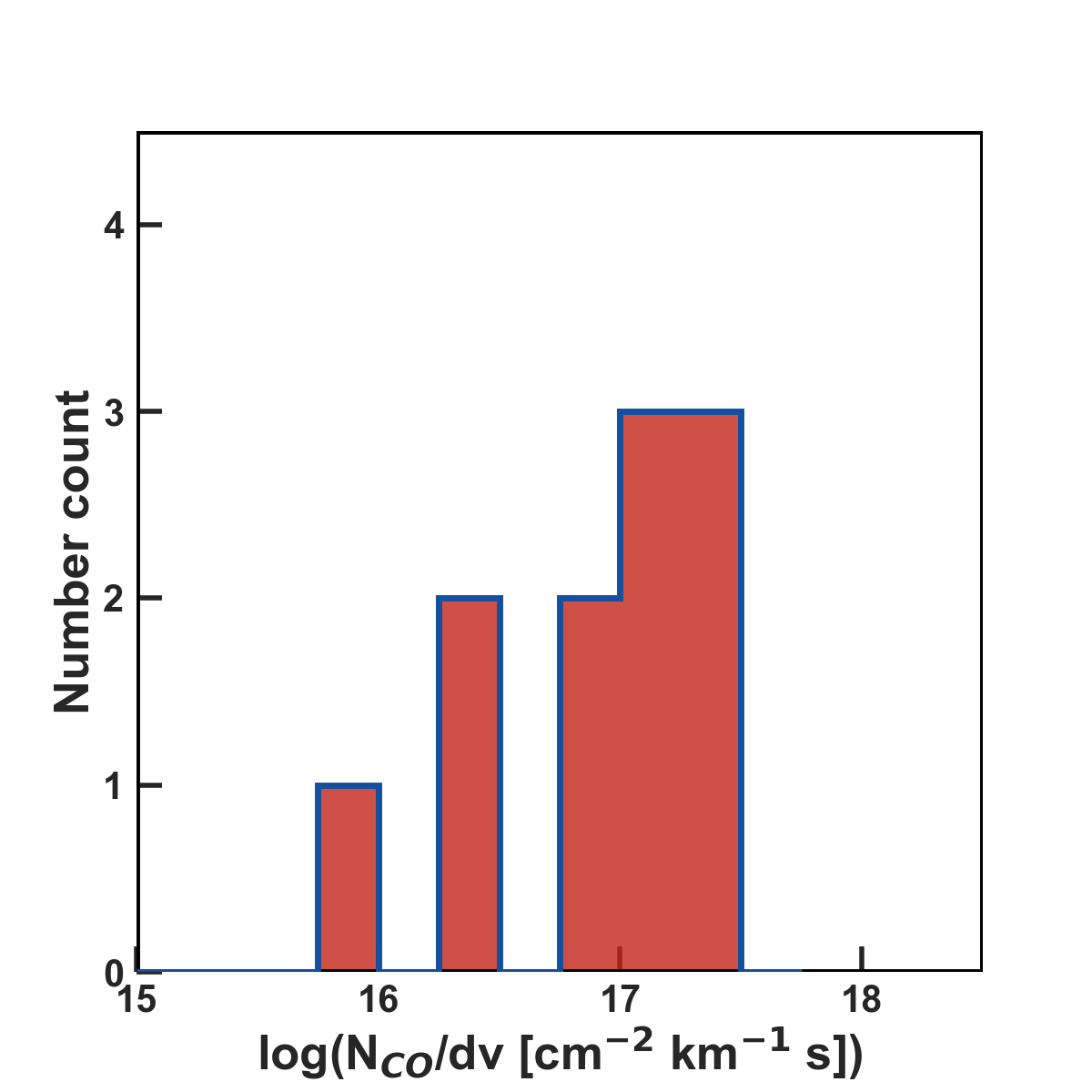}
\caption{Results of the gas excitation analysis with {\tt RADEX}. Red histograms show the median values of the marginal 
probability distributions of $n_{\rm H_2}$, $T_{\rm k}$ and $N_{\rm CO}/dv$ obtained by modeling the CO SLEDs with a single 
excitation component (see Table~\ref{tab:radout}). Blue histograms illustrate how results change when removing the 
CO(1--0) fluxes from the analysis.}
\label{fig:radout}
\end{figure*}

The CO(1--0) line emission in high-redshift dusty star-forming galaxies may have contributions from gas components not seen in 
the mid-$J$ CO lines \citep[e.g.,][]{ivison10}, and this transition appears to be a good proxy of the extended and low-density 
molecular gas reservoirs that are not directly related to star formation. We reproduced the single-component analysis 
without the CO(1--0) fluxes in order to determine how the gas density and kinetic temperature are affected by this transition 
and to avoid the difficulty of constraining possible differential lensing effects between $J_{\rm up}=1$ and higher-$J$ 
transitions. The comparison shown in Figure~\ref{fig:radout} only indicates minor differences in the inferred molecular gas 
properties, suggesting that this single data point can not bias the fit of our SLEDs, well-sampled between $J_{\rm up}=3$ and 
$J_{\rm up}=7$.

In some cases our simple one-component model poorly reproduces the CO ladder. For example, a significant discrepancy is found 
at $J_{\rm up} \geq 6$ for PLCK\_G244.8+54.9 (Fig.~\ref{fig:g244}), and both the CO(1--0) and $J_{\rm up} > 7$ fluxes are 
underestimated for PLCK\_G092.5+42.9 and PLCK\_G113.7+61.0 (Fig.~\ref{fig:sled2}). This suggests that the 
CO SLEDs in those GEMS trace at least two gas phases with distinct physical properties, an extended and cold gas phase with 
low excitation and the more compact and warmer gas reservoirs with higher excitation. Previously, several studies also used 
multicomponent LVG models to adequately describe the CO excitation, both at low \citep[e.g.,][]{weiss05} and high redshift 
\citep[e.g.,][]{daddi15}, including for distant obscured starbursts \citep[e.g.,][]{danielson11,hodge13,yang17}. Obtaining
reliable detections of low-excitation gas phases in SMGs remains nonetheless challenging because this component is intrinsically
weaker than in local normal star-forming galaxies \citep{rosenberg15}, and because spatially-integrated CO SLEDs are sensitive 
to luminosity-weighted parameters and therefore dominated by the less massive, but highly excited dense gas reservoirs 
\citep{kamenetzky18}. The CO SLEDs of a majority of six out of 11 GEMS are not sufficiently sampled to identify such multiple 
gas phases.

We used two-excitation component models with the same approach, and assigned different parameters to each excitation 
component within common physical boundaries. Most importantly, the model assumes two additional priors on the relative sizes
and temperatures within the Bayesian analysis, requiring that the low-excitation gas component is cooler and more extended than 
the high-excitation component \citep[as supported by multi-$J$ CO observations in SMGs,][]{ivison11,casey14}. For PLCK\_G092.5+42.9
and PLCK\_G244.8+54.9, the presence of the two gas phases with different levels of excitation is highlighted by the two apparent 
peaks and the best-fitting solution from the two-component model provides a much better fit to the overall SLED (see 
Figs.~\ref{fig:g244} and \ref{fig:sled2}). The two components are also apparent in PLCK\_G113.7+61.0, PLCK\_G138.6+62.0, and 
PLCK\_G165.7+67.0, where the single-component model significantly underestimates the CO(1--0) and CO(3--2) fluxes (see 
Fig.~\ref{fig:sled2}). The overall SLEDs of these sources are nevertheless nearly constant up to $J_{\rm up}=6$ and dominated by 
the excited gas phase. For this reason we checked that the second lower excitation phase is also detected after removing the 
CO(1--0) lines from the analysis, and found that the CO(3--2) fluxes remain underestimated by the new models. We therefore conclude 
that the detection of multiple gas phases in PLCK\_G113.7+61.0, PLCK\_G138.6+62.0, and PLCK\_G165.7+67.0 is robust and that the 
properties of the low excitation component inferred with and without the CO(1--0) fluxes are consistent within 1$\sigma$.

Our two-component models favors low-excitation gas phases peaking between $J_{\rm up}=2$ and 5, while the high-excitation
components peak at higher $J$ and consequently have elevated densities and temperatures (best-fitting results summarized in 
Table~\ref{tab:radout}). For PLCK\_G092.5+42.9 and PLCK\_G244.8+54.9, the high excitation component even reaches 
log$(n_{\rm H_2}/{\rm cm}^{-3}) = 4.49^{+0.87}_{-1.17}$ and $3.83^{+0.79}_{-0.94}$, for kinetic temperatures of about 70 and 200~K, 
respectively. In these two sources, the cooler and more extended gas phase covers the main peak at $J_{\rm up}=4$--6 and is 
thereby significantly excited, with conditions similar to those of the highly excited gas phase in the Cosmic Eyelash 
\citep{danielson11}, or to the bulk of the gas reservoirs in low-redshift compact ULIRGs and nearby starbursts 
\citep[e.g.,][]{bradford03}. Interestingly, for PLCK\_G113.7+61.0, PLCK\_G138.6+62.0, and PLCK\_G165.7+67.0, the best-fitting 
lower excitation component peaks at $J_{\rm up}=2$--3, similarly to the CO SLEDs observed for the inner Galactic disk 
\citep[as illustrated in Fig.~\ref{fig:normsled},][]{fixsen99} or local spirals \citep[e.g.,][]{braine92}. We measure 
molecular hydrogen densities of about $10^{2.8}$ and $10^{2.4}$~cm$^{-3}$ for PLCK\_G113.7+61.0 and PLCK\_G138.6+62.0, respectively, 
suggesting that this Milky Way-like component is diffuse and might trace extended gas reservoirs not directly fueling star 
formation, as already postulated for other high-redshift dusty starbursts \citep[e.g., the Cosmic Eyelash,][]{danielson11}.

\begin{table}
\centering
\begin{tabular}{lccc}
\hline
\hline 
\rule{0pt}{2ex} Source & log($n_{\rm H_2}$) & log($T_{\rm k}$) & log($N_{\rm CO}/dv$) \\
 & [cm$^{-3}$] & [K] & [cm$^{-2}$ km$^{-1}$ s] \\
\hline \\[-0.4em]
 & \multicolumn{3}{l}{Single component} \\[+0.3em]
PLCK\_G045.1+61.1 & $3.61^{+0.70}_{-1.02}$ & $2.17^{+0.25}_{-0.27}$ & $16.45^{+0.96}_{-0.69}$ \\[+0.3em]
PLCK\_G080.2+49.8 & $3.10^{+0.68}_{-0.73}$ & $2.20^{+0.33}_{-0.30}$ & $16.79^{+0.68}_{-0.73}$ \\[+0.3em]
PLCK\_G092.5+42.9 & $3.20^{+0.11}_{-0.80}$ & $2.98^{+0.02}_{-0.03}$ & $15.93^{+1.28}_{-0.33}$ \\[+0.3em]
PLCK\_G102.1+53.6 & $4.06^{+1.34}_{-1.27}$ & $1.51^{+0.44}_{-0.24}$ & $17.43^{+1.16}_{-1.05}$ \\[+0.3em]
PLCK\_G113.7+61.0 & $2.59^{+0.26}_{-0.20}$ & $2.88^{+0.09}_{-0.19}$ & $17.20^{+0.12}_{-0.20}$ \\[+0.3em]
PLCK\_G138.6+62.0 & $2.67^{+0.41}_{-0.32}$ & $2.74^{+0.19}_{-0.39}$ & $17.20^{+0.21}_{-0.25}$ \\[+0.3em]
PLCK\_G145.2+50.9 & $3.65^{+1.27}_{-1.09}$ & $1.60^{+0.57}_{-0.27}$ & $17.40^{+1.07}_{-0.83}$ \\[+0.3em]
PLCK\_G165.7+67.0 & $2.75^{+0.52}_{-0.48}$ & $2.40^{+0.35}_{-0.35}$ & $17.15^{+0.47}_{-0.31}$ \\[+0.3em]
PLCK\_G200.6+46.1 & $2.63^{+0.45}_{-0.42}$ & $2.71^{+0.18}_{-0.23}$ & $16.28^{+0.54}_{-0.53}$ \\[+0.3em]
PLCK\_G231.3+72.2 & $2.95^{+0.43}_{-0.58}$ & $2.69^{+0.21}_{-0.35}$ & $16.77^{+0.50}_{-0.80}$ \\[+0.3em]
PLCK\_G244.8+54.9 & $2.80^{+0.08}_{-0.07}$ & $2.97^{+0.02}_{-0.04}$ & $17.48^{+0.05}_{-0.08}$ \\[+0.5em]
\hline \\[-0.4em]
 & \multicolumn{3}{l}{Two components} \\[+0.3em]
PLCK\_G092.5+42.9 L & $3.97^{+0.84}_{-1.44}$ & $1.21^{+0.06}_{-0.05}$ & $17.27^{+1.22}_{-0.60}$ \\[+0.3em]
PLCK\_G092.5+42.9 H & $4.49^{+0.87}_{-1.17}$ & $1.82^{+0.18}_{-0.15}$ & $18.09^{+0.87}_{-0.75}$ \\[+0.3em]
PLCK\_G113.7+61.0 L & $2.81^{+0.80}_{-0.82}$ & $1.44^{+0.31}_{-0.24}$ & $16.54^{+0.83}_{-1.04}$ \\[+0.3em]
PLCK\_G113.7+61.0 H & $3.22^{+0.72}_{-0.90}$ & $2.46^{+0.34}_{-0.33}$ & $17.48^{+0.91}_{-0.58}$ \\[+0.3em]
PLCK\_G138.6+62.0 L & $2.42^{+0.81}_{-0.64}$ & $1.55^{+0.27}_{-0.29}$ & $16.18^{+0.98}_{-1.10}$ \\[+0.3em]
PLCK\_G138.6+62.0 H & $2.87^{+0.96}_{-0.88}$ & $2.33^{+0.40}_{-0.39}$ & $17.66^{+0.89}_{-0.72}$ \\[+0.3em]
PLCK\_G165.7+67.0 L & $3.22^{+0.74}_{-1.08}$ & $1.69^{+0.25}_{-0.34}$ & $17.14^{+0.91}_{-1.01}$ \\[+0.3em]
PLCK\_G165.7+67.0 H & $2.95^{+1.30}_{-0.97}$ & $2.20^{+0.47}_{-0.38}$ & $17.57^{+1.00}_{-1.07}$ \\[+0.3em]
PLCK\_G244.8+54.9 L & $3.06^{+0.70}_{-1.09}$ & $1.82^{+0.19}_{-0.17}$ & $17.54^{+0.90}_{-0.59}$ \\[+0.3em]
PLCK\_G244.8+54.9 H & $3.83^{+0.79}_{-0.94}$ & $2.30^{+0.21}_{-0.22}$ & $18.05^{+0.87}_{-0.66}$ \\[+0.2em]
\hline
\end{tabular}
\caption{Molecular gas properties of the {\it Planck}'s dusty GEMS inferred from the MCMC sampling of the LVG model 
parameter space using {\tt RADEX}. Values quoted for each parameter are the median and $\pm 1 \sigma$ uncertainties 
of the marginal probability distribution functions. All sources are modeled with a single excitation component and those
for which there is a significant mismatch between the best-fit model and observed fluxes of low and/or high-$J$ CO lines
are also modeled with two excitation components. In these cases, ``L'' and ``H'' indicate the low and high-excitation 
components, respectively.}
\label{tab:radout}
\end{table}

\subsubsection{Individual spectral components}

We used the spectral component separation described in Sect.~\ref{ssec:cosled} and the individual CO SLEDs shown in 
Fig.~\ref{fig:cogrid} to study the variations of the gas excitation within each kinematic component. The new LVG models 
were derived using the same approach and the same number of gas excitation components as for the spectrally-integrated 
CO spectra (see Fig.~\ref{fig:sled4}). The best-fitting models were then extrapolated down to low-$J$ to predict CO(1--0) 
fluxes and derive gas masses for the blue and red components, assuming $\alpha_{\rm CO}=0.8$~M$_{\odot}$/(K~km~s$^{-1}$~pc$^2$)
(see Table~\ref{tab:mass}).

For PLCK\_G045.1+61.1, the two modeled CO SLEDs peak between $J_{\rm up}=4$ and $J_{\rm up}=5$ similar to that obtained for the 
global spectra in Fig.~\ref{fig:sled3}. The density and temperature of the single gas excitation components are therefore 
consistent, within the uncertainties, with log$(n_{\rm H_2}/{\rm cm}^{-3})=3.61^{+0.70}_{-1.02}$ and log$(T_{\rm k}/{\rm K})=2.17^{+0.25}_{-0.27}$, 
as measured on the overall SLED. Comparing the CO(1--0) fluxes extrapolated from the best-fitting models shown in 
Fig.~\ref{fig:sled4} with those extrapolated from the LVG models of the overall SLED (Fig.~\ref{fig:sled3}) suggests that the 
blue and red components each comprise about half of the total gas mass in PLCK\_G045.1+61.1.

The two spectral components in PLCK\_G092.5+42.9 also exhibit similar levels of excitation, including a cool and extended gas phase 
peaking at $J_{\rm up}=3$--5 and a warmer and more compact phase producing a secondary peak at $J_{\rm up}=7$--9. The red kinematic 
component more closely resembles the global SLED and strongly peaks at $J_{\rm up}=4$--5, although the flux drop measured for 
the CO(6--5) transition is affected by large uncertainties on the component separation, and possibly by differential magnification 
effects (see discussion in Sect.~\ref{sssec:diff}). Its properties are fully consistent with the best-fit parameters of 
Table~\ref{tab:radout}. The kinetic temperatures of the blue kinematic component are also similar and its densities are 
log$(n_{\rm H_2}/{\rm cm}^{-3})=2.17^{+0.96}_{-0.53}$ and $3.20^{+0.81}_{-0.95}$ for the low and high-excitation phases, respectively, which 
is somewhat lower, but consistent within 1$\sigma$ with those measured on the overall SLED. Models of the blue and red components 
extrapolated to $J_{\rm up}=1$ show that they contain about 30\% and 20\% of the gas mass inferred from the CO(1--0) line detection, 
respectively, suggesting that gas components not seen in mid-$J$ CO lines also contribute to the CO(1--0) fluxes.

In PLCK\_G244.8+54.9, the two components have comparable amplitudes over the entire SLED, and the CO(1--0) fluxes extrapolated 
from the two-component gas excitation models suggest that they contain similar gas masses. The properties of the red component are
consistent with those listed in Table~\ref{tab:radout}. However, the SLED of the blue component rises from $J_{\rm up}=8$ to 
$J_{\rm up}=10$, which given the large uncertainties, provides marginal evidence that the warm gas phase in the blue component is 
more excited, with molecular gas densities about one order of magnitude higher than those determined from the integrated SLED.

\subsection{CO-inferred molecular gas masses}
\label{ssec:mgas}

Detailed modeling of well-sampled CO SLEDs provide robust estimates of the molecular gas masses from mid-$J$ CO lines. The 
measured CO line luminosities can be converted to those of CO(1--0), $L'_{\rm CO(1-0)}$, and to the total molecular hydrogen 
masses, $M^{\rm CO}({\rm H_2})$, using the best-fitting CO excitation models and the relation 
$M^{\rm CO}({\rm H_2})=\alpha_{\rm CO}L'_{\rm CO(1-0)}$. For the 11 {\it Planck}'s dusty GEMS, we extrapolated the best-fitting LVG 
models inferred exclusively from the $J_{\rm up}>3$ CO flux measurements, and predicted the CO(1--0) fluxes and total masses 
of gas embedded within these excited reservoirs, directly related to the on-going star formation. This method does not 
include the possible additional contribution from diffuse ISM components extended over roughly kpc scales and spatially 
segregated from the excited gas phases that have already been detected in some SMGs \citep[see e.g.,][]{harris10}, and it is 
not strongly sensitive to differential magnification since this effect is not significantly affecting the overall shape of 
the CO SLEDs (see Sect.~\ref{sssec:diff}). We used the two-excitation components model for the five sources with CO SLEDs that
are poorly fit by a single combination of $n_{\rm H_2}$ and $T_{\rm k}$, and the single-component model for the six others (see 
Figs.~\ref{fig:sled2} and \ref{fig:sled3}). The predicted CO(1--0) fluxes were then converted into line luminosities and 
molecular hydrogen masses using $\alpha_{\rm CO}=0.8$~M$_{\odot}$/(K~km~s$^{-1}$~pc$^2$), following the discussion in 
Sect.~\ref{ssec:aCO}.

We obtain the total gas masses listed in Table~\ref{tab:mass}, which are spatially-integrated over the source components 
falling within the beam of the IRAM 30-m telescope and corrected for the gravitational magnification factors, $\mu_{\rm gas}$, 
of Table~\ref{tab:prop}. $M^{\rm CO}({\rm H_2})$ broadly ranges between 10$^{10}$ and 10$^{11}$~M$_{\odot}$, with a mean value of 
$4.3 \times 10^{10}$~M$_{\odot}$, akin to the masses obtained for other samples of lensed or unlensed high-redshift SMGs with 
single-dish CO(1--0) detections or well-sampled CO SLEDs \citep[e.g.,][]{harris12,bothwell13}. This shows that the GEMS have 
global gas contents comparable with the overall SMG population, despite being extremely bright on the sky due to their strong 
gravitational magnifications. PLCK\_G145.2+50.9 is almost a factor two more massive than other sources in the sample, suggesting 
that this is an extremely gas-rich starburst or that our estimate of the magnification factor may be too low, for example due 
to the presence of large-scale structures at different redshifts along the line of sight. The best-fitting models 
in Figs.~\ref{fig:g244} and \ref{fig:sled2} indicate that the low-excitation component dominates the overall mass budget in 
PLCK\_G092.5+42.9, PLCK\_G113.7+61.0, PLCK\_G138.6+ 62.0, and PLCK\_G244.8+54.9, while the mass of both components are comparable 
in PLCK\_G165.7+67.0. For these five sources, the molecular gas masses would be systematically lower if we had used the single 
component models instead.

We then used the dust masses from SED fitting in C15 and the luminosity-weighted magnification factors of the dust 
continuum, $\mu_{\rm dust}$, presented in Table~\ref{tab:prop}, to infer the global gas-to-dust mass ratios, $\delta_{\rm GDR}$
(see Table~\ref{tab:mass}). We obtained an average gas-to-dust mass ratio of about 150 over the sample, consistent with other 
high-redshift SMGs \citep[e.g.,][]{ivison11} and local ULIRGs \citep[e.g.,][]{solomon97}.

The best-fitting {\tt RADEX} models of the CO $J_{\rm up} \geq 3$ ladder systematically underestimate the CO(1--0) fluxes for 
the five sources detected with the GBT in \citet{harrington18}, regardless of our assumption of the number of excitation 
components. If this reflects the intrinsic ratio of the gas mass probed in these transitions, then it implies that some of 
the {\it Planck}'s dusty GEMS contain low-density gas reservoirs segregated from the excited components traced by the mid- to 
high-$J$ CO lines observed with EMIR. As discussed above, Figures~\ref{fig:g244} and \ref{fig:sled2} show that the cooler gas 
phase in the two-component LVG models of PLCK\_G092.5+42.9, PLCK\_G165.7+67.0, and PLCK\_G244.8+54.9 is still significantly 
excited, with peaks in the range $J_{\rm up}=3$--5 implying molecular gas densities of 10$^3$--10$^4$~cm$^{-3}$. This suggests that 
these GEMS might not only contain the two excited gas phases with different conditions, but also additional reservoirs of gas 
with lower levels of excitation that are not currently included in the analysis and would explain the excess of CO(1--0) 
emission. Assuming that the differences between the measured CO(1--0) fluxes and those predicted by the two-component models 
are indeed associated with such diffuse and low-excitation components, we find that these reservoirs enclose about 20--50\% of 
the total molecular hydrogen mass in PLCK\_G092.5+42.9, PLCK\_G165.7+67.0, and PLCK\_G244.8+54.9. This is comparable with the 
diffuse mass fraction of 50\% measured in the Cosmic Eyelash \citep{danielson11}. These fractions would rise up to 80\% if we 
instead assumed the single-component excitation models despite their poor description of the CO SLEDs in the high-$J$ regime. 

If the CO(1--0) emission arises partially from gas not probed in the higher-$J$ lines, then differential magnification can 
be an issue, provided that the relative calibration between the GBT and IRAM observations is robust. Several authors 
\citep[e.g.,][]{rybak15,spilker15} have already pointed out that CO(1--0) line luminosities can have different magnification 
factors than higher-$J$ CO lines, due to the different spatial distributions of the underlying gas reservoirs. Quantifying this 
effect would require subarcsec resolution CO(1--0) interferometry. 
However, we showed in Sect.~\ref{ssec:lens} that differential magnification effects between the mid-$J$ CO emission and dust 
continuum remain below 30\%. The excess of CO(1--0) emission observed for these GEMS is also more significant than the upper 
limit on the differential lensing effect between low- and mid-$J$ CO lines of about 30\% predicted by \citet{serjeant12}, assuming 
conservatively that the cold and warm CO phases follow very different distributions (thus ignoring massive star-forming disks with 
well-mixed gas phases). Moreover, attributing the enhanced CO(1--0) emission in the GEMS solely to differential lensing would 
imply $\mu_{\rm CO(1-0)} > \mu_{\rm mid-J\ CO}$, consistently for all sources. This would correspond to caustic line positions 
preferentially magnifying the extended low-density regions emitting CO(1--0), with respect to the compact star-forming clouds 
emitting the bulk of the mid-$J$ CO lines. Although such configurations are not ruled out, for high magnification factors 
$\mu \simeq 20$, the compact regions are more likely to be more strongly magnified than the extended ones \citep{hezaveh12}, 
which strongly disfavors this interpretation. 

For PLCK\_G113.7+61.0 and PLCK\_G138.6+62.0, the two other GEMS with CO(1--0) measurements, the lower excitation component 
included in the LVG analysis peaks at $J_{\rm up}=2$--3. This is characteristic of diffuse gas, as already discussed in 
Sect.~\ref{ssec:lvg}, and includes between 50\% and 80\% of the total molecular gas mass in the systems, depending on whether 
CO(1--0) is included in the fit or not. We note that these mass fractions are highly uncertain since the SLEDs are flat in 
the $J_{\rm up}>4$ regime, which complicates the component separation.

\subsection{Photon-dominated region models}
\label{ssec:pdr}

In star-forming galaxies, the physical properties and chemical composition of gas clouds affected by the surrounding 
intense radiation field from the newly formed massive stars have been extensively described by photon-dominated region 
(PDR) models \citep[e.g.,][]{kaufman99,lepetit06,meijerink07}. These models are now proposing a coherent picture of the 
structure of molecular clouds, in the intermediate column density regime between \ion{H}{ii} regions and prestellar cores, 
where atomic hydrogen remains neutral and the gas material is predominantly heated by the external FUV field. Atomic carbon 
is ionized in the outer layers and the CO molecules form deeper within the clouds where the ionizing FUV radiation is 
sufficiently attenuated \citep[e.g.,][]{hollenbach97}. In this section, we derive simple PDR models to infer the gas 
densities and the strengths of the incident radiation fields from the spatially-integrated line fluxes.

\subsubsection{Method}

\begin{figure}
\centering
\includegraphics[width=0.48\textwidth]{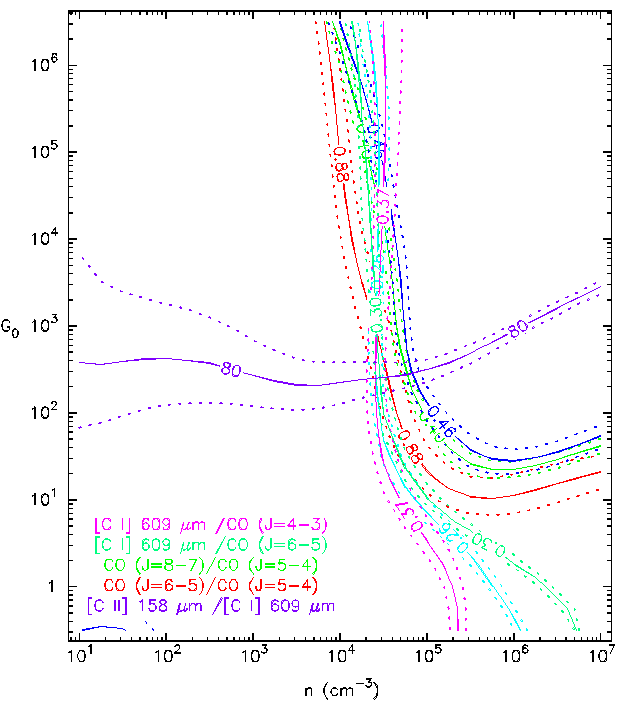}
\caption{Example of constraints on the number densities of hydrogen nuclei, $n_{\rm PDR}$, and the FUV radiation fields, $G_{\rm 0}$, 
obtained for PLCK\_G045.1+61.1 using the PDR models from \citet{kaufman99,kaufman06}. Contours show the CO, [\ion{C}{i}] and 
[\ion{C}{ii}] luminosity ratios, as labeled in the caption, with solid lines showing the average ratios and dotted lines their 
1$\sigma$ uncertainties.}
\label{fig:pdrex}
\end{figure}

The PDR diagnostics of the GEMS were derived using the standard one-dimensional models of \citet{kaufman99,kaufman06}, which have 
been widely applied in ISM studies of high-redshift SMGs \citep[e.g.,][]{rawle14,gullberg15,wardlow17}. We derived the combination 
of cloud density, $n_{\rm PDR}$, and FUV radiation field, $G_{\rm 0}$, that best reproduce the observed line luminosity ratios using 
the PDR {\sc Toolbox} \citep{pound08}, a publicly available implementation of the model diagnostics. We followed this approach to 
infer the physical conditions in the GEMS, under the assumption that they comprise a single giant molecular cloud \citep{kaufman99}. 
Although this simple scenario cannot truly do justice to the intrinsic complexity of the ISM configurations, it allows us to 
perform a simple and uniform treatment of the integrated properties and to derive the luminosity-weighted, spatially-averaged gas 
properties. Unlike for the LVG analysis, the average PDR densities are expressed in terms of number densities of hydrogen nuclei 
and vary in the range 1~$<$~log($n_{\rm PDR}$/cm$^{-3}$)~$<$~7. The strengths of the radiation fields produced by the surrounding 
young stellar populations (given in units of the Habing field, i.e., 1.6~$\times$~10$^{-3}$~erg~cm$^{-2}$~s$^{-1}$) are in the range 
$-$0.5~$<$~log($G_{\rm 0}$)~$<$~6.5.

We selected the line ratios that provide complementary constraints on $n_{\rm PDR}$ and $G_{\rm 0}$ and assumed, for a given 
source, that all spatially-integrated line luminosities arise from a unique PDR that covers the same surface area at all
frequencies. High-resolution dust and gas interferometry of high-redshift dusty star-forming galaxies show that this population 
usually hosts a range of giant molecular cloud complexes \citep[e.g.,][]{thomson15} associated with distinct PDRs, as observed 
in the local Universe, and this approximation therefore implies that the output parameters will be luminosity weighted toward
the intrinsically brightest regions.

\begin{table*}
\centering
\begin{tabular}{lcccc}
\hline
\hline
\rule{0pt}{2ex} Source & log($n_{\rm PDR}$) & log($G_{\rm 0}$) & $T$ & $R_{\rm PDR}$ \\
 & [cm$^{-3}$] & [Habing units] & [K] & [kpc] \\
\hline
PLCK\_G045.1+61.1 & 4.44 $\pm$ 0.19 & 2.50 $\pm$ 0.17 & $100^{+20}_{-12}$ & 3.3 \\[+0.3em]
PLCK\_G080.2+49.8 & 3.98 $\pm$ 0.11 & 2.25 $\pm$ 0.22 & $140^{+30}_{-30}$ & 3.7 \\[+0.3em]
PLCK\_G092.5+42.9 & 4.84 $\pm$ 0.19 & 4.18 $\pm$ 0.73 & $800^{+550}_{-520}$ & 0.9 \\[+0.3em]
PLCK\_G102.1+53.6 & 4.80 $\pm$ 0.69 & 3.50 $\pm$ 0.85 & $320^{+250}_{-180}$ & 1.1 \\[+0.3em]
PLCK\_G113.7+61.0 & 4.28 $\pm$ 0.09 & 3.43 $\pm$ 0.25 & $300^{+70}_{-50}$ & 1.6 \\[+0.3em]
PLCK\_G138.6+62.0 & 4.53 $\pm$ 0.08 & 3.70 $\pm$ 0.41 & $360^{+100}_{-90}$ & 0.8 \\[+0.3em]
PLCK\_G145.2+50.9 & 4.47 $\pm$ 0.10 & 2.84 $\pm$ 0.38 & $120^{+40}_{-20}$ & 5.8 \\[+0.3em]
PLCK\_G165.7+67.0 & 4.45 $\pm$ 0.08 & 3.40 $\pm$ 0.12 & $260^{+40}_{-60}$ & 1.1 \\[+0.3em]
PLCK\_G200.6+46.1 & 4.16 $\pm$ 0.36 & 2.17 $\pm$ 0.34 & $110^{+30}_{-20}$ & 4.5 \\[+0.3em]
PLCK\_G231.3+72.2 & 4.40 $\pm$ 0.26 & 3.15 $\pm$ 0.71 & $180^{+180}_{-80}$ & 2.3 \\[+0.3em]
PLCK\_G244.8+54.9 & 5.09 $\pm$ 0.12 & 3.73 $\pm$ 0.31 & $540^{+240}_{-170}$ & 1.4 \\[+0.3em]
\hline
\end{tabular}
\caption{Best-fitting values of the average number density of hydrogen nuclei, $n_{\rm PDR}$, and FUV radiation field, 
$G_{\rm 0}$, obtained for the {\it Planck}'s dusty GEMS using the publicly available implementation of the PDR models from 
\citet{kaufman99,kaufman06}. The global PDR conditions in PLCK\_G045.1+61.1, obtained from the line ratios illustrated in 
Figure~\ref{fig:pdrex}, are consistent with those in the brightest of the four multiple images \citep{nesvadba16}. Average 
PDR surface temperatures are deduced from the source position in the $n_{\rm PDR}$--$G_{\rm 0}$ parameter space. The last 
column lists rough estimates of the PDR sizes following \citet{wolfire90} and \citet{stacey10}, discussed further in 
Sect.~\ref{ssec:size}.}
\label{tab:pdr}
\end{table*}

The CO $J/(J-1)$ line luminosity ratios essentially constrain the range of gas density in the PDRs, even for the GEMS with 
high-$J$ CO detections. Diagnostics combining CO and atomic carbon transitions from our companion paper \citepalias{nesvadba18} 
also lead to degenerate solutions spanning several orders of magnitude in $G_{\rm 0}$. In order to better constrain the gas conditions 
we therefore included lose constraints from [\ion{C}{ii}], expected from the ensemble average properties of high-redshift galaxies.
In intense starbursts, the bulk of [\ion{C}{ii}] emission arises from PDRs \citep[e.g.,][]{stacey10,rigopoulou14} and, given its 
low critical density, this line preferentially traces the surface of PDRs. Consequently, line ratios involving [\ion{C}{ii}] 
efficiently probe the total energy budget from the external FUV radiation field \citep[e.g.,][]{goldsmith12}. Since most of the 
GEMS do not have available [\ion{C}{ii}] line measurements, we used the distribution of source-integrated [\ion{C}{ii}] line 
luminosities measured in \citet{gullberg15} for 17 strongly lensed SMGs from the South-Pole Telescope (SPT) survey. We used the 
average [\ion{C}{ii}]/FIR luminosity ratio for this sample, which we rescaled to the $L_{\rm FIR}$ values of each GEMS in the 
rest-frame range 42--500~$\mu$m \citepalias[using the photometry of][]{canameras15}. The resulting [\ion{C}{ii}] luminosities 
are consistent with our only available line detection for PLCK\_G045.1+61.1, $L_{\rm [\ion{C}{ii}]} \sim 5.4 \times 10^{10}$~L$_{\odot}$, 
spatially-integrated over the source components by rescaling the resolved ALMA line luminosity of \citet{nesvadba16}. We also used
the [\ion{C}{i}](1--0) line detections with EMIR for some GEMS \citepalias{nesvadba18}, and converted the [\ion{C}{i}](2--1) line 
luminosities to [\ion{C}{i}](1--0) using the average line ratio over the sample for the sources where the $^3$P$_{\rm 1}$--$^3$P$_{\rm 0}$ 
transition of atomic carbon falls outside the observing band.

The best-fitting values of $n_{\rm PDR}$ and $G_{\rm 0}$ listed in Table~\ref{tab:pdr} were inferred from these complementary PDR 
diagnostics of CO, [\ion{C}{i}] and [\ion{C}{ii}]. We did not distinguish either the different excitation components identified 
with our LVG models in Sect.~\ref{ssec:lvg}, or the spectral components \citep[as done, for instance, in][]{rawle14}, although 
deblending the PDR properties of individual velocity components within the sources should become feasible by combining the high 
S/N EMIR CO and [\ion{C}{i}] spectra with follow-up [\ion{C}{ii}] observations. We computed parameter uncertainties using Monte 
Carlo simulations by randomly drawing each line luminosity from a Gaussian distribution with $\sigma$ equal to the measurement 
error, deducing the line ratios and best-fitting values of $n_{\rm PDR}$ and $G_{\rm 0}$ from the PDR diagnostics, and taking the 
median and 1$\sigma$ errors on each parameter after 500 iterations.

Line ratios involving CO(1--0) from \citet{harrington18} occupy different regions in the $n_{\rm PDR}$--$G_{\rm 0}$ parameter space 
than those from mid-$J$ CO, [\ion{C}{i}](1--0), and [\ion{C}{ii}], and point toward FUV radiation fields being systematically lower 
than ratios involving other species, although CO(1--0) and [\ion{C}{i}](1--0) have similar critical densities and should both arise 
from the outer layers of PDRs. We found that the discrepancy vanishes when using the CO(1--0) luminosities extrapolated from our LVG 
models, except for PLCK\_G165.7+67.0. The multiple solutions obtained for this source might be due to blending of multiple spatial 
components in the beam of the IRAM 30-m telescope (see more details in Ca\~nameras et al., 2018a, A\&A accepted), or to 
different distributions of [\ion{C}{i}](1--0) and CO(1--0) emission on small
scales. Given these uncertainties and hints of distinct low density gas components, we did not consider the single-dish 
CO(1--0) measurements in the PDR analysis. We also put aside the high-$J$ CO transitions with $J_{\rm up}>7$ that trace the 
dense molecular gas of $n \gtrsim 10^5$~cm$^{-3}$ at the inner transition region between the PDRs and molecular clouds, where 
the external FUV radiation is significantly attenuated and other processes such as cosmic rays could have a major contribution 
to the gas heating \citep[e.g.,][]{papadopoulos10}. Furthermore, we verified that the gas densities are well constrained in the 
process and not dominated by our assumptions on the [\ion{C}{ii}] line luminosities by reproducing the PDR analysis only with the 
CO and [\ion{C}{i}] line ratios. The resulting values of $n_{\rm PDR}$ are consistent within 1$\sigma$ with the best-fitting results 
of Table~\ref{tab:pdr}. Finally, we also investigated how the results on $G_{\rm 0}$ would be affected if a significant fraction 
of the [\ion{C}{ii}] emission in the GEMS originates outside PDRs, for example from \ion{H}{ii} regions or diffuse gas reservoirs 
\citep{madden93,gerin15}. For the worst case scenario where non-PDR [\ion{C}{ii}] emission is about 50\% 
\citep[see also][]{abel06,decarli14}, a factor of two decrease in the [\ion{C}{ii}]/[\ion{C}{i}](1--0) luminosity leads to FUV 
radiation fields that are 0.5--0.7~dex lower than those presented in Table~\ref{tab:pdr}.

Recent studies including \citet{bothwell17} have advocated the need to account for the influence of cosmic rays on PDR
models, since this process becomes an important driver of gas excitation deeply within the molecular clouds shielded from the 
external FUV radiation \citep[][]{papadopoulos10,kazandjian15}. Cosmic ray radiation fields in intensely star-forming galaxies 
such as the GEMS are expected to be significantly enhanced compared to those in the Milky Way, due to efficient particle 
acceleration in supernova shockwaves. \citet{bothwell17} argue that including this gas heating source is crucial to avoid 
underestimating the PDR densities. However, quantifying the exact influence of cosmic rays on the gas conditions at high
attenuation requires the use of dense gas diagnostics such as line ratios of HCN, HNC, and HCO$^+$, which is beyond the scope 
of this paper. Here we use ratios of low- to mid-$J$ CO, [\ion{C}{i}] and [\ion{C}{ii}] lines, which are expected to vary little 
in the presence of enhanced cosmic ray ionization rates in intense starbursts, for sufficiently high surrounding FUV radiation 
fields \citep[][]{meijerink11}. For these reasons, we ignored this heating source during the analysis. Furthermore, we
emphasize that modifying other assumptions in the models, for instance on the overall PDR geometry, could affect the 
resulting PDR densities and FUV radiation fields to a similar extent.

\subsubsection{Properties of the PDRs}

\begin{figure}
\centering
\includegraphics[width=0.48\textwidth]{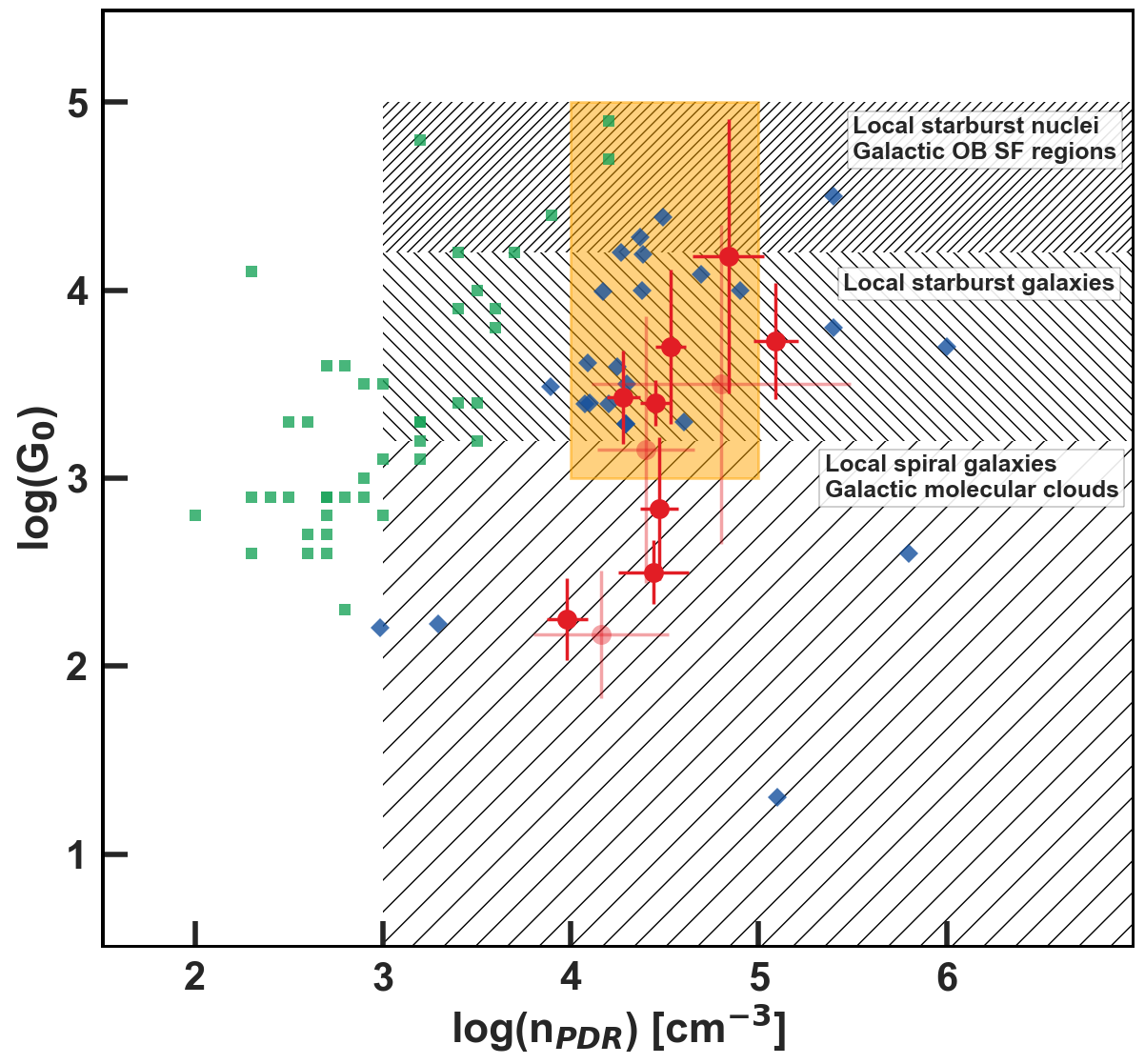}
\caption{Results of the PDR modeling of the {\it Planck}'s dusty GEMS using line luminosity ratios of CO, [\ion{C}{i}], and [\ion{C}{ii}] 
(red points, see further details in the text), with error bars showing the 1$\sigma$ uncertainties on the best-fitting values. The 
average PDR densities, $n_{\rm PDR}$, correspond to number densities of hydrogen nuclei and the FUV radiation fields, $G_{\rm 0}$, 
are given in Habing units (1.6~$\times$~10$^{-3}$~erg~cm$^{-2}$~s$^{-1}$). The sources without [\ion{C}{i}](1--0) detection are plotted 
in light red. We compare with other SMGs in the literature 
\citep[blue points,][]{cox11,danielson11,valtchanov11,alaghband13,huynh14,rawle14,bothwell17}, with the regimes of local normal 
star-forming galaxies from \citet[][green points]{malhotra01}, and local ULIRGs \citep[orange region,][]{davies03}, and with the 
average position of nearby galaxy populations and Galactic regions inferred in \citet{stacey91} using [\ion{C}{ii}]/CO(1--0) line 
ratios (hatched regions). The physical conditions in the GEMS closely resemble those of local ULIRGs and starbursts, with very 
similar densities in the range $n_{\rm PDR} = 10^4$--$10^5$~cm$^{-3}$. The PDR models also suggest that some {\it Planck}'s dusty GEMS 
are illuminated by lower FUV radiation fields, more typical of local spiral galaxies and molecular clouds in the Milky Way.}
\label{fig:pdr}
\end{figure}

We followed this approach to 
infer the best-fitting densities and radiation field strengths of the GEMS within the PDR scenario, and converted these quantities 
to the corresponding range of PDR surface temperatures using figure~2 of \citet{kaufman99}. All values are presented in 
Table~\ref{tab:pdr} and Fig.~\ref{fig:pdr}. The starbursts cover a small range in gas density, $n_{\rm PDR} = 10^4$--$10^5$~cm$^{-3}$, 
and about two orders of magnitude in external FUV field, $G_{\rm 0} = 10^2$--$10^4$ Habing fields. These PDR conditions are very similar 
to those in the Cosmic Eyelash \citep[log($n_{\rm PDR}$/cm$^{-3}$) $\simeq$ 4.1 and log($G_{\rm 0}$) $\simeq$ 3.6,][]{danielson11} and
unresolved studies of other high-redshift SMGs in the literature \citep[][]{cox11,valtchanov11,alaghband13,huynh14,rawle14}.
In Figure~\ref{fig:pdr}, the only points with PDR densities significantly above $10^{4-5}$~cm$^{-3}$ are the strongly lensed 
SMGs from the SPT survey presented in \citet{bothwell17}, likely due to the use of a different PDR code or to their 
implementation of enhanced cosmic ray ionization rates. In a similar sample drawn from the SPT, \citet{gullberg15} find 
log($G_{\rm 0}$)~=~2--4, comparable
with our measurements, and log($n_{\rm PDR}$/cm$^{-3}$)~=~2--5, using spatially-integrated [\ion{C}{ii}] and CO(1--0) line emission. 
Since [\ion{C}{ii}] arises from a wide range of environments, correcting the integrated [\ion{C}{ii}] flux for emission arising
from outside PDRs \citep[e.g. from \ion{H}{ii} regions, diffuse gas reservoirs,][]{madden93} would lower the [\ion{C}{ii}]/FIR 
luminosity ratios of that sample and increase the densities, closer to those measured in the GEMS (essentially from the mid-$J$ 
CO and [\ion{C}{i}] lines, see Fig.~\ref{fig:pdrex}).

These results compare well with the values $G_{\rm 0}=10^{2.2}$--$10^{4.5}$ obtained for $z=1$--4 SMGs in \citet{wardlow17}, 
by modeling the PDR conditions from stacked spectra of the [\ion{O}{i}]63$\mu$m and [\ion{Si}{ii}]34$\mu$m fine-structure lines in 
eight to 37 sources, and measurements of [\ion{C}{ii}] from the literature. However, this study finds average gas densities of 
10--1000~cm$^{-3}$, about two orders of magnitude lower than for other SMGs. This difference could be partly due to selection 
biases, systematic uncertainties from the stacking analysis that involves both lensed and unlensed SMGs, or possibly because these
fine-structure lines probe different gas regions than CO and \ion{C}{i} in these high-redshift environments, leading to complex 
geometric effects. As noted in \citet{wardlow17}, the difficulty in correctly assessing the level of [\ion{O}{i}]63$\mu$m 
self-absorption might also influence the resulting PDR densities \citep[see also][]{vasta10}.

In Figure~\ref{fig:pdr}, we compare the best-fitting values of $n_{\rm PDR}$ and $G_{\rm 0}$ with the regimes covered by local 
populations of star-forming galaxies. The PDRs in the {\it Planck}'s dusty GEMS are denser than those in the representative sample 
of normal star-forming galaxies with [\ion{C}{ii}] and [\ion{O}{i}] detections from \citet{malhotra01}. The PDR densities in the
GEMS correspond very well to those obtained in the nuclei of local ULIRGs by modeling the ratios of near-infrared molecular 
hydrogen lines \citep{davies03}, and are slightly lower than those within the dense cores of Galactic giant molecular clouds 
\citep[e.g.,][]{bergin96}. In addition, we show the FUV radiation fields derived from [\ion{C}{ii}]/CO(1--0) line luminosity ratios 
in \citet{stacey91} for local starbursts and non-starburst spiral galaxies, starburst cores, and giant molecular clouds and 
denser OB star-forming regions in the Milky Way. For most of the GEMS, the $G_{\rm 0}$ values closely resemble those measured 
in nearby ULIRGs and bluer starbursts. The models of PLCK\_G045.1+61.1, PLCK\_G080.2+49.8, PLCK\_G145.2+50.9, and 
PLCK\_G200.6+46.1 nonetheless indicate that the PDRs in these GEMS are illuminated by less intense radiation fields, more 
typical to the regimes of local spiral galaxies and Galactic molecular clouds \citep{stacey91}, but still 2--3 orders of 
magnitude stronger than the average value in the local Galactic ISM. Moreover, the conditions in the GEMS imply average PDR 
surface temperatures of 100--800~K.

The results of this PDR analysis generally suggest that the most intense dust-obscured starbursts at $z=2$--4 have ISM properties 
akin to various low-redshift environments, ranging from the central regions of ULIRGs triggered by major mergers and of bluer 
starburst galaxies, to normal star-forming galaxies and the dense Galactic star-forming clouds illuminated by young stars. High 
spatial resolution interferometric observations of multiple gas tracers combined with high magnifications are required to resolve 
individual giant molecular clouds in gravitationally lensed high-redshift dusty starbursts and further investigate the conditions 
within individual PDRs, as well as constraining their sizes and finding their local counterparts.

\section{Discussion}
\label{sec:discu}

\subsection{Gas density estimates}

Our density estimates from the LVG and PDR models presented in Tables~\ref{tab:radout} and \ref{tab:pdr} differ by about one 
order of magnitude, and average PDR densities, $n_{\rm PDR} = n_{\rm H} + 2 \times n_{\rm H_2}$, are higher than $n_{\rm H_2}$ measured on 
the integrated CO SLEDs \citep[akin to results of][for the Cosmic Eyelash]{danielson11}. For half of the GEMS\footnote{in particular 
for PLCK\_G138.6+62.0, PLCK\_G165.7+67.0, PLCK\_G200.6+46.1, PLCK\_G231.3+72.2 and PLCK\_G244.8+54.9.}, the molecular hydrogen 
densities derived with the relation $n_{\rm H_2} \simeq n_{\rm PDR}/2$ are significantly higher than those obtained from the best-fitting 
LVG models, either with one or two excitation components. Our temperature estimates do not exhibit such discrepancy, since the PDR 
surface temperatures of Table~\ref{tab:pdr} are within 1$\sigma$ of the kinetic temperatures from the LVG analysis.

The difference must therefore lie in the different assumptions inherent to each density estimate. Firstly, LVG model parameters are 
recovered with typical uncertainties of 0.5~dex and up to 1~dex, as also shown by \citet{tunnard16}, with possibly some additional 
dependence on the choice of the relation between escape probability and optical depth \citep{vandertak07}. Secondly, the outputs of 
PDR models vary for different configurations of the incident radiation field, cloud geometries and density distributions throughout 
the PDR \citep{spaans96}. Our assumption of a unique, homogeneous PDR with a single external FUV source is certainly simplistic and, 
interestingly, \citet{kaufman99} and \citet{wolfire90} note that applying these models to entire galaxies rather than individual PDRs 
leads to density and $G_{\rm 0}$ estimates biased to high values. Nonetheless, although lacking detailed constraints on the internal 
structure of gas clouds only allows us to derive rough estimates of the physical ISM conditions in the GEMS, averaging over the sample 
provides valuable comparisons between low and high-redshift populations, which is the main scope of this analysis.

It is also possible that additional heating mechanisms contribute. For example mechanical heating can harden the UV radiation 
fields even in photon-dominated regions \citep[e.g.,][]{kazandjian12}, and would predominantly boost the high-$J$ lines, thereby 
mimicking elevated gas densities \citep[][]{kazandjian15}. This is an interesting possibility for the GEMS given their intense star 
formation and broad line widths, perhaps a sign of strong turbulence \citep[as found for the Ruby,][]{canameras17b}. However, a 
detailed analysis is beyond the scope of this paper.

\subsection{Estimates of the PDR sizes}
\label{ssec:size}

We followed \citet{danielson11} in deriving rough estimates of the azimuthally-averaged sizes of the GEMS, using the total 
CO-inferred masses of molecular gas directly related to the on-going star formation from Sect.~\ref{ssec:mgas} and the 
best-fitting molecular gas densities from the PDR analysis ($n_{\rm H_2} \simeq n_{\rm PDR}/2$, assuming negligible atomic 
hydrogen densities). This calculation assumes that all the molecular gas is uniformly distributed over the galaxies, which 
is known to be a strong assumption, since the intense dust-obscured star formation in the GEMS and other high-redshift SMGs 
has proven to be irregular, either due to mergers \citep[e.g.,][]{engel10}, or clumpy gas distributions over extended disks 
\citep[e.g.,][]{swinbank11}. 

Spheres of mass equal to the measured $M^{\rm CO}({\rm H_2})$ would have typical radii, $R$, ranging between 130 and 340~pc to 
match the best-fitting gas densities. Alternatively, using a 100-pc thick disk model we obtain radii between 150 and 720~pc. 
For the Ruby, where we have direct, resolved constraints on scales of 100~pc from ALMA dust continuum and CO(4--3) interferometry
\citep{canameras17b}, we obtained radii of 130~pc and 165~pc for the sphere and disk models, respectively, slightly larger than the 
intrinsic size of individual clumps, but smaller than the total starburst size. Since these simple estimates result from the PDR 
densities, which are constrained by mid-$J$ CO lines and from the gas masses extrapolated from the LVG excitation analysis, they 
correspond to the extent of the gas reservoirs fueling star formation and suggest that this gas phase is more compact than the 
overall dust continuum emission (taken as the system size, see Table~\ref{tab:prop}).

Alternatively, the relation $G_{\rm 0} \propto L_{\rm FIR}/R_{\rm PDR}^2$ from \citet{wolfire90} allows us to convert the incident FUV 
radiation field for PDRs randomly distributed within a galaxy of total FIR luminosity, $L_{\rm FIR}$, to the total size of the 
area covered by the PDRs, $R_{\rm PDR}$. We assumed that $R_{\rm PDR}$ is equal to the total size of the gas reservoirs, including 
the extended and low density component suggested by the CO SLED for some sources, and we followed \citet{stacey10} in using this 
relation, together with measurements available in the literature for the local starburst M82, to infer the sizes of the GEMS 
\citep[see also][]{gullberg15,wardlow17}. We obtained the source radii listed in Table~\ref{tab:pdr}, which range between 0.8 and
5.8~kpc, akin to the typical Gaussian half-width at half maximum sizes of the molecular gas reservoirs in high-redshift compact 
starbursts and SMGs of about 0.5--5~kpc \citep[e.g.,][]{ivison11,bussmann13,ikarashi15}. Despite the large uncertainties inherent 
to this scaling relation, the resulting PDR sizes are remarkably similar to the intrinsic dimensions of PLCK\_G165.7+67.0 and 
PLCK\_G244.8+54.9 in Table~\ref{tab:prop}, the two GEMS with high-resolution SMA or ALMA interferometry, providing robust measurements 
of the delensed dust sizes. This suggests that the overall gas reservoirs extend over kpc scales, comparable to those covered by 
the cold dust continuum and about ten times larger than the intense star-forming clumps identified in some of the sources.

\subsection{Constraints on additional heating mechanisms}
\label{ssec:agn}

Studies simulating the chemical composition of photon and X-ray dominated regions (XDR) as a function of their physical properties 
have claimed that luminosity ratios of CO lines can be used to distinguish effectively the gas clouds heated by the FUV radiation 
field from star formation (PDR) from those primarily irradiated by X-ray emission from an AGN (XDR). We can therefore use our CO line
survey to search for signatures of AGN heating on the molecular gas reservoirs in the GEMS. We used the grid of PDR and XDR models
from \citet{meijerink05} and \citet{meijerink07}, which suggest that CO line ratios, in particular those between high-$J$ transitions
($J_{\rm up} \geq 10$) and CO(1--0), are significantly higher in XDRs. The reason is that, for a given energy injection from external 
radiation fields, CO emission will arise from smaller physical scales and warmer environments in XDRs. We compared the observed line
ratios with those simulated in \citet{meijerink07} for the high density model, the most representative for spatially-integrated gas
conditions in SMGs, which also covers the PDR properties in the GEMS. For $n=10^4$--$10^{6.5}$~cm$^{-3}$ and $G_{\rm 0}=10^2$--$10^5$ Habing 
fields, the high-$J$ CO versus CO(1--0) line ratios increase with the external radiation fields and are significantly higher in XDRs.

The ratios measured in the GEMS are systematically lower than those predicted by the model over this density range. This also holds 
when using only the highest and lowest-$J$ CO lines detected with EMIR. For instance, we find $L_{\rm CO(10-9)}/L_{\rm CO(1-0)} = 25 \pm 11$ 
for PLCK\_G092.5+42.9, while \citet{meijerink07} obtain a lower limit on this luminosity ratio of approximately 400 in XDRs. Although 
atomic carbon diagnostics suggest that PLCK\_G138.6+62.0 falls in the regime covered both by PDRs and XDRs \citepalias{nesvadba18}, 
we find $L_{\rm CO(7-6)}/L_{\rm CO(1-0)}$ and $L_{\rm CO(7-6)}/L_{\rm CO(3-2)}$ to be lower than those in the grid of XDR models by a factor of 
about 3--4. Consequently, there is no evidence that gas heating in the GEMS can be attributed to incident X-ray radiation fields 
from a central AGN above the lower value of 1.6~erg~s$^{-1}$~cm$^{-2}$ considered in the models. This is consistent with the minor 
AGN contribution to the overall dust heating obtained in \citetalias{canameras15} and our choice of describing the integrated line 
emission with PDR models in Sect.~\ref{ssec:pdr}. Other ratios involving dense gas tracers (e.g., HCN/HCO$^+$) would further constrain 
the upper limits on the incident X-ray fields.

\section{Summary}
\label{sec:summary}

In this paper, we have presented an analysis of the physical properties of highly-excited molecular gas reservoirs in the {\it Planck}'s 
dusty GEMS, a small set of the brightest strongly lensed high-redshift dusty star-forming galaxies on the extragalactic sky identified
with the {\it Planck} and {\it Herschel} satellites, using an extensive CO emission-line survey with EMIR on the IRAM 30-m telescope. 
We detected 45 CO rotational lines from $J_{\rm up}=3$ to $J_{\rm up}=11$ in the 11 submillimeter sources, ranging between $z=2.2$ 
and $z=3.6$, with velocity-integrated fluxes up to 37~$\mu^{-1}$~Jy~km~s$^{-1}$. The line profiles are broad, with FWHM~=~200--750~km~s$^{-1}$, 
and they are well-fitted with single Gaussians for seven sources while the remaining four show evidence for double velocity components 
over multiple $J_{\rm up}$ values.

Firstly, using the well-sampled CO spectral-line energy distributions from mid- to high-$J$ with up to eight transitions per source and 
published CO(1--0) observations, we performed a detailed analysis of the CO gas excitation to shed light on the conditions of the 
interstellar medium within these intense dusty starbursts. The peak of the CO ladder falls between $J_{\rm up}=4$ and $J_{\rm up}=7$ for 
nine out of 11 sources. Moreover, the two brightest GEMS, PLCK\_G092.5+42.9 and PLCK\_G244.8+54.9, exhibit double peaks with a highly-excited, 
warm gas component reaching a maximum at $J_{\rm up}>7$. These results are globally consistent with the broad range of gas excitations 
found amongst the population of lensed and unlensed SMGs, and with the inner regions of low-redshift starburst galaxies such as Arp~220 
and M82.

Our detailed lensing models from resolved dust continuum and mid-$J$ CO line interferometry, and the EMIR CO line profiles, suggest 
that differential lensing does not play a major role in this analysis. In the worst case, this effect might induce uncertainties 
comparable to those from other assumptions (e.g., on the spatial configuration of the gas reservoirs), and we ignored its impact on 
the CO SLEDs and resulting gas properties.

Secondly, we characterized the gas excitation from radiative transfer LVG models of the spatially-integrated CO flux ratios, following a 
similar approach as in \citet{yang17}. For a single excitation component, we obtained average densities $n_{\rm H_2} = 10^{2.6}$--$10^{4.1}$~cm$^{-3}$, 
kinetic temperatures $T_{\rm k} = 30$--1000~K and column densities of CO normalized per unit velocity gradient 
$N_{\rm CO}/dv = 10^{16}$--$10^{17.5}$~cm$^{-2}$~km$^{-1}$~s for the bulk of the molecular gas reservoirs in the {\it Planck}'s dusty GEMS.
For five sources, our well-sampled CO ladders highlight two gas phases with different properties and we reproduced our analysis using 
two excitation components. We found elevated densities, $n_{\rm H_2} = 10^{3.0}$--$10^{4.5}$~cm$^{-3}$, and temperatures of about 70--320~K 
for the warm gas reservoirs with high excitation. The properties of the cooler and more extended low-excitation component in 
PLCK\_G092.5+42.9 and PLCK\_G244.8+54.9 are similar to average conditions in local ULIRGs but, for PLCK\_G113.7+61.0 and 
PLCK\_G138.6+62.0 this component is diffuse, with $n_{\rm H_2} = 10^{2.4}$--$10^{2.8}$~cm$^{-3}$. This suggests that some high-redshift  
dusty starbursts contain a cool, low-density gas phase, comparable to that over the Galactic disk. Moreover, deblending individual 
kinematic components in three sources provides hints of varying CO excitations in two of them.

Thirdly, the intrinsic molecular gas masses derived from the best-fitting excitation models of the $J_{\rm up} > 3$ CO ladder and 
$\alpha_{\rm CO}=0.8$~M$_{\odot}$/(K~km~s$^{-1}$~pc$^2$) range from 0.6 to $12 \times 10^{10}$~M$_{\odot}$, implying an average gas-to-dust 
ratio of 150. These models predict lower CO(1--0) fluxes than those measured with the GBT for five sources. We interprete these low 
$r_{J \geq {\rm 3,1}}$ ratios as additional evidence for a diffuse gas phase, rather than differential magnification, and find that 
20 to 50\% of the total gas masses are embedded within these reservoirs, provided that the relative IRAM/GBT flux calibrations are 
robust.

Lastly, the CO line-luminosity ratios are consistent with those predicted by models of photon-dominated regions and disfavor scenarios in 
which the gas clouds are irradiated by intense X-ray fields from AGNs. We combined these transitions with single-dish [\ion{C}{i}] line 
detections presented in a companion paper as well as other [\ion{C}{ii}] measurements from the literature to derive PDR models 
\citep{kaufman99} and infer the global ISM conditions of the {\it Planck}'s dusty GEMS. Our EMIR line detections provide robust constraints 
on the PDR gas densities, $n_{\rm PDR} = 10^4$--$10^5$~cm$^{-3}$, higher than in local normal star-forming galaxies. This results in 
molecular hydrogen densities greater than those obtained with the LVG models, possibly due to geometric effects or contributions from 
mechanical heating. The FUV radiation fields from newly formed stellar populations are intense, from 10$^2$ to 10$^4$ times that of 
the Milky Way disk, although we caution that for some GEMS, $G_{\rm 0}$ depends strongly on our choice of the [\ion{C}{ii}] luminosity. These 
spatially-averaged conditions are consistent with other high-redshift SMGs and cover various low-redshift environments, ranging from the 
cores of ULIRGs, to bluer starbursts and dense Galactic molecular clouds. PDR radii are of order of 1--6~kpc, showing that the overall gas 
reservoirs and delensed dust continuum sizes are comparable, and nearly one order of magnitude larger than for individual star-forming 
clumps. 

Our study demonstrates the need to perform extensive line surveys to fully characterize the CO excitation and number of ISM phases 
in high-redshift SMGs, and that spatially-averaged properties of this population cover a range of low-redshift environments. In the 
future, combining high spatial resolution interferometry of multiple gas tracers and strong gravitational magnifications will be an
ideal way to probe the range of physical conditions within individual giant molecular clouds in high-redshift dusty starbursts, and 
constrain the local mechanisms setting the star-formation efficiency.

\section*{Acknowledgements}

RC would like to thank Raphael Gavazzi for useful discussions. The authors would like to thank the anonymous referee, whose comments
were helpful in improving the paper, and the telescope staff at the IRAM 30-m telescope for their excellent support during 
observations. We are also very grateful to the former director of IRAM, P. Cox, for his generous attribution of Director's Discretionary 
Time early on during the program, and to Daniel Dicken for sharing telescope time. This work is based on observations carried out under 
projects number 082-12, D09-12, 065-13, 094-13, 223-13, 108-14, and 217-14 with the IRAM 30-m telescope. IRAM is supported by INSU/CNRS 
(France), MPG (Germany) and IGN (Spain). This work was supported by the Programme National Cosmology et Galaxies (PNCG) of CNRS/INSU 
with INP and IN2P3, co-funded by CEA and CNES. RC was supported by DFF -- 4090-00079. ML acknowledges CNRS and CNES for support. CY was 
supported by an ESO Fellowship.

\bibliography{copaper}

\begin{appendix}

\section{Profiles and measured properties of the CO rotational lines}

\begin{figure*}
\includegraphics[width=0.30\textwidth]{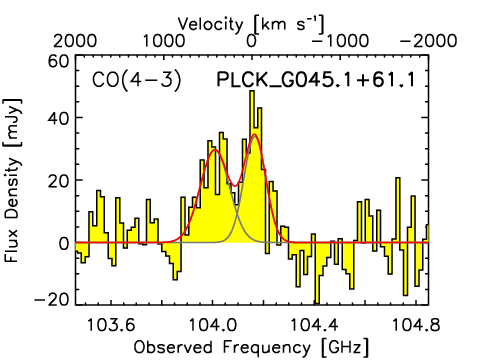}
\includegraphics[width=0.30\textwidth]{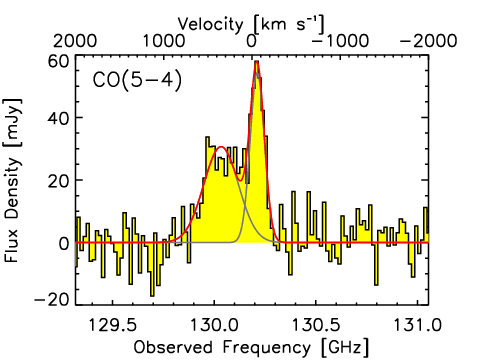}
\includegraphics[width=0.30\textwidth]{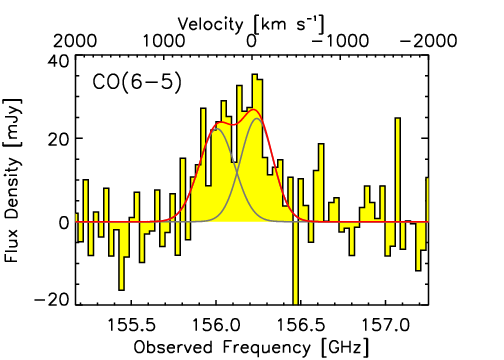}

\includegraphics[width=0.30\textwidth]{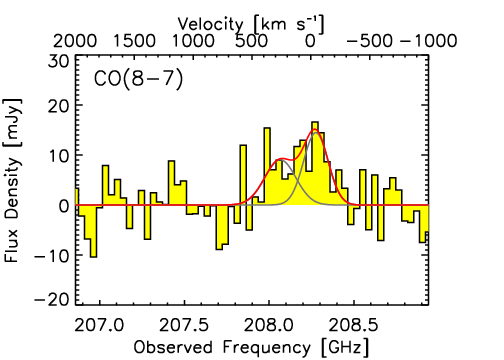}
\includegraphics[width=0.30\textwidth]{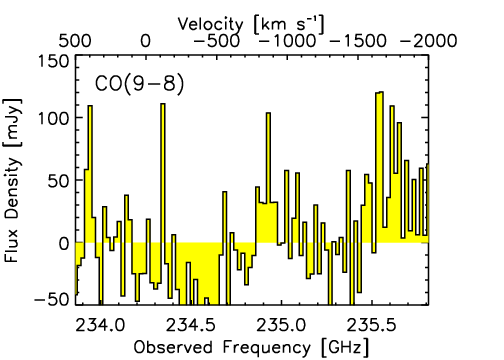}

\includegraphics[width=0.30\textwidth]{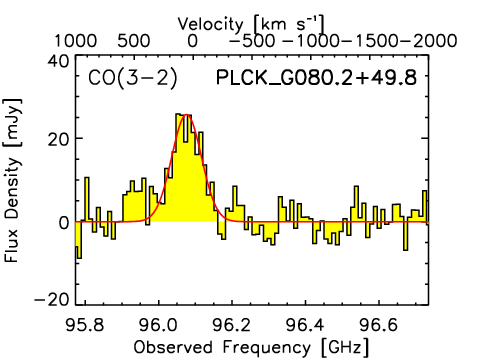}
\includegraphics[width=0.30\textwidth]{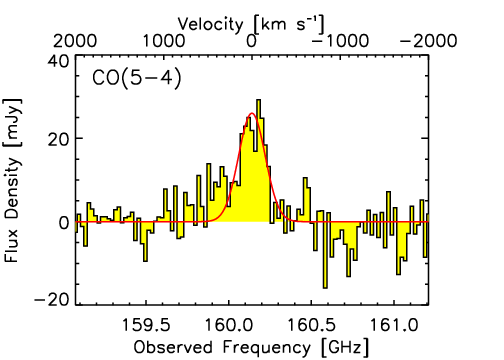}
\includegraphics[width=0.30\textwidth]{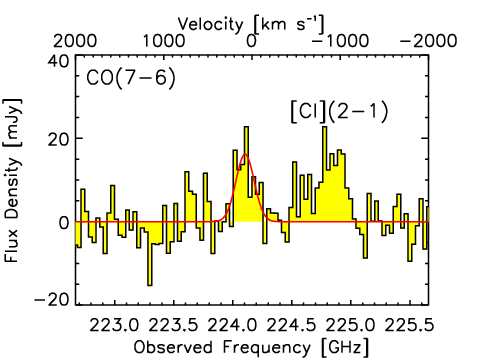}

\includegraphics[width=0.30\textwidth]{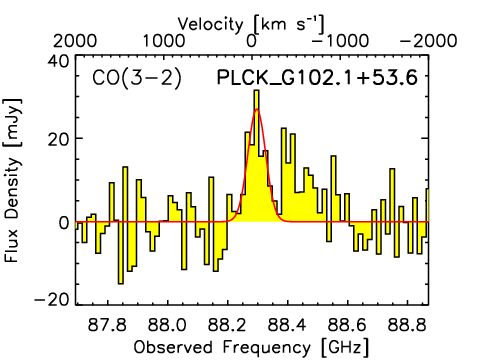}
\includegraphics[width=0.30\textwidth]{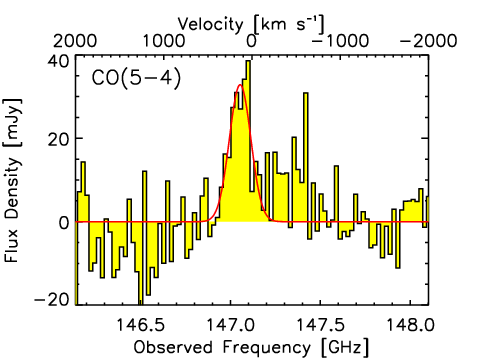}
\includegraphics[width=0.30\textwidth]{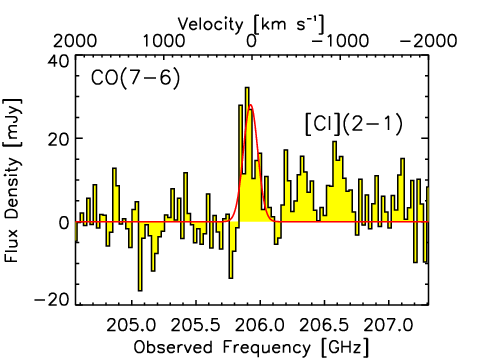}
\caption{Continuum-subtracted and binned spectra of PLCK\_G045.1+61.1 (top two rows), PLCK\_G080.2+49.8 (third row),
and PLCK\_G102.1+53.6 (bottom row), fitted with one or two Gaussian components. See further details in Fig.~\ref{fig:lines1} 
caption.}
\label{fig:lines2}
\end{figure*}

\begin{figure*}
\includegraphics[width=0.30\textwidth]{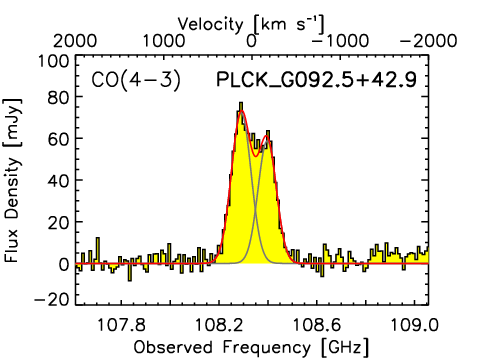}
\includegraphics[width=0.30\textwidth]{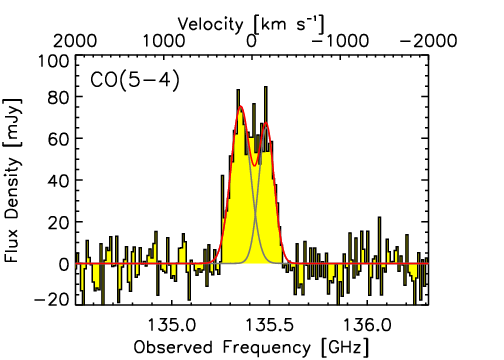}
\includegraphics[width=0.30\textwidth]{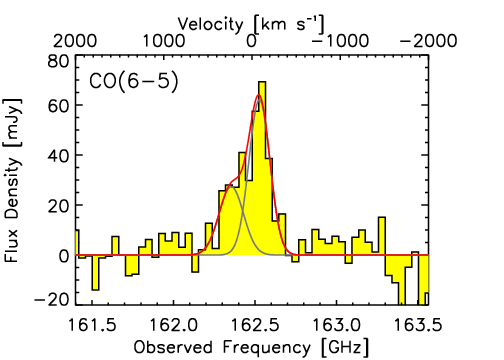}

\includegraphics[width=0.30\textwidth]{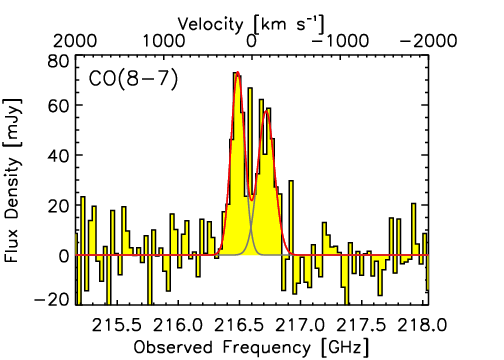}
\includegraphics[width=0.30\textwidth]{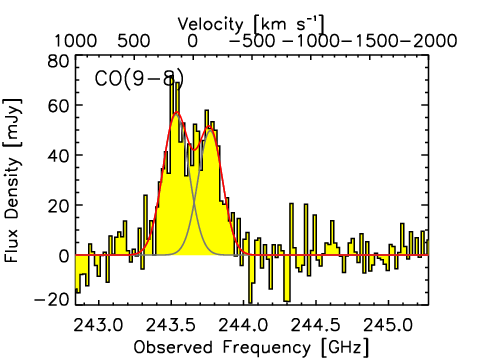}
\includegraphics[width=0.30\textwidth]{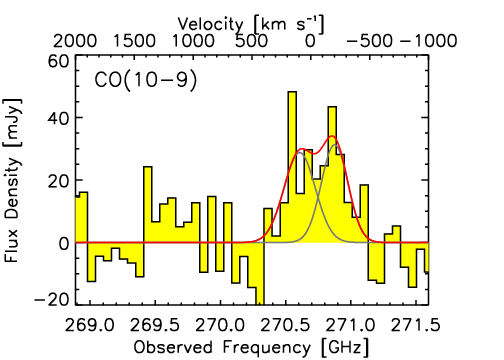}

\includegraphics[width=0.30\textwidth]{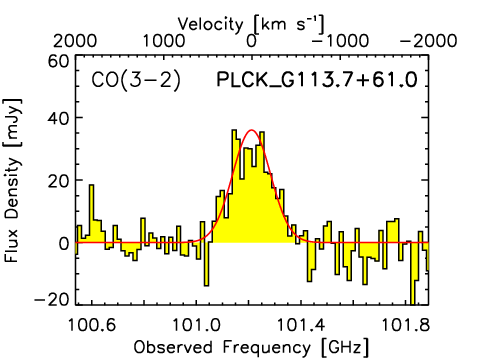}
\includegraphics[width=0.30\textwidth]{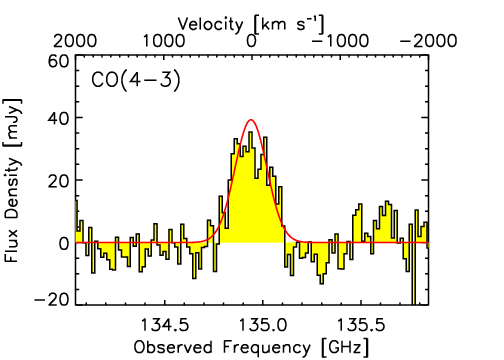}
\includegraphics[width=0.30\textwidth]{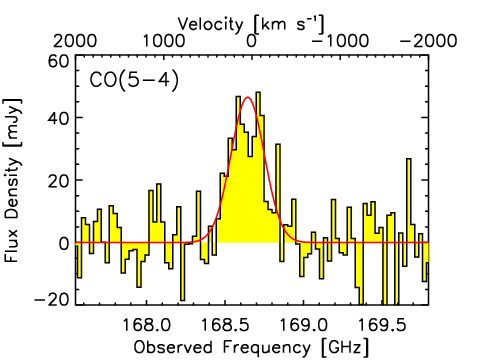}

\includegraphics[width=0.30\textwidth]{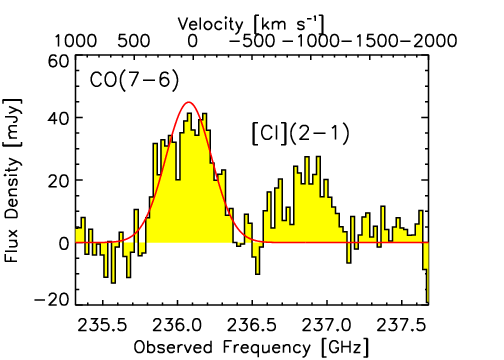}
\includegraphics[width=0.30\textwidth]{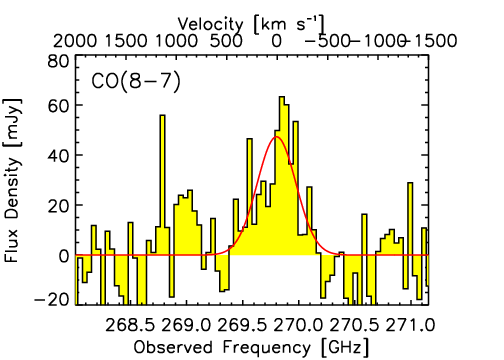}

\includegraphics[width=0.30\textwidth]{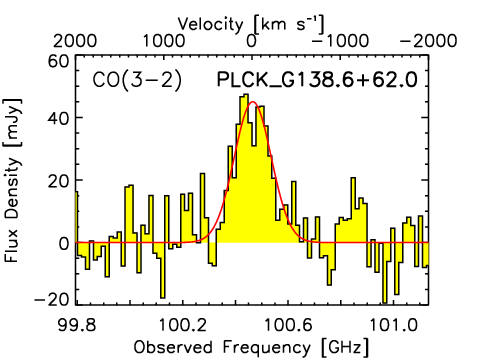}
\includegraphics[width=0.30\textwidth]{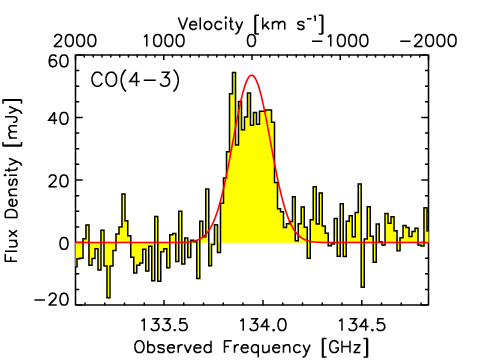}
\includegraphics[width=0.30\textwidth]{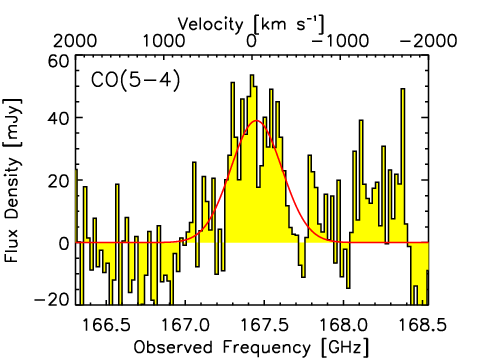}

\includegraphics[width=0.30\textwidth]{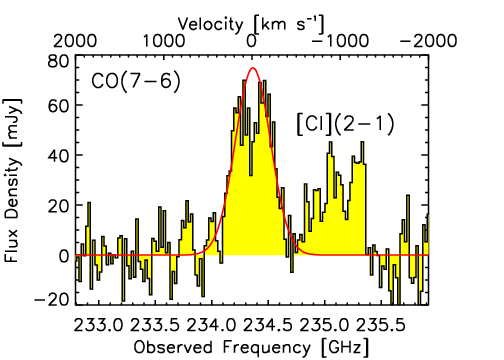}
\caption{Continuum-subtracted and binned spectra of PLCK\_G092.5+42.9 (top two rows), PLCK\_G113.7+61.0 (third and fourth 
rows), and PLCK\_G138.6+62.0 (bottom two rows), fitted with one or two Gaussian components. See further details 
in Fig.~\ref{fig:lines1} caption.}
\label{fig:lines3}
\end{figure*}

\begin{figure*}
\includegraphics[width=0.30\textwidth]{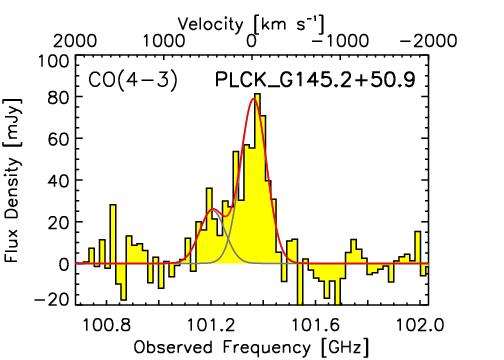}
\includegraphics[width=0.30\textwidth]{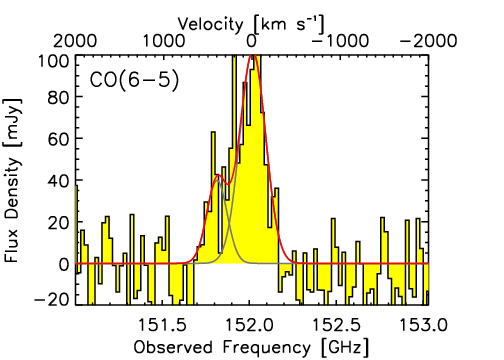}

\includegraphics[width=0.30\textwidth]{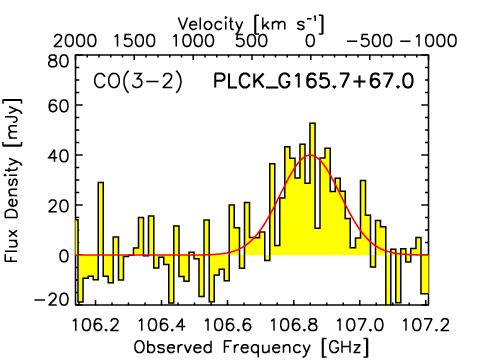}
\includegraphics[width=0.30\textwidth]{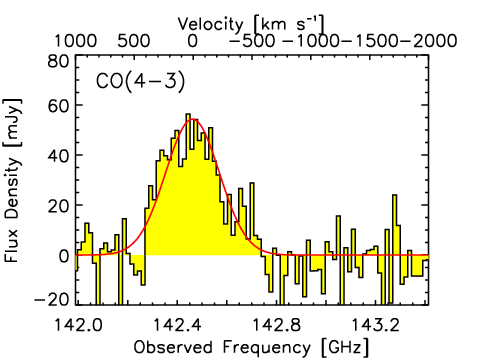}
\includegraphics[width=0.30\textwidth]{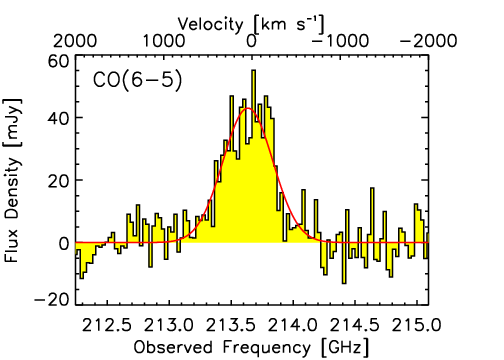}

\includegraphics[width=0.30\textwidth]{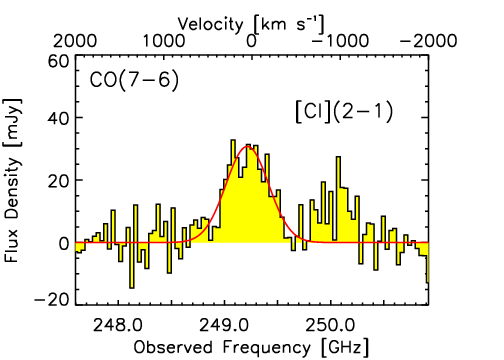}

\includegraphics[width=0.30\textwidth]{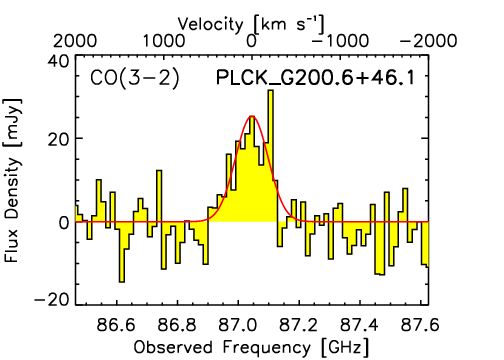}
\includegraphics[width=0.30\textwidth]{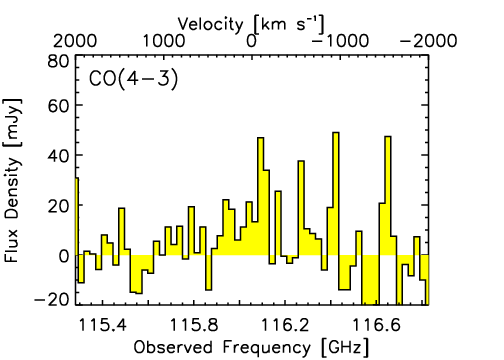}
\includegraphics[width=0.30\textwidth]{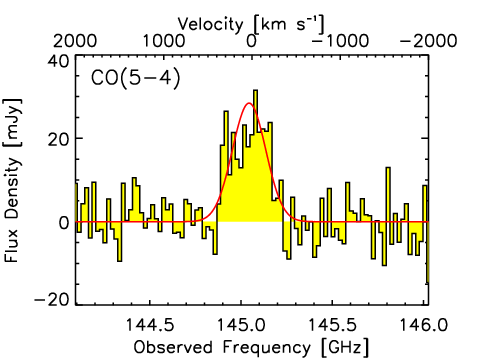}

\includegraphics[width=0.30\textwidth]{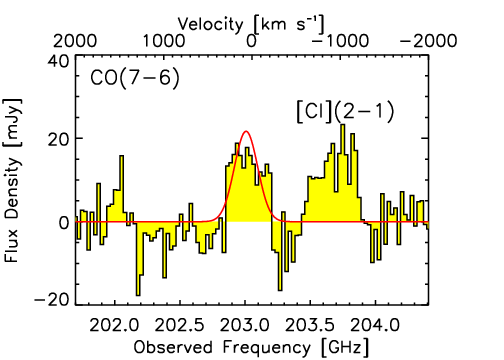}
\includegraphics[width=0.30\textwidth]{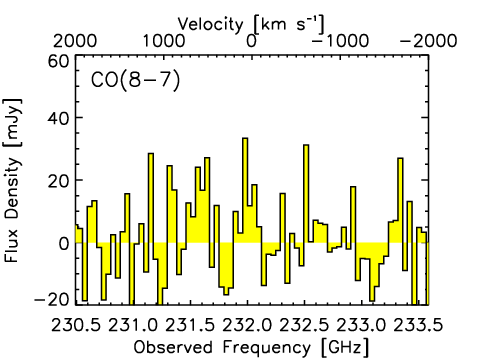}

\includegraphics[width=0.30\textwidth]{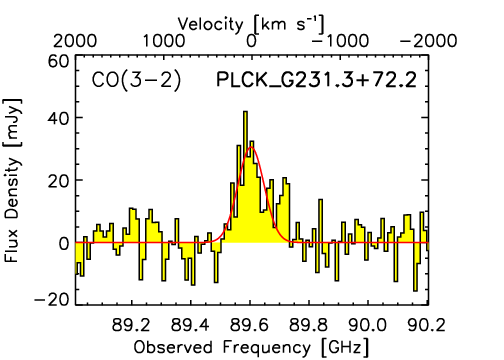}
\includegraphics[width=0.30\textwidth]{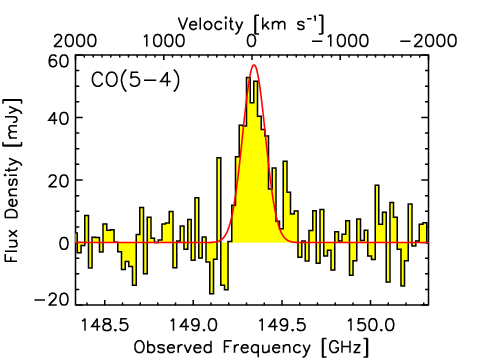}
\includegraphics[width=0.30\textwidth]{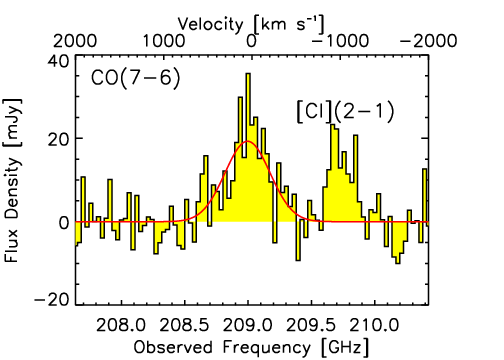}
\caption{Continuum-subtracted and binned spectra of PLCK\_G145.2+50.9 (first row), PLCK\_G165.7+67.0 (second and third rows), 
PLCK\_G200.6+46.1 (fourth and fifth rows), and PLCK\_G231.3+72.2 (bottom row), fitted with one or two Gaussian components. See 
further details in Fig.~\ref{fig:lines1} caption.}
\label{fig:lines4}
\end{figure*}

Tables~\ref{tab:obslog} and \ref{tab:linefit} present the observation log with EMIR and the properties of each CO emission line, 
respectively. The spatially-integrated and binned spectra of all observed CO lines are shown in Figs.~\ref{fig:lines2}, \ref{fig:lines3}, 
and \ref{fig:lines4}, after subtracting the continuum baselines fitted on line-free spectral channels. All spectra from a given source
were fitted consistently with a single or two Gaussian components.

\begin{table*}
\centering
\begin{tabular}{lcccccccc}
\hline
\hline
\rule{0pt}{2ex} Source & Line & Project ID & Observing date & Band & $\nu_{\rm tuning}$ & $t_{\rm obs}$ & rms noise & S/T$^*_a$ \\
 & & & [dd/mm/yy] & & [GHz] & [hr] & [mJy] & [Jy K$^{-1}$] \\
\hline
PLCK\_G045.1+61.1 & CO(4--3) & 094-13 & 07, 10/06/13 & E090 & 101.90 & 1.4 & 10.0 & 6.0 \\
 & CO(5--4) & 094-13 & 10/06/13 & E150 & 130.20 & 1.6 & 7.3 & 6.2 \\
 & CO(6--5) & 223-13, 108-14 & 04/02/14 \& 17/09/14 & E150 & 156.40 & 3.4 & 7.9 & 6.3 \\
 & CO(8--7) & 223-13 & 19/04/14 & E230 & 207.00 & 2.4 & 4.6 & 7.2 \\
 & CO(9--8) & 094-13 & 09/06/13 & E230 & 232.00 & 0.4 & 43.4 & 7.9 \\
PLCK\_G080.2+49.8 & CO(3--2) & 082-12 & 28/10/2012 & E090 & 97.40 & 1.6 & 6.0 & 6.0 \\
 & CO(5--4) & D09-12 & 30/11/2012 & E150 & 160.12 & 4.6 & 6.4 & 6.6 \\
 & CO(7--6) & 223-13 & 03/02/14 & E230 & 224.40 & 3.2 & 6.1 & 7.6 \\
PLCK\_G092.5+42.9 & CO(4--3) & 094-13, 108-14 & 10/04/13 \& 20, 21, 22/06/14 & E090 & 104.70 & 13.8 & 4.7 & 6.0 \\
 & CO(5--4) & 094-13 & 05/06/13 & E150 & 143.68 & 2.0 & 12.6 & 6.1 \\
 & CO(6--5) & 108-14 & 17/09/14 & E150 & 162.47 & 0.8 & 10.4 & 6.4 \\
 & CO(8--7) & 094-13 & 06/06/13 & E230 & 216.00 & 0.8 & 18.4 & 7.4 \\
 & CO(9--8) & 223-13 & 02/02/14 & E230 & 244.60 & 2.4 & 9.0 & 8.0 \\
 & CO(10--9) & 223-13 & 19/04/14 & E230 & 270.67 & 2.2 & 12.0 & 8.6 \\
PLCK\_G102.1+53.6 & CO(3--2) & 094-13 & 07, 09/06/13 & E090 & 89.40 & 0.9 & 7.0 & 5.9 \\
 & CO(5--4) & 094-13 & 09/06/13 & E150 & 148.00 & 1.0 & 9.3 & 6.4 \\
 & CO(7--6) & 223-13 & 03/02/14 & E230 & 206.00 & 1.6 & 8.5 & 7.2 \\
PLCK\_G113.7+61.0 & CO(3--2) & 094-13 & 09/04/13 & E090 & 101.90 & 2.4 & 7.0 & 6.0 \\
 & CO(4--3) & 094-13 & 07/06/13 & E150 & 134.50 & 2.2 & 7.5 & 6.3 \\
 & CO(5--4) & 223-13 & 19/04/14 & E150 & 168.80 & 1.2 & 10.2 & 6.5 \\
 & CO(7--6) & 217-14 & 18, 19/02/15 & E230 & 236.67 & 4.0 & 15.5 & 7.9 \\
 & CO(8--7) & 223-13 & 18/04/14 & E230 & 270.01 & 2.2 & 22.2 & 8.6 \\
PLCK\_G138.6+62.0 & CO(3--2) & 094-13 & 09/04/13 & E090 & 101.90 & 1.6 & 10.1 & 6.0 \\
 & CO(4--3) & 094-13 & 07/06/13 & E150 & 134.50 & 1.6 & 8.6 & 6.3 \\
 & CO(5--4) & 094-13 & 07/06/13 & E150 & 168.00 & 0.8 & 16.1 & 6.7 \\
 & CO(7--6) & 217-14 & 19/02/15 & E230 & 231.30 & 1.6 & 14.5 & 7.8 \\
PLCK\_G145.2+50.9 & CO(4--3) & 094-13 & 09/04/13 & E090 & 101.90 & 1.2 & 12.4 & 6.0 \\
 & CO(6--5) & 094-13 & 05/06/13 & E150 & 151.97 & 0.8 & 17.6 & 6.5 \\
PLCK\_G165.7+67.0 & CO(3--2) & 094-13 & 09/04/13 & E090 & 101.90 & 2.4 & 14.0 & 6.0 \\
 & CO(4--3) & 094-13 & 05/06/13 & E150 & 143.68 & 0.8 & 11.4 & 6.4 \\
 & CO(6--5) & 223-13 & 02, 03/02/14 & E230 & 216.70 & 1.6 & 9.7 & 7.3 \\
 & CO(7--6) & 217-14 & 19, 20/02/15 & E230 & 245.50 & 2.4 & 8.8 & 8.2 \\
PLCK\_G200.6+46.1 & CO(3--2) & 094-13 & 08, 10/06/13 & E090 & 88.00 & 1.6 & 6.7 & 5.9 \\
 & CO(4--3) & 065-13 & 31/08/13 & E090 & 115.90 & 1.4 & 31.9 & 6.0 \\
 & CO(5--4) & 065-13 & 31/08/13 & E150 & 145.00 & 2.4 & 6.4 & 6.2 \\
 & CO(7--6) & 217-14 & 20, 21/02/15 & E230 & 206.28 & 3.2 & 8.0 & 7.1 \\
 & CO(8--7) & 094-13 & 09/06/13 & E230 & 232.00 & 0.8 & 18.3 & 7.8 \\
PLCK\_G231.3+72.2 & CO(3--2) & 094-13 & 08/06/13 & E090 & 89.40 & 1.6 & 5.9 & 5.9 \\
 & CO(5--4) & 094-13 & 09/06/13 & E150 & 148.00 & 0.8 & 10.6 & 6.4 \\
 & CO(7--6) & 217-14 & 20/02/15 & E230 & 209.10 & 2.4 & 15.1 & 7.2 \\
PLCK\_G244.8+54.9 & CO(3--2) & 094-13, 108-14 & 05/04/13 \& 17, 18, 19/06/14 & E090 & 89.00 & 9.0 & 3.1 & 5.9 \\
 & CO(4--3) & 094-13, 223-13 & 06/06/13 \& 31/01/14 & E090 & 115.11 & 3.4 & 13.0 & 6.0 \\
 & CO(5--4) & 094-13 & 05/06/13 & E150 & 143.68 & 1.6 & 8.8 & 6.4 \\
 & CO(6--5) & 094-13, 223-13 & 06/06/13 \& 31/01/14 & E150 & 172.30 & 1.6 & 16.8 & 6.6 \\
% & CO(7--6) & 217-14 & 19/02/15 & E230 & 205.20 & 2.4 & 9.3 & 7.1 \\
 & CO(8--7) & 094-13 & 06/06/13 & E230 & 230.50 & 1.4 & 15.6 & 7.8 \\
 & CO(9--8) & 223-13 & 01, 02/02/14 & E230 & 259.40 & 1.6 & 11.7 & 8.4 \\
 & CO(10--9) & 223-13 & 02/02/14 & E330 & 289.70 & 1.4 & 11.5 & 9.1 \\
 & CO(11--10) & 223-13 & 02/02/14 & E330 & 317.00 & 3.8 & 15.3 & 10.4 \\
\hline
\end{tabular}
\caption{Observation log with EMIR on the IRAM 30-m telescope. Here $\nu_{\rm tuning}$ is the tuning frequency of the EMIR receivers 
that was used to observe a given CO transition. Total exposure times $t_{\rm obs}$ do not include the bad scans discarded for the 
optimal spectrum reduction. The spectrum rms values were measured on baseline channels with {\tt CLASS}. Telescope efficiencies 
S/T$^*_a$ were extrapolated from the calibration tables and used to convert the line fluxes to Jy~km~s$^{-1}$. We note that the 
CO(3--2) and CO(4--3) transitions in PLCK\_G244.8+54.9 and PLCK\_G092.5+42.9, respectively, were included in a backup sideband 
while observing HCN and HCO$^+$ transitions in these sources with long integrations (program 108-14).} 
\label{tab:obslog}
\end{table*}

\begin{table*}
\centering
\begin{tabular}{lccccccc}
\hline
\hline
\rule{0pt}{2ex} Source & Line & $\nu_{\rm obs}$ & Redshift & FWHM$_{\rm line}$ & $\mu I_{\rm line}$ & $\mu L_{\rm line}$ & $\mu L'_{\rm line}$ \\
 & & [GHz] & & [km~s$^{-1}$] & [Jy~km~s$^{-1}$] & [10$^{8}$~L$_{\odot}$] & [10$^{10}$~K~km~s$^{-1}$ pc$^2$] \\
\hline
PLCK\_G045.1+61.1 & CO(4--3) & 104.16 $\pm$ 0.01 & 3.4261 $\pm$ 0.0003 & 224 $\pm$ 51 & 10.6 $\pm$ 2.0 & 10.7 $\pm$ 2.0 & 34.1 $\pm$ 6.4 \\
 & & 104.01 $\pm$ 0.01 & 3.4326 $\pm$ 0.0005 & 397 $\pm$ 78 & 12.7 $\pm$ 2.2 & 12.8 $\pm$ 2.2 & 40.8 $\pm$ 7.1 \\
 & CO(5--4) & (130.20) & (3.4261) & 191 $\pm$ 16 & 10.9 $\pm$ 0.8 & 13.8 $\pm$ 1.0 & 22.5 $\pm$ 1.6 \\
 & & (130.01) & (3.4326) & 484 $\pm$ 41 & 15.7 $\pm$ 1.2 & 19.8 $\pm$ 1.5 & 32.4 $\pm$ 2.5 \\
 & CO(6--5) & (156.23) & (3.4261) & 369 $\pm$ 107 & 9.2 $\pm$ 2.6 & 13.9 $\pm$ 3.9 & 13.2 $\pm$ 3.7 \\  
 & & (156.00) & (3.4326) & 554 $\pm$ 115 & 11.4 $\pm$ 2.9 & 17.2 $\pm$ 4.4 & 16.4 $\pm$ 4.2 \\
 & CO(8--7) & (208.26) & (3.4261) & 241 $\pm$ 62 & 3.5 $\pm$ 0.9 & 7.1 $\pm$ 1.8 & 2.8 $\pm$ 0.7 \\
 & & (207.96) & (3.4326) & 304 $\pm$ 107 & 2.7 $\pm$ 0.9 & 5.4 $\pm$ 1.8 & 2.2 $\pm$ 0.7 \\ 
 & CO(9--8) & -- & -- & (500) & $<$ 23.1 & $<$ 52.4 & $<$ 14.7 \\
PLCK\_G080.2+49.8 & CO(3--2) & 96.10 $\pm$ 0.01 & 2.5984 $\pm$ 0.0001 & 319 $\pm$ 35 & 7.9 $\pm$ 0.6 & 3.8 $\pm$ 0.3 & 28.6 $\pm$ 2.2 \\
 & CO(5--4) & 160.13 $\pm$ 0.01 & 2.5987 $\pm$ 0.0003 & 431 $\pm$ 84 & 10.5 $\pm$ 1.0 & 8.4 $\pm$ 0.8 & 13.7 $\pm$ 1.3 \\
 & CO(7--6) & 224.17 $\pm$ 0.02 & 2.5984 $\pm$ 0.0003 & 245 $\pm$ 52 & 4.1 $\pm$ 0.8 & 4.6 $\pm$ 0.9 & 2.7 $\pm$ 0.5 \\
PLCK\_G092.5+42.9 & CO(4--3) & 108.39 $\pm$ 0.01 & 3.2535 $\pm$ 0.0001 & 267 $\pm$ 21 & 15.6 $\pm$ 1.6 & 14.5 $\pm$ 1.5 & 46.2 $\pm$ 4.7 \\
 & & 108.29 $\pm$ 0.01 & 3.2575 $\pm$ 0.0001 & 270 $\pm$ 18 & 21.3 $\pm$ 1.6 & 19.8 $\pm$ 1.5 & 63.1 $\pm$ 4.7 \\
 & CO(5--4) & (135.48) & (3.2535) & 234 $\pm$ 35 & 16.0 $\pm$ 1.4 & 18.6 $\pm$ 1.6 & 30.3 $\pm$ 2.7 \\
 & & (135.35) & (3.2575) & 253 $\pm$ 35 & 19.6 $\pm$ 1.5 & 22.8 $\pm$ 1.7 & 37.2 $\pm$ 2.8 \\
 & CO(6--5) & (162.57) & (3.2535) & 257 $\pm$ 49 & 16.4 $\pm$ 2.7 & 22.9 $\pm$ 3.8 & 21.6 $\pm$ 3.6 \\ 
 & & (162.41) & (3.2575) & 280 $\pm$ 92 & 9.3 $\pm$ 2.8 & 13.0 $\pm$ 3.9 & 12.3 $\pm$ 3.7 \\
 & CO(8--7) & (216.72) & (3.2535) & 270 $\pm$ 34 & 15.5 $\pm$ 1.8 & 28.8 $\pm$ 3.3 & 11.5 $\pm$ 1.3 \\ 
 & & (216.51) & (3.2575) & 182 $\pm$ 23 & 13.4 $\pm$ 1.5 & 24.9 $\pm$ 2.8 & 9.9 $\pm$ 1.1 \\
 & CO(9--8) & (243.78) & (3.2535) & 229 $\pm$ 18 & 12.4 $\pm$ 0.9 & 25.9 $\pm$ 1.9 & 7.3 $\pm$ 0.5 \\
 & & (243.55) & (3.2575) & 258 $\pm$ 20 & 15.5 $\pm$ 1.0 & 32.4 $\pm$ 2.1 & 9.1 $\pm$ 0.6 \\
 & CO(10--9) & (270.83) & (3.2535) & 261 $\pm$ 114 & 9.0 $\pm$ 2.9 & 20.9 $\pm$ 6.7 & 4.3 $\pm$ 1.4 \\
 & & (270.58) & (3.2575) & 255 $\pm$ 59 & 10.6 $\pm$ 2.5 & 24.6 $\pm$ 5.8 & 5.0 $\pm$ 1.2 \\
PLCK\_G102.1+53.6 & CO(3--2) & 88.29 $\pm$ 0.01 & 2.9166 $\pm$ 0.0004 & 215 $\pm$ 58 & 5.1 $\pm$ 1.8 & 3.0 $\pm$ 1.0 & 22.4 $\pm$ 7.9 \\
 & CO(5--4) & 147.12 $\pm$ 0.01 & 2.9171 $\pm$ 0.0002 & 324 $\pm$ 89 & 11.0 $\pm$ 1.8 & 10.7 $\pm$ 1.7 & 17.4 $\pm$ 2.9 \\
 & CO(7--6) & 205.93 $\pm$ 0.01 & 2.9172 $\pm$ 0.0002 & 186 $\pm$ 42 & 5.0 $\pm$ 0.9 & 6.8 $\pm$ 1.2 & 4.0 $\pm$ 0.7 \\
PLCK\_G113.7+61.0 & CO(3--2) & 101.21 $\pm$ 0.01 & 2.4166 $\pm$ 0.0002 & 537 $\pm$ 44 & 18.2 $\pm$ 1.4 & 7.7 $\pm$ 0.6 & 58.2 $\pm$ 4.5 \\
 & CO(4--3) & 134.94 $\pm$ 0.01 & 2.4166 $\pm$ 0.0002 & 541 $\pm$ 43 & 20.5 $\pm$ 1.5 & 11.6 $\pm$ 0.8 & 36.9 $\pm$ 2.7 \\
 & CO(5--4) & 168.66 $\pm$ 0.01 & 2.4168 $\pm$ 0.0003 & 484 $\pm$ 54 & 21.2 $\pm$ 2.1 & 15.0 $\pm$ 1.5 & 24.4 $\pm$ 2.4 \\
 & CO(7--6) & 236.09 $\pm$ 0.01 & 2.4167 $\pm$ 0.0001 & 482 $\pm$ 26 & 20.7 $\pm$ 1.1 & 20.4 $\pm$ 1.1 & 12.2 $\pm$ 0.6 \\
 & CO(8--7) & 269.83 $\pm$ 0.03 & 2.4162 $\pm$ 0.0004 & 426 $\pm$ 85 & 21.6 $\pm$ 3.6 & 24.4 $\pm$ 4.1 & 9.7 $\pm$ 1.6 \\
PLCK\_G138.6+62.0 & CO(3--2) & 100.46 $\pm$ 0.01 & 2.4420 $\pm$ 0.0002 & 487 $\pm$ 56 & 22.4 $\pm$ 2.1 & 9.7 $\pm$ 0.9 & 73.0 $\pm$ 6.8 \\
 & CO(4--3) & 133.95 $\pm$ 0.01 & 2.4420 $\pm$ 0.0002 & 537 $\pm$ 35 & 26.9 $\pm$ 1.6 & 15.5 $\pm$ 0.9 & 49.3 $\pm$ 2.9 \\
 & CO(5--4) & 167.44 $\pm$ 0.02 & 2.4416 $\pm$ 0.0005 & 630 $\pm$ 103 & 28.1 $\pm$ 3.6 & 20.2 $\pm$ 2.6 & 33.0 $\pm$ 4.2 \\
 & CO(7--6) & 234.37 $\pm$ 0.01 & 2.4418 $\pm$ 0.0002 & 472 $\pm$ 33 & 26.5 $\pm$ 1.9 & 26.7 $\pm$ 1.9 & 15.9 $\pm$ 1.1 \\
PLCK\_G145.2+50.9 & CO(4--3) & 101.38 $\pm$ 0.01 & 3.5477 $\pm$ 0.0002 & 397 $\pm$ 79 & 25.6 $\pm$ 1.8 & 27.4 $\pm$ 1.9 & 87.2 $\pm$ 6.1 \\
 & & 101.20 $\pm$ 0.02 & 3.5557 $\pm$ 0.0009 & 355 $\pm$ 28 & 10.3 $\pm$ 1.9 & 11.0 $\pm$ 2.0 & 35.2 $\pm$ 6.5 \\
 & CO(6--5) & (152.04) & (3.5477) & 370 $\pm$ 39 & 32.4 $\pm$ 3.6 & 51.9 $\pm$ 5.8 & 49.0 $\pm$ 5.4 \\
 & & (151.78) & (3.5557) & 290 $\pm$ 70 & 8.9 $\pm$ 3.0 & 14.2 $\pm$ 4.8 & 13.5 $\pm$ 4.6 \\
PLCK\_G165.7+67.0 & CO(3--2) & 106.85 $\pm$ 0.01 & 2.2363 $\pm$ 0.0004 & 655 $\pm$ 86 & 26.8 $\pm$ 3.2 & 9.9 $\pm$ 1.2 & 75.0 $\pm$ 9.0 \\
 & CO(4--3) & 142.47 $\pm$ 0.01 & 2.2362 $\pm$ 0.0002 & 547 $\pm$ 40 & 29.2 $\pm$ 2.0 & 14.4 $\pm$ 1.0 & 46.0 $\pm$ 3.1 \\
 & CO(6--5) & 213.64 $\pm$ 0.01 & 2.2367 $\pm$ 0.0002 & 640 $\pm$ 39 & 29.8 $\pm$ 1.5 & 22.1 $\pm$ 1.1 & 20.9 $\pm$ 1.0 \\
 & CO(7--6) & 249.23 $\pm$ 0.02 & 2.2365 $\pm$ 0.0002 & 580 $\pm$ 48 & 18.6 $\pm$ 1.3 & 16.1 $\pm$ 1.1 & 9.6 $\pm$ 0.7 \\
PLCK\_G200.6+46.1 & CO(3--2) & 87.04 $\pm$ 0.01 & 2.9726 $\pm$ 0.0004 & 505 $\pm$ 57 & 13.3 $\pm$ 1.4 & 8.0 $\pm$ 0.8 & 60.4 $\pm$ 6.4 \\
 & CO(4--3) & -- & -- & (500) & $<$ 17.0 & $<$ 13.6 & $<$ 43.4 \\
 & CO(5--4) & 145.06 $\pm$ 0.01 & 2.9727 $\pm$ 0.0003 & 481 $\pm$ 45 & 12.7 $\pm$ 1.2 & 12.7 $\pm$ 1.2 & 20.7 $\pm$ 2.0 \\
 & CO(7--6) & 203.00 $\pm$ 0.02 & 2.9735 $\pm$ 0.0003 & 333 $\pm$ 41 & 6.6 $\pm$ 0.9 & 9.2 $\pm$ 1.3 & 5.5 $\pm$ 0.8 \\
 & CO(8--7) & -- & -- & (500) & $<$ 9.7 & $<$ 15.5 & $<$ 6.2 \\
PLCK\_G231.3+72.2 & CO(3--2) & 89.61 $\pm$ 0.01 & 2.8589 $\pm$ 0.0003 & 445 $\pm$ 62 & 12.9 $\pm$ 1.3 & 7.3 $\pm$ 0.7 & 54.9 $\pm$ 5.5 \\
 & CO(5--4) & 149.32 $\pm$ 0.01 & 2.8592 $\pm$ 0.0002 & 350 $\pm$ 36 & 18.4 $\pm$ 1.5 & 17.3 $\pm$ 1.4 & 28.2 $\pm$ 2.3 \\ 
 & CO(7--6) & 209.02 $\pm$ 0.02 & 2.8592 $\pm$ 0.0003 & 469 $\pm$ 71 & 12.4 $\pm$ 1.4 & 16.3 $\pm$ 1.8 & 9.7 $\pm$ 1.1 \\
\hline
\end{tabular}
\caption{Properties of the CO emission lines obtained by fitting the continuum-subtracted spectra using a single or two 
Gaussian components with {\tt CLASS}. Columns are: source name; CO transition; observed frequency; redshift of the line or 
spectral component; line FWHM; velocity-integrated flux density $\mu I_{\rm line}$, used for the line-excitation analysis and 
uncorrected for lensing magnification; observed line luminosity in solar luminosities and in K~km~s$^{-1}$~pc$^2$. For 
the sources fitted with two Gaussians, we fixed the central velocity of each spectral component to those measured on the 
lowest-$J$ CO transition.}
\label{tab:linefit}
\end{table*}

\begin{table*}
\centering
\begin{tabular}{lccccccc}
\hline 
\hline
\rule{0pt}{2ex} Source & Line & $\nu_{\rm obs}$ & Redshift & FWHM$_{\rm line}$ & $\mu I_{\rm line}$ & $\mu L_{\rm line}$ & $\mu L'_{\rm line}$ \\
 & & [GHz] & & [km~s$^{-1}$] & [Jy~km~s$^{-1}$] & [10$^{8}$~L$_{\odot}$] & [10$^{10}$~K~km~s$^{-1}$ pc$^2$] \\
\hline
PLCK\_G244.8+54.9 & CO(3--2) & 86.38 $\pm$ 0.01 & 3.0033 $\pm$ 0.0004 & 282 $\pm$ 56 & 8.3 $\pm$ 0.8 & 5.1 $\pm$ 0.5 & 38.3 $\pm$ 3.7 \\ 
 & & 86.28 $\pm$ 0.02 & 3.0078 $\pm$ 0.0008 & 403 $\pm$ 75 & 9.5 $\pm$ 1.2 & 5.8 $\pm$ 0.7 & 44.0 $\pm$ 5.6 \\
 & CO(4--3) & (115.17) & (3.0033) & 293 $\pm$ 51 & 13.3 $\pm$ 2.2 & 10.8 $\pm$ 1.8 & 34.5 $\pm$ 5.7 \\
 & & (115.04) & (3.0078) & 316 $\pm$ 60 & 13.6 $\pm$ 2.3 & 11.1 $\pm$ 1.9 & 35.4 $\pm$ 6.0 \\
 & CO(5--4) & (143.95) & (3.0033) & 283 $\pm$ 23 & 13.2 $\pm$ 1.2 & 13.5 $\pm$ 1.2 & 21.9 $\pm$ 2.0 \\
 & & (143.79) & (3.0078) & 525 $\pm$ 47 & 20.5 $\pm$ 1.6 & 20.9 $\pm$ 1.6 & 34.1 $\pm$ 2.7 \\
 & CO(6--5) & (172.73) & (3.0033) & 323 $\pm$ 98 & 14.2 $\pm$ 3.7 & 17.4 $\pm$ 4.5 & 16.4 $\pm$ 4.3 \\ 
 & & (172.53) & (3.0078) & 667 $\pm$ 180 & 22.7 $\pm$ 3.8 & 27.7 $\pm$ 4.6 & 26.3 $\pm$ 4.4 \\
 & CO(8--7) & (230.26) & (3.0033) & 461 $\pm$ 125 & 12.1 $\pm$ 3.0 & 19.7 $\pm$ 4.9 & 7.9 $\pm$ 1.9 \\ 
 & & (230.00) & (3.0078) & 640 $\pm$ 131 & 18.5 $\pm$ 3.1 & 30.1 $\pm$ 5.1 & 12.0 $\pm$ 2.0 \\
 & CO(9--8) & (259.01) & (3.0033) & 321 $\pm$ 35 & 14.2 $\pm$ 1.5 & 26.0 $\pm$ 2.8 & 7.3 $\pm$ 0.8 \\   
 & & (258.72) & (3.0078) & 466 $\pm$ 51 & 18.9 $\pm$ 1.9 & 34.6 $\pm$ 3.5 & 9.7 $\pm$ 1.0 \\
 & CO(10--9) & (287.76) & (3.0033) & 297 $\pm$ 34 & 18.1 $\pm$ 1.9 & 36.9 $\pm$ 3.9 & 7.5 $\pm$ 0.8 \\ 
 & & (287.44) & (3.0078) & 341 $\pm$ 46 & 13.5 $\pm$ 1.8 & 27.5 $\pm$ 3.7 & 5.6 $\pm$ 0.8 \\
 & CO(11--10) & 316.22 $\pm$ 0.06 & 3.0066 $\pm$ 0.0008 & 579 $\pm$ 89 & 24.8 $\pm$ 4.3 & 55.5 $\pm$ 9.6 & 8.5 $\pm$ 1.5 \\
\hline
\end{tabular}
\caption{Continued.}
\label{tab:linefit2}
\end{table*}

\section{Results of the RADEX LVG modeling}

\begin{figure*}
\centering
\includegraphics[width=.4\textwidth]{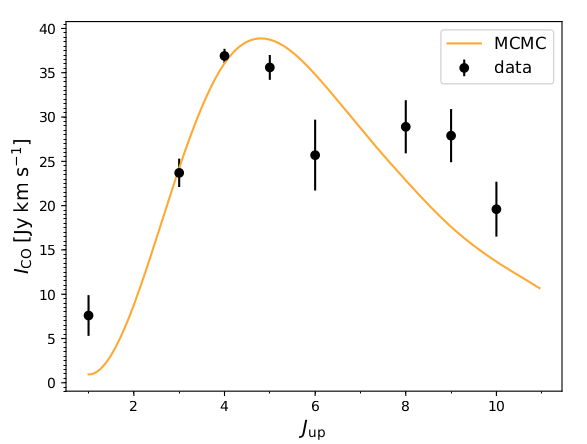}
\includegraphics[width=.4\textwidth]{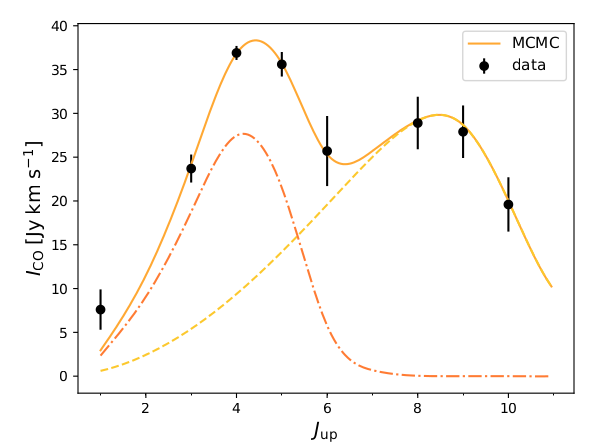}

\includegraphics[width=.4\textwidth]{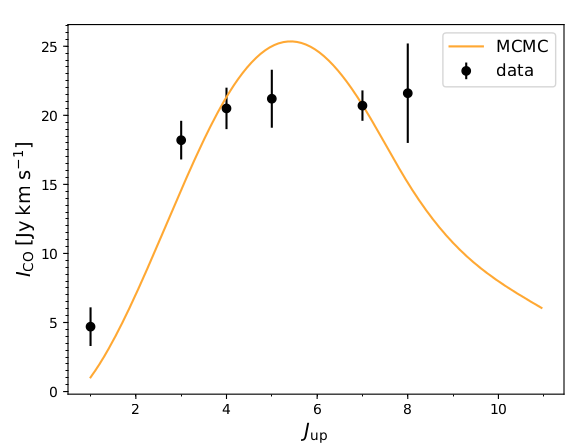}
\includegraphics[width=.4\textwidth]{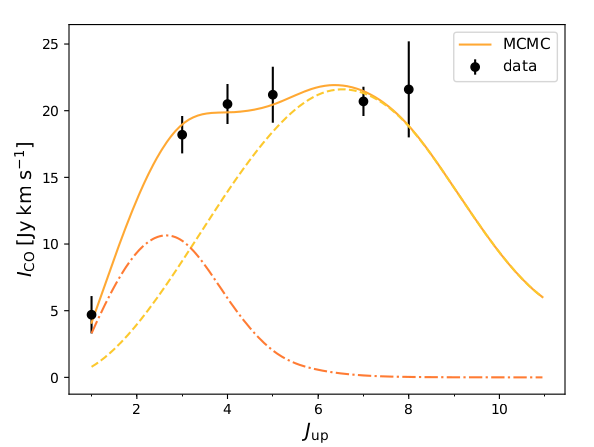}

\includegraphics[width=.4\textwidth]{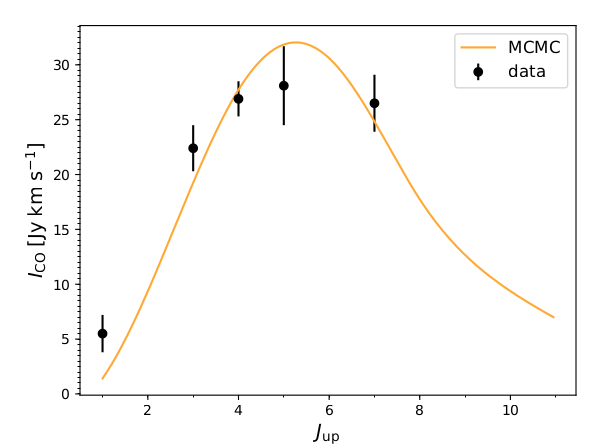}
\includegraphics[width=.4\textwidth]{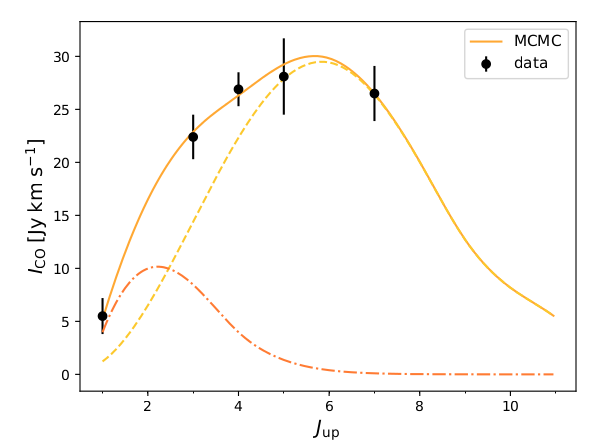}

\includegraphics[width=.4\textwidth]{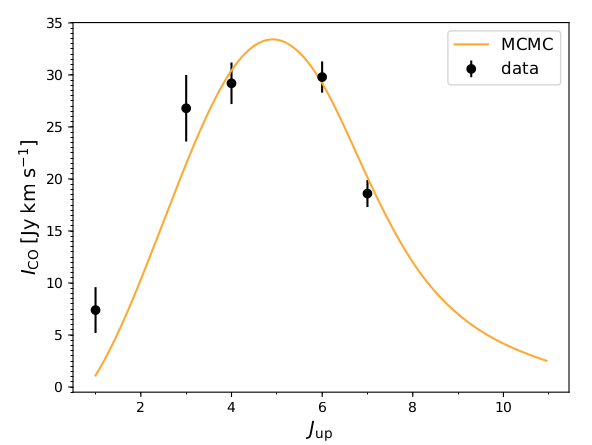}
\includegraphics[width=.4\textwidth]{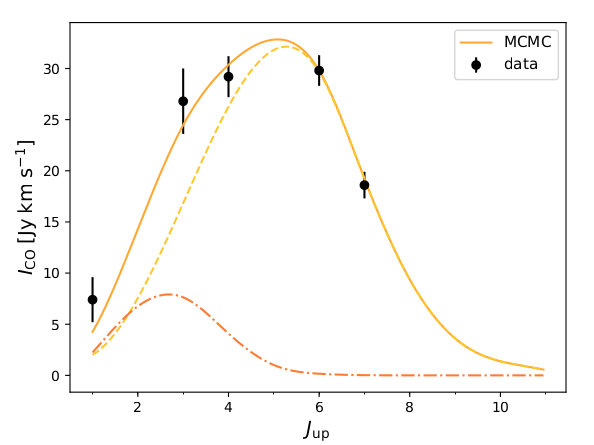}
\caption{Best-fit LVG models using {\tt RADEX} for PLCK\_G092.5+42.9 (top), PLCK\_G113.7+61.0 (center-top), PLCK\_G138.6+62.0 
(center-bottom), and PLCK\_G165.7+67.0 (bottom), for a single gas excitation component (left) and two components (right).}
\label{fig:sled2}
\end{figure*}

\begin{figure*}
\centering
\includegraphics[width=.42\textwidth]{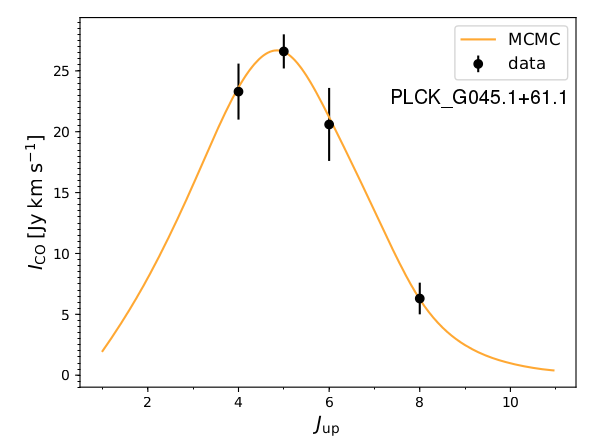}
\includegraphics[width=.42\textwidth]{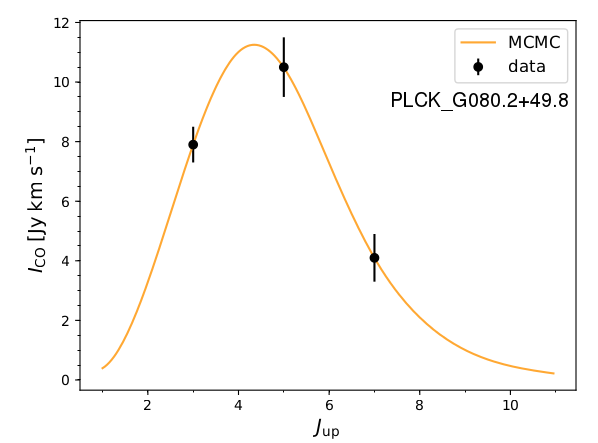}

\includegraphics[width=.42\textwidth]{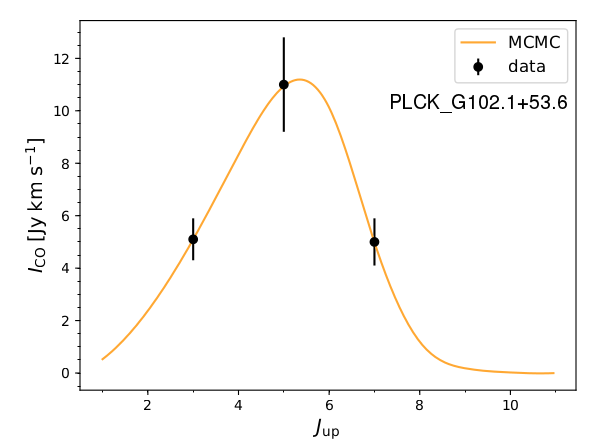}
\includegraphics[width=.42\textwidth]{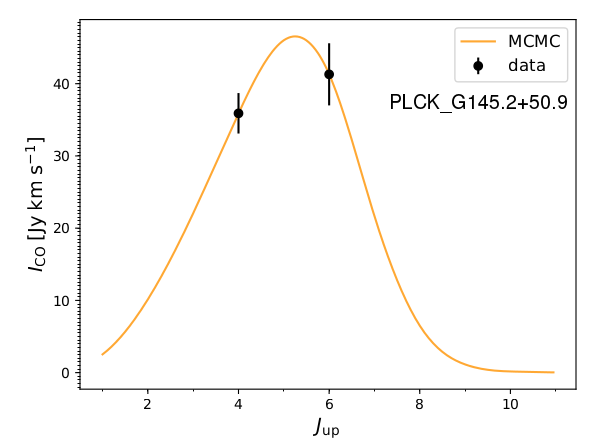}

\includegraphics[width=.42\textwidth]{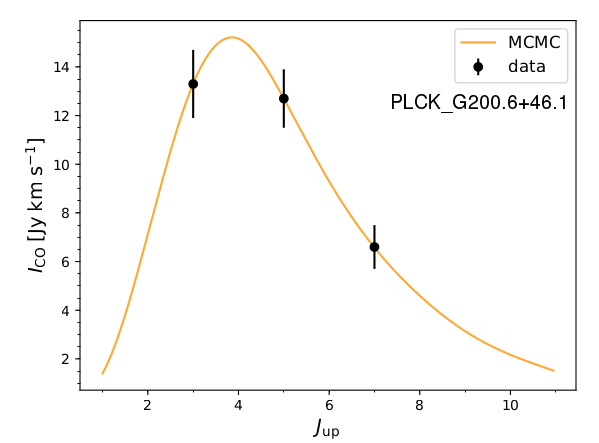}
\includegraphics[width=.42\textwidth]{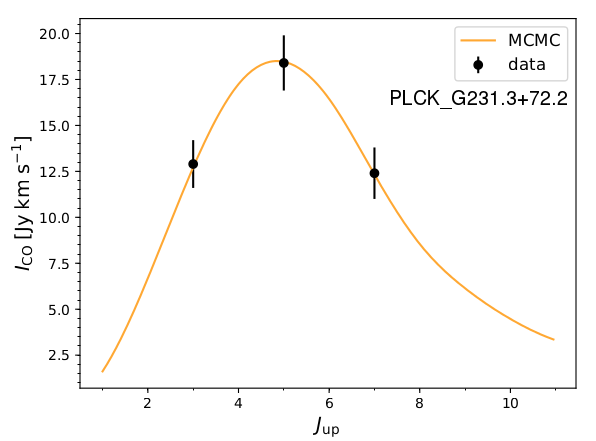}
\caption{Single-component models of the gas excitation in the remaining six {\it Planck}'s dusty GEMS (see Fig.~\ref{fig:g244} 
caption).}
\label{fig:sled3}
\end{figure*}

\begin{figure*}
\centering
\includegraphics[width=.42\textwidth]{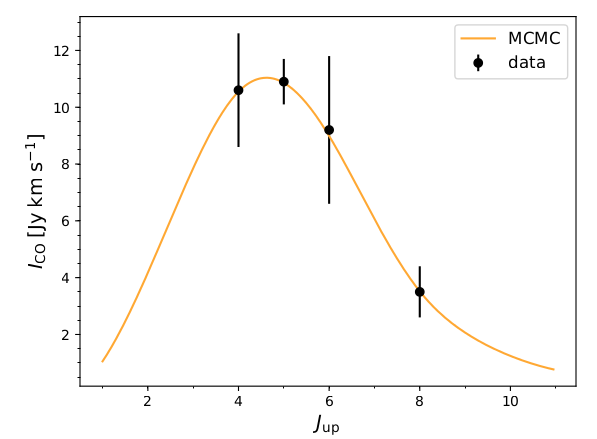}
\includegraphics[width=.42\textwidth]{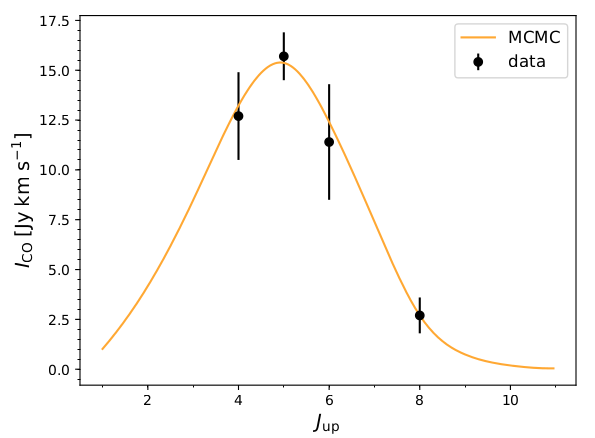}

\includegraphics[width=.42\textwidth]{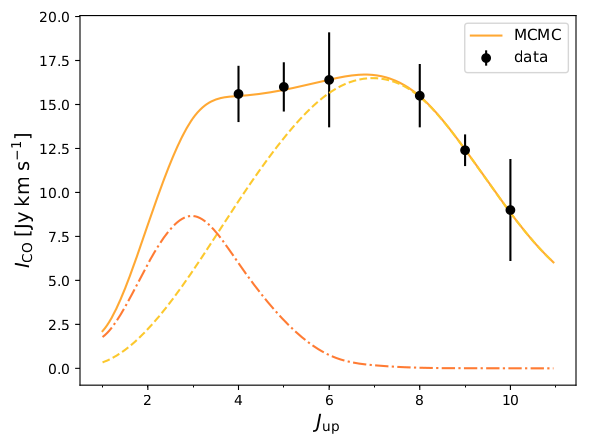}
\includegraphics[width=.42\textwidth]{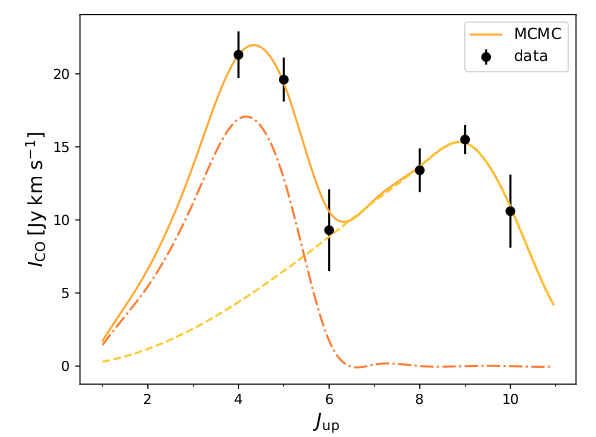}

\includegraphics[width=.42\textwidth]{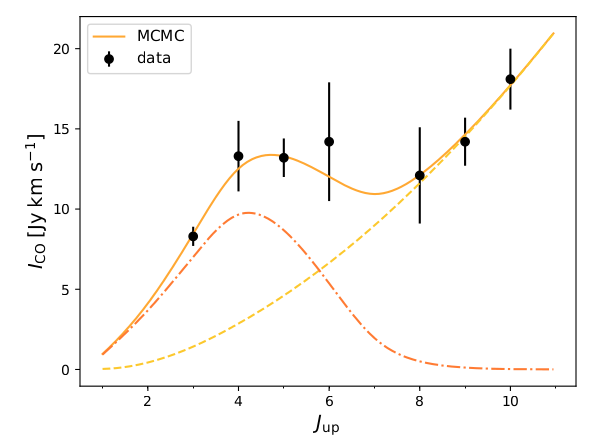}
\includegraphics[width=.42\textwidth]{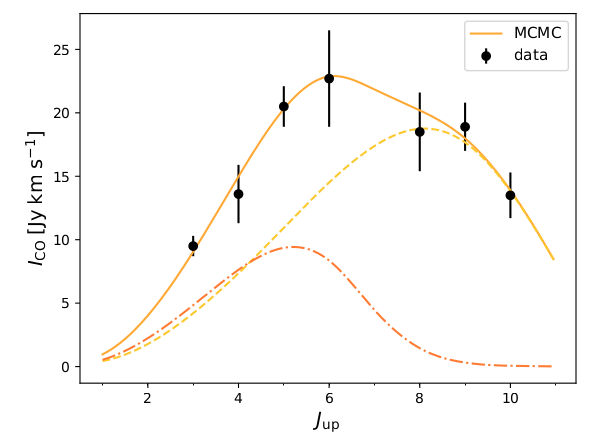}
\caption{Best-fit single or double-component LVG models of the gas excitation in the individual blue (left) and red (right) 
kinematic components of PLCK\_G045.1+61.1 (top), PLCK\_G092.5+42.9 (center), and PLCK\_G244.8+54.9 (bottom). See further 
details in Fig.~\ref{fig:g244} caption.}
\label{fig:sled4}
\end{figure*}

We show the CO SLEDs resulting from our analysis of the gas excitation, using the MCMC implementation of the LVG models
from {\tt RADEX} \citep{vandertak07,yang17}. Figure~\ref{fig:sled2} presents the single and two-component CO excitation models 
of the four {\it Planck}'s dusty GEMS for which we obtain evidence of two distinct gas phases with different properties.
Figure~\ref{fig:sled3} shows the remaining six sources that are conveniently fitted with a single component. The SLEDs plotted 
in Figure~\ref{fig:sled4} are the best-fit LVG models for individual kinematic components deblended from the spectra of 
PLCK\_G045.1+61.1, PLCK\_G092.5+42.9, and PLCK\_G244.8+54.9, using the same number of excitation components as for the 
spectrally-integrated SLEDs.

\end{appendix}

\end{document}